\documentclass[iop,portrait,letter]{emulateapj}

\usepackage[latin1]{inputenc}
\usepackage[english]{babel}
\usepackage{amsmath}
\usepackage{amssymb}
\usepackage{xspace}
\usepackage{natbib}
\usepackage{ifthen}
\newcommand{\salpha}{0.12}                   % Best fit stretch alpha
\newcommand{\cbeta}{2.51}                    % Best fit color beta

\newcommand{\bestfitw}{-0.997}               % Best fit w for a flat Universe with systematics
\newcommand{\bestfitwustaterr}{+0.050}       % Statistical error
\newcommand{\bestfitwdstaterr}{-0.054}       % Systematic error
\newcommand{\bestfitwusyserr}{+0.077}
\newcommand{\bestfitwdsyserr}{-0.082}

\newcommand{\bestfitcurvew}{-1.035} 
\newcommand{\bestfitcurvewustaterr}{+0.055}
\newcommand{\bestfitcurvewdstaterr}{-0.059}
\newcommand{\bestfitcurvewusyserr}{+0.093}
\newcommand{\bestfitcurvewdsyserr}{-0.097}

\newcommand{\totalsne}{557}                  % Total number of SNe in

\newcommand{\wmeancolourgroundnir}{%         % Mean weighted color, two here, one Tonry
  0.01\pm0.07}
\newcommand{\wmeanoffsetgroundnir}{%         % Mean weighted offset
  0.03\pm0.10}   

\newcommand{\wmeancolourNICMOShiz}{0.06\pm0.03} % Riess, above redshift 1.1
\newcommand{\wmeanoffsetNICMOShiz}{-0.01\pm0.06}

\newcommand{\wmeancolourptthreetooneptone}{0.02\pm0.01}

\newcommand{\intdisp}{0.15}                  % Intrinsic disp 
\newcommand{\zs}{z_{\star}}

\newcommand{\saltmodelriessdrops}{two\xspace}

\newboolean{salttwo}
\setboolean{salttwo}{true}

\newcommand{\compname}{Union2\xspace}
\newcommand{\ang}{\AA\xspace}
\newcommand{\etal}{et~al.\xspace}
\newcommand{\scp}{SCP\xspace}
\newcommand{\snls}{SNLS\xspace}
\newcommand{\essence}{ESSENCE\xspace}
\newcommand{\sn}{SN\xspace}

\newcommand{\sne}{SNe\xspace}
\newcommand{\snia}{SN~Ia\xspace}
\newcommand{\sneia}{SNe~Ia\xspace}
\newcommand{\salt}{SALT\xspace}
\newcommand{\saltii}{SALT2\xspace}
\newcommand{\sed}{SED\xspace}

\newcommand{\psf}{PSF\xspace}
\newcommand{\hst}{HST\xspace}
\newcommand{\wfpc}{WFPC2\xspace}
\newcommand{\vlt}{VLT\xspace}
\newcommand{\ntt}{NTT\xspace}
\newcommand{\ctio}{CTIO\xspace}
\newcommand{\cfht}{CFHT\xspace}
\newcommand{\gemini}{Gemini\xspace}
\newcommand{\subaru}{Subaru\xspace}
\newcommand{\focas}{FOCAS\xspace}
\newcommand{\fors}{FORS1\xspace}
\newcommand{\forstwo}{FORS2\xspace}
\newcommand{\niri}{NIRI\xspace}
\newcommand{\cfhk}{CFH12k\xspace}
\newcommand{\susi}{SuSI2\xspace}
\newcommand{\suprimecam}{SuprimeCam\xspace}
\newcommand{\mosaic}{MOSAIC II\xspace}
\newcommand{\isaac}{ISAAC\xspace}
\newcommand{\pc}{PC\xspace}
\newcommand{\acs}{ACS\xspace}
\newcommand{\stsci}{STScI\xspace}
\newcommand{\stsdas}{STSDAS\xspace}
\newcommand{\crrej}{\texttt{crrej}\xspace}

\newcommand{\iraf}{IRAF\xspace}

\newcommand{\ccd}{CCD\xspace}
\newcommand{\cte}{CTE\xspace}
\newcommand{\sdss}{SDSS\xspace}

\newcommand{\rsdss}{r_\mathrm{\sdss}}
\newcommand{\isdss}{i_\mathrm{\sdss}}

\newcommand{\sncw}{2001cw\xspace}
\newcommand{\snhb}{2001hb\xspace}
\newcommand{\sngn}{2001gn\xspace}
\newcommand{\sngq}{2001gq\xspace}
\newcommand{\sngo}{2001go\xspace}
\newcommand{\sngy}{2001gy\xspace}

\newcommand{\snfk}{1999fk\xspace}

\newcommand{\ie}{i.e.\xspace}
\newcommand{\eg}{e.g.\xspace}

\newcommand{\ah}{^\mathrm{h}}
\newcommand{\am}{^\mathrm{m}}
\newcommand{\as}{^\mathrm{s}}
\newcommand{\dd}{^\circ}
\newcommand{\dm}{'}
\newcommand{\ds}{''}

\newcommand{\om}{\Omega_M}
\renewcommand{\ol}{\Omega_\Lambda}

\newcommand{\fov}[2]{#1$\times$#2}
\newcommand{\ccdconf}[2]{#1$\times$#2}
\newcommand{\ccdkconf}[2]{\ccdconf{#1k}{#2k}}
\newcommand{\detector}[1]{\parbox[t]{3.5cm}{\raggedright\footnotesize#1}}

\slugcomment{Submitted to ApJ on December 7th, 2009}
\slugcomment{{\sc Accepted to ApJ:} April 9, 2010} 
\shorttitle{Spectra and Light Curves of Six \sneia and \compname}
\shortauthors{Amanullah \etal}

\begin{document}

\title{Spectra and HST Light Curves of Six Type Ia Supernovae at
  $0.511<z<1.12$ and the \compname Compilation%
  \footnote{Based in part on observations made with the NASA/ESA {\em
      Hubble Space Telescope}, obtained from the data archive at the
    Space Telescope Science Institute (STScI). STScI is operated by
    the Association of Universities for Research in Astronomy (AURA),
    Inc. under the NASA contract NAS 5-26555. The observations are
    associated with programs HST-GO-08585 and HST-GO-09075. Based, in
    part, on observations obtained at the ESO La Silla Paranal
    Observatory (ESO programs 67.A-0361 and 169.A-0382). Based, in
    part, on observations obtained at the Cerro Tololo Inter-American
    Observatory (CTIO), National Optical Astronomy Observatory (NOAO).
    Based on observations obtained at the Canada-France-Hawaii
    Telescope (CFHT). Based, in part, on observations obtained at the
    Gemini Observatory (Gemini programs GN-2001A-SV-19 and
    GN-2002A-Q-31). Based, in part on observations obtained at the
    Subaru Telescope. Based, in part, on data that were obtained at
    the W.M. Keck Observatory. } }
%\title{HST Light Curves of Six Type Ia Supernovae $0.511<z<1.12$ and
%  the Updated Union Compilation}
%\title{Light curves and spectra of six Type Ia supernovae at $0.511<z<1.12$}
%\title{Six well measured Type Ia supernovae at high redshift}
%\title{Six high-redshift type Ia supernovae observed with the Hubble
%  Space Telescope}

\author{%
  R.~Amanullah\altaffilmark{1,2},
  C.~Lidman\altaffilmark{2},
  D.~Rubin\altaffilmark{4,6},
  G.~Aldering\altaffilmark{4},
  P.~Astier\altaffilmark{5},
  K.~Barbary\altaffilmark{4,6},
  M.~S.~Burns\altaffilmark{7},
  A.~Conley\altaffilmark{8},
  K.~S.~Dawson\altaffilmark{24},
  S.~E.~Deustua\altaffilmark{9},
  M.~Doi\altaffilmark{10},
  S.~Fabbro\altaffilmark{11},
  L.~Faccioli\altaffilmark{4,12},
  H.~K.~Fakhouri\altaffilmark{4,6},
  G.~Folatelli\altaffilmark{13},
  A.~S.~Fruchter\altaffilmark{9},
  H.~Furusawa\altaffilmark{26},
  G.~Garavini\altaffilmark{1},
  G.~Goldhaber\altaffilmark{4,6},
  A.~Goobar\altaffilmark{1,2},
  D.~E.~Groom\altaffilmark{4},
  I.~Hook\altaffilmark{14,25},
  D.~A.~Howell\altaffilmark{3,22},
  N.~Kashikawa\altaffilmark{26},
  A.~G.~Kim\altaffilmark{4},
  R.~A.~Knop\altaffilmark{15},
  M.~Kowalski\altaffilmark{23},
  E.~Linder\altaffilmark{12},
  J.~Meyers\altaffilmark{4,6},
  T.~Morokuma\altaffilmark{26,27},
  S.~Nobili\altaffilmark{1,2}, 
  J.~Nordin\altaffilmark{1,2},
  P.~E.~Nugent\altaffilmark{4},
  L.~\"Ostman\altaffilmark{1,2},
  R.~Pain\altaffilmark{5},
  N.~Panagia\altaffilmark{9,17,18},
  S.~Perlmutter\altaffilmark{4,6},
  J.~Raux\altaffilmark{5},
  P.~Ruiz-Lapuente\altaffilmark{16},
  A.~L.~Spadafora\altaffilmark{4},
  M. ~Strovink\altaffilmark{4,6},
  N.~Suzuki\altaffilmark{4},
  L.~Wang\altaffilmark{19},
  W.~M.~Wood-Vasey\altaffilmark{20},
  N.~Yasuda\altaffilmark{21}
  (The Supernova Cosmology Project)}
\altaffiltext{1}{Department of Physics, Stockholm University, 
  Albanova University Center, SE--106~91 Stockholm, Sweden} %
\altaffiltext{2}{The Oskar Klein Centre for Cosmoparticle Physics,
  Department of Physics, AlbaNova, Stockholm University, SE--106~91
  Stockholm, Sweden} %
\altaffiltext{3}{Las Cumbres Observatory Global Telescope Network,
  6740 Cortona Dr.,Suite 102, Goleta, CA 93117, USA} %
\altaffiltext{4}{E. O. Lawrence Berkeley National Laboratory, 1
  Cyclotron Rd., Berkeley, CA 94720, USA } %
\altaffiltext{5}{LPNHE, Université Pierre et Marie Curie Paris 6, 
  Université Paris Diderot Paris 7, CNRS-IN2P3,
  4 place Jussieu, 75005 Paris, France} %
%\altaffiltext{5}{LPNHE, CNRS--IN2P3, University of Paris VI \& VII,
%  Paris, France } %
\altaffiltext{6}{Department of Physics, University of California
  Berkeley, Berkeley, 94720--7300 CA, USA} %
\altaffiltext{7}{Colorado College, 14 East Cache La Poudre St.,
  Colorado Springs, CO 80903} %
\altaffiltext{8}{Center for Astrophysics and Space Astronomy,
  University of Colorado, 389 UCB, Boulder, CO 80309}
%\altaffiltext{8}{Department of Astronomy and Astrophysics, University  
%  of Toronto, 60 St. George St., Toronto, Ontario M5S 3H8, Canada}
\altaffiltext{9}{Space Telescope Science Institute, 3700 San Martin
  Drive, Baltimore, MD 21218, USA} %
\altaffiltext{10}{Institute of Astronomy, School of Science,
  University of Tokyo, Mitaka, Tokyo, 181--0015, Japan} %
\altaffiltext{11}{Department of Physics and Astronomy, University of
  Victoria, PO Box 3055, Victoria, BC V8W 3P6, Canada} %
\altaffiltext{12}{Space Sciences Laboratory, University of California
  Berkeley, Berkeley, CA 94720, USA} %
\altaffiltext{13}{Observatories of the Carnegie Institution of
  Washington, 813 Santa Barbara St., Pasadena, CA 9110, USA} %
\altaffiltext{14}{Sub-Department of Astrophysics, University of
  Oxford, Denys Wilkinson Building, Keble Road, Oxford OX1 3RH, UK} %
\altaffiltext{15}{Meta-Institute of Computational Astronomy,
  \texttt{\url{http://www.physics.drexel.edu/mica}}} %
\altaffiltext{16}{Department of Astronomy, University of Barcelona,
  Barcelona, Spain } %
\altaffiltext{17}{INAF --- Osservatorio Astrofisico di Catania, Via S.
  Sofia 78, I--95123, Catania, Italy} %
\altaffiltext{18}{Supernova Ltd, OYV \#131, Northsound Road, Virgin
  Gorda, British Virgin Islands} %
\altaffiltext{19}{Department of Physics, Texas A\&M University,
  College Station, TX 77843, USA} %
\altaffiltext{20}{Department of Physics and Astronomy, 3941 O'Hara St,
  University of Pittsburgh, Pittsburgh, PA 15260, USA} %
\altaffiltext{21}{Institute for Cosmic Ray Research, University of
  Tokyo, Kashiwa, 277 8582 Japan} %
\altaffiltext{22}{Department of Physics, University of California,
  Santa Barbara, Broida Hall, Mail Code 9530, Santa Barbara, CA
  93106--9530, USA} %
\altaffiltext{23}{Physikalisches Institut, Universit\"at Bonn,
  Nussallee 12, D-53115 Bonn} %
\altaffiltext{24}{Department of Physics and Astronomy, University of
  Utah, Salt Lake City, UT 84112, USA } %
\altaffiltext{25}{INAF - Osservatorio Astronomico di Roma, via
  Frascati 33, 00040 Monteporzio (RM), Italy} %
\altaffiltext{26}{National Astronomical Observatory of Japan, 2-21-1
  Osawa, Mitaka, Tokyo 181-8588, Japan} %
\altaffiltext{27}{Research Fellow of the Japan Society for the
  Promotion of Science} %
%\altaffiltext{28}{Université Paris Diderot, CNRS-IN2P3, Paris, France}

\begin{abstract}
  \noindent %
  We report on work to increase the number of well-measured Type Ia
  supernovae (\sneia) at high redshifts. Light curves, including high
  signal-to-noise HST data, and spectra of six \sneia that were
  discovered during 2001 are presented. Additionally, for the two \sne
  with $z>1$, we present ground-based $J$-band photometry
  from Gemini and the VLT. These are among the most distant
  \sneia for which ground based near-IR observations have been
  obtained. %, having different sensitivity and possibly other
  %systematic uncertainties than space measurements. 
  We add these six \sneia together with other data sets that have
  recently become available in the literature to the Union compilation
  \citep{2008ApJ...686..749K}. We have made a number of refinements to
  the Union analysis chain, the most important ones being the
  refitting of all light curves with the \saltii fitter and an
  improved handling of systematic errors. We call this new
  compilation, consisting of $\totalsne$ supernovae, the \compname
  compilation. The flat concordance $\Lambda$CDM model remains an
  excellent fit to the \compname data with the best fit constant
  equation of state parameter
  $w=\bestfitw^{\bestfitwustaterr}_{\bestfitwdstaterr}\mathrm{(stat)}^{\bestfitwusyserr}_{\bestfitwdsyserr}
  \mathrm{(stat+sys\ together)}$ for a flat universe, or
  $w=\bestfitcurvew^{\bestfitcurvewustaterr}_{\bestfitcurvewdstaterr}\mathrm{(stat)}^{\bestfitcurvewusyserr}_{\bestfitcurvewdsyserr}
  \mathrm{(stat+sys\ together)}$ with curvature. We also present
  improved constraints on $w(z)$. While no significant change in $w$
  with redshift is detected, there is still considerable room for
  evolution in $w$. The strength of the constraints depend strongly on
  redshift. In particular, at $z \gtrsim 1$, the existence and nature
  of dark energy are only weakly constrained by the data.
%   The value of $w$ at $z \gtrsim 1$ is well-enough constrained
%    to suggest the existence of dark energy.
\end{abstract}

\section{Introduction}
Type Ia supernovae (\sneia) are an excellent tool for probing the
expansion history of the Universe. %
% It was not long after it was discovered that these objects could be
% used as standard candles that 
About a decade ago, combined observations of nearby and distant \sneia
led to the discovery of the accelerating universe
\citep{1998Natur.391...51P,1998ApJ...509...74G,1998ApJ...507...46S,
  1998AJ....116.1009R,1999ApJ...517..565P}.

Following these pioneering efforts, the combined work of several
different teams during the past decade has provided an impressive
increase in both the total number of \sneia and the quality of the
individual measurements. At the high redshift end ($z \gtrsim 1$), the
Hubble Space Telescope (\hst) has played a key role. It has
successfully been used for high-precision optical and infrared
follow-up of \sne discovered from the ground
\citep{2003ApJ...598..102K,2003ApJ...594....1T,2004ApJ...602..571B,%
  2009ApJ...700.1415N}, and, by using the Advanced Camera for Surveys
(\acs), to carry out both search and follow-up from space
\citep{2004ApJ...607..665R,2007ApJ...659...98R,2008ApJ...673..981K,dawson:2009}.
At the same time, several large-scale ground-based projects have been
populating the Hubble Diagram at lower redshifts. The Katzman
Automatic Imaging Telescope \citep{2001ASPC..246..121F}, the Nearby
Supernova Factory \citep{2006NewAR..50..436C}, the Center for
Astrophysics \sn group \citep{2009ApJ...700..331H}, the Carnegie
Supernova Project \citep{2006PASP..118....2H,2010AJ....139..120F}, and
the Palomar Transient Factory \citep{2009PASP..121.1395L} are
conducting searches and/or follow-up for \sne at low redshifts ($z <
0.1$). The \sn Legacy Survey (\snls) \citep{2006AA...447...31A} and
\essence \citep{2007ApJ...666..674M,2007ApJ...666..694W} are building
\sn samples over the redshift interval $0.3 < z < 1.0$, and the \sdss
\sn Survey \citep{2008AJ....136.2306H,2009ApJS..185...32K} is building
a \sn sample over the redshift interval $0.1 < z < 0.3$, a redshift
interval that has been relatively neglected in the past. These
projects have discovered $\sim700$~well-measured \sne. The number of
well-measured \sne beyond $z\sim1$ is approximately 20 and is
comparatively small.
% Defined as z>1 and surviving the Kowalski et al. cuts.

\citet{2008ApJ...686..749K} (hereafter K08) provided a framework to
analyze these and future datasets in a homogeneous manner and created
a compilation, called the ``Union'' \sneia compilation, of what was
then the world's SN data sets. Recently, \citet{2009ApJ...700.1097H}
(hereafter H09) added a significant number of nearby \sne to a subset
of the ``Union'' set to create a new compilation, and similarly the
\sdss \sn survey \citep{2009ApJS..185...32K} (hereafter KS09) carried
out an analysis of a compilation including their large
intermediate-$z$ data set \citep{2008AJ....136.2306H}. When combined
with baryon acoustic oscillations \citep{2005ApJ...633..560E}, the H09
compilation leads to an estimate of the equation of state parameter
that is consistent with a cosmological constant while KS09 get
significantly different results depending on which light curve fitter
they use.

An important role for \sneia beyond $z \sim 1$, in addition to
constraining the time evolution of $w$, is their power to constrain
astrophysical effects that would systematically bias cosmological
fits. Most evolutionary effects are expected to monotonically change
with redshift and are not expected to mimic dark energy over the
entire redshift interval over which SNe Ia can be observed.
Evolutionary effects might also have additional detectable
consequences, such as a shift in the average color of \sneia or a
change in the intrinsic dispersion about the best fit cosmology.

Interestingly, the most distant \sneia in the Union compilation
(defined here as \sneia with $z\gtrsim 1.1$) are almost all redder
than the average color of \sneia over the redshift interval $0.3 < z <
1.1$. The result is unexpected as bluer \sneia at lower redshifts are
also brighter \citep{1998A&A...331..815T,2005A&A...443..781G} and
should therefore be easier to detect at higher redshifts. Possible
explanations for the redder than average colors of very distant \sneia
range from the technical, such as an incomplete understanding of the
calibration of the instruments used for obtaining the high redshift
data, to the more astrophysically interesting, such as a real lack of
bluer \sne at high redshifts.
% We discuss why they could be redder, they still seem to land on the
% the Hubble diagram, so I do not think we should mention that this
% could have anything to do with evolution of the color-mag relation.
%
%% or evolution in the relationships used to standardize observed \sne
%% magnitudes.

The underlying assumption in using \sneia in cosmology is that the
luminosity of both near and distant events can be standardized with
the same luminosity versus color and luminosity versus light curve
shape relationships. While drifts in \snia populations are expected
from a combination of the preferential discovery of brighter \sneia
and changes in the mix of galaxy types with redshift
\citep{2007ApJ...667L..37H} --- effects that will affect different
surveys by differing amounts --- a lack of evolution in these
relationships with redshift has not been convincingly demonstrated
given the precision of current data sets. This assumption needs to be
continuously examined as larger and more precise \snia data sets
become available.

% 
%% This is not true since they are still red after these two technical
%% issues have been corrected.
%% 
% As we will discuss later in the paper, a combination of two technical
% issues made these SNe redder that they really were.

%% Compute the mean and standard deviation of the offset. A common
%% denominator is the NICMOS calibration. Mention this and the general
%% opt-IR offset.

% Clearly, the reasons for the paucity of blue SNe at $z\gtrsim 1.1$ is
% needed if \sne are to play an important role in future dark energy
% experiments. The first step is to verify the veracity of the effect,
% and this requires additional and independently measured sets of very
% distant \sneia.

In this paper, we report on work to increase the number of
well-measured distant \sneia by presenting \sneia that were discovered
in ground based searches during 2001 and then followed with WFPC2 on
\hst. Two of the new \sneia are at $z \sim 1.1$ and have high-quality
ground-based infrared observations that were obtained with \isaac on
the \vlt and \niri on \gemini.  
%These data can be used to shed light on the reasons for the unusual
%red colors of \sneia at $z \sim 1.1$, because they do not rely on the
%calibration of NICMOS, as all other \sneia at $z \sim 1.1$ do. 
This paper is the first paper in a series of papers that will provide
a comparable sample of $z>1$ \sneia to the \sne now available in the
literature. The \sneia in this series of papers were discovered in
2001 (this paper), 2002 (Suzuki \etal in preparation) and from 2005 to
2006 during the Supernova Cosmology Project (\scp) cluster survey
\citep{dawson:2009}.

%Currently, no single \snia survey contributes more than the whole, so
%combining the results from all \snia surveys still leads to an
%improvement in the constraints that can be set on the properties of
%dark energy. In addition to performing a homogeneous analysis of the
%world sample of \snia, K08 also studied the tension between different
%surveys. It is a recipe that we will follow in this paper.

%In this paper we continue the work of building a high precision Hubble
%Diagram at high redshift, by presenting six \sneia that were
%discovered in ground based searches and then followed with \hst. We
%also present ground based infrared observations of the \sneia at
%$z>1$. We will publish additional \sneia at $z>1$ discovered by the
%\scp in \citet{dawson:2009} and Suzuki et al. (in preparation).

%Not only is it important to discover \sneia at high redshifts, but
%also to be able to do obtain high precision follow-up photometry in
%multiple filters. The statistical weight of each individual \sn
%depends on how accurate its maximum magnitude, light curve shape and
%color are measured. Currently, the color uncertainty is the dominating
%the error budget when these parameters are combined to form the
%quantity that is used for probing cosmology.

The paper is organized as follows. In Section \ref{sec:search}, we
describe the \sn search and the spectroscopic confirmation, while
Sections \ref{sec:followup}, \ref{sec:groundphoto} and
\ref{sec:spacephoto} contain a description of the follow-up imaging
and the \sn photometry. The light curve fitting is described in
Section \ref{sec:lcfitting}. In Section \ref{sec:union} we update the
K08 analysis both by adding new data and by improving the analysis
chain. The paper ends with a discussion and a summary.

%The paper is outlined as follows: In Section~\ref{sec:search} we
%describe the search and spectroscopic confirmation, while
%Section~\ref{sec:followup} covers the follow-up data.
%Sections~\ref{sec:groundphoto} and~\ref{sec:spacephoto} present the
%ground and space based photometry respectively. The light curve
%fitting is described in section~\ref{sec:lcfitting} and added to the
%K08 sample in~\ref{sec:union}. In Section~\ref{sec:salt2} we discuss
%the uncertainties from light curve fitting, and the paper is concluded
%in section~\ref{sec:conclusions}. % and~\ref{sec:summary}.

\section{Search, discovery and spectroscopic confirmation}
\label{sec:search}

The \sne were discovered during two separate high-redshift \sn search
campaigns that were conducted during the Northern Spring of 2001. The
first campaign (hereafter Spring 2001) consisted of searches with the
\cfhk \citep{cuillandre:2000} camera on the 3.4~m Canada-France-Hawaii
Telescope (\cfht) and the \mosaic \citep{1998SPIE.3355..577M} camera
on the 4.0~m Cerro Tololo Inter-American Observatory (\ctio) Blanco
telescope.
%The \cfhk search was carried out in the
%VIRMOS deep imaging survey 1400+05 field \citep{2004A&A...417..839L}.
The second campaign (hereafter Subaru 2001) was done with \suprimecam
\citep{2002PASJ...54..833M} on the 8.2~m Subaru telescope. All searches were
``classical'' searches
\citep{1995ApJ...440L..41P,1997ApJ...483..565P}, i.e., the survey
region was observed twice with a delay of approximately one month
between the two observations, and the two epochs were then analyzed to
find transients. Details of the search campaigns can be found in
\citet{2005A&A...430..843L} and \citet{2010PASJ...62...19M}.

The Spring 2001 data were processed to find transient objects and the
most promising candidates were given an internal \scp name and a
priority. The priority was based on a number of factors: the
significance of the detection, the relative increase in the
brightness, the distance from the center of the apparent host, the
brightness of the candidate and the quality of the subtraction. Note
however that these factors were not applied independently of each
other, but the priorities were rather based on a combination of
factors. For example, candidates on core were only avoided if they had
a small relative brightness increase over the span of 1~month. The AGN
structure function shows that AGNs rarely have strong changes over
1~month.

The candidates discovered in the Spring 2001 campaign were distributed
to teams working at Keck and Paranal observatories for spectroscopic
confirmation. The distribution was based on the likely redshifts.
Candidates that were likely to be SNe~Ia at $z\gtrsim 0.7$ were sent
to Keck, while candidates that were thought to be nearer
were sent to the \vlt. In later years, when \forstwo was upgraded with
a CCD with increased red sensitivity, the most distant candidates
would also be sent to the VLT. All Subaru 2001 candidates were sent to
\focas (Faint Object Camera And Spectrograph) on Subaru for
spectroscopic confirmation \citep{2010PASJ...62...19M}. 

In total, four instruments (\fors on the ESO \vlt, ESI and LRIS on
Keck, and \focas on \subaru) were used to determine redshifts and to
spectroscopically confirm the SN type. The dates of the spectroscopic
runs are listed in Table \ref{tab:spectroruns} and the observations of
individual candidates are listed in Table \ref{tab:SpecObservations}
%while the spectra are presented in Appendix~\ref{sec:spectra}.

\begin{figure*}[p]
  \centering
  \includegraphics[width=\textwidth]{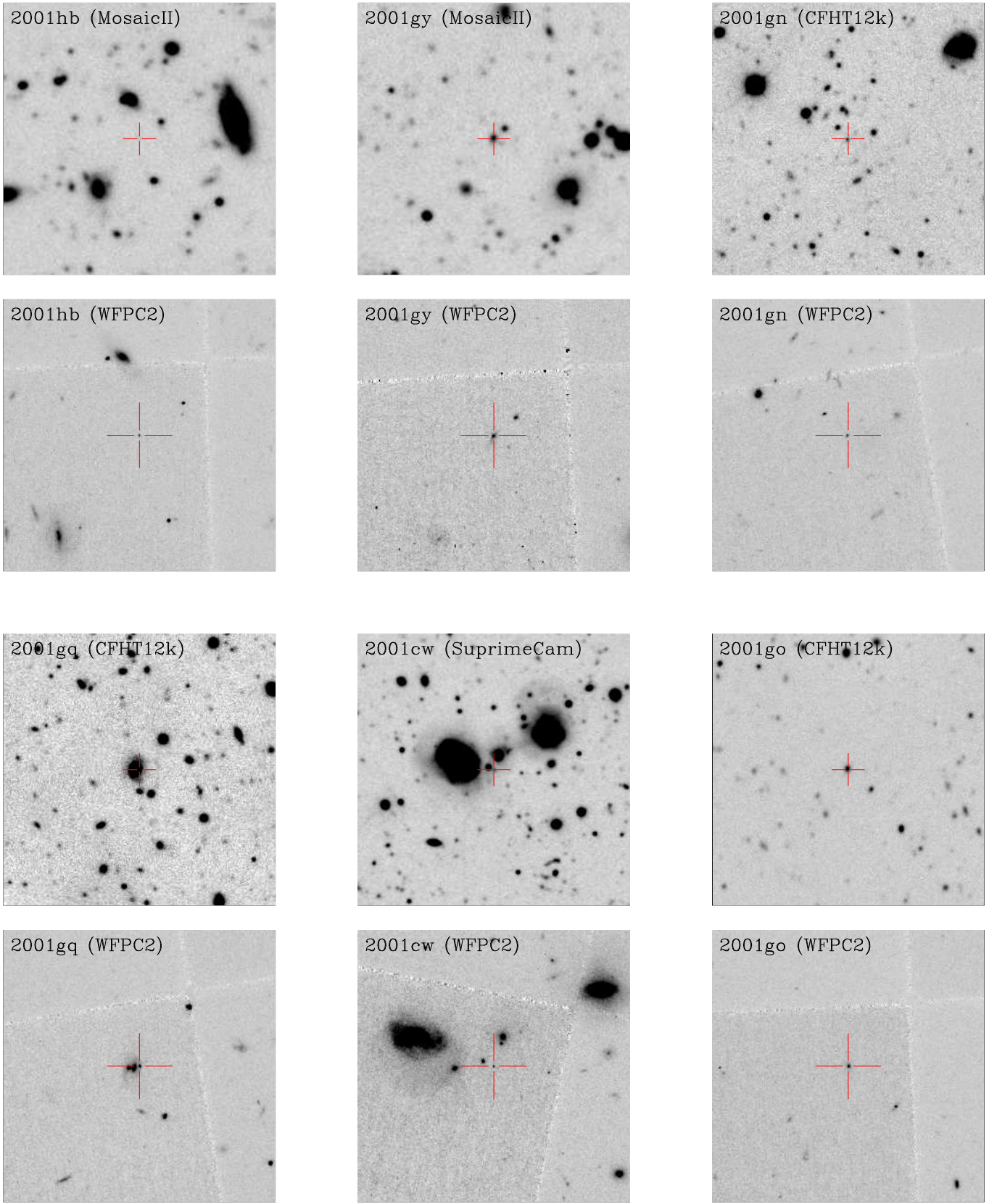}
  \caption{\sn finding charts. North is up and East is to the
    left. The \sne are marked with red cross hairs. The images have
    been created by stacking all $I$-band data taken with the search
    instrument and all F814W data obtained with \wfpc. The patch
    widths are $1'$ and $0.5'$ for the ground-based and \hst images
    respectively.%
    \label{fig:FC}}
\end{figure*}

Only those candidates that were confirmed as \sneia were then
scheduled for follow-up observations from the ground and with \hst. In
total, six \sne were sent for HST follow-up: one from the Subaru 2001
campaign and five from the Spring 2001 campaign. The \sne are listed
in Tables~\ref{tab:SpecObservations} and \ref{tab:SpecResults}.
Finding charts are provided in Figure~\ref{fig:FC}.

%Distant SNe~Ia are faint, and much of the flux lies beyond the
%8000\ang region where the efficiency of CCD detectors is generally low
%and the subtraction of the bright OH lines is problematic for
%spectroscopy. Telescopes with large collecting areas and instruments
%with red sensitive CCDs are required. 

\begin{deluxetable*}{lllll}
  \tablecaption{List of the instruments and
    telescopes used for spectroscopy.
    \label{tab:spectroruns}}
  \tablehead{
    \colhead{Search}      & 
    \colhead{Instr./Tel.} & 
    \colhead{Detector} & 
    \colhead{Resolution} & 
    \colhead{Observing dates}}
  \startdata
Spring 2001 & \fors/Antu           & \detector{Tektronix\ \ccdkconf{2}{2} CCD} & 500  &   2001 April 21 -- 22          \\
Spring 2001 & LRIS/Keck I          & \detector{Tektronix\ \ccdkconf{2}{2} CCD} & 850  &   2001 April 20               \\
Spring 2001 & ESI/Keck II          & \detector{MIT-LL\ \ccdkconf{2}{4}  CCD}   & 5000 &   2001 April 21 -- 24          \\
Subaru 2001 & FOCAS/Subaru         & \detector{SITe\ \ccdkconf{2}{4} CCD}      & 1000 &   2001 May 26 -- 27            \\
\enddata
\end{deluxetable*}

\begin{deluxetable*}{llllllll}
\tabletypesize{\footnotesize}
\tablecaption{Summary of the spectroscopic
  observations.  For completeness, the observations that are
  reported in \citet{2005A&A...430..843L} and
  \citet{2010PASJ...62...19M} are also included. The Galactic
  extinction, $E(B-V)$, from \citet{1998ApJ...500..525S},
  is presented together with the coordinates for each \sn.
  \label{tab:SpecObservations}}
\tablehead{
& & \multicolumn{1}{c}{$\alpha (2000)$} & \multicolumn{1}{c}{$\delta (2000)$} & 
\multicolumn{1}{c}{$E(B-V)$}        & \multicolumn{1}{c}{MJD} &
\multicolumn{1}{c}{Instrument and}  & 
\multicolumn{1}{c}{Exp.}\\
\multicolumn{1}{c}{IAU name}        & \multicolumn{1}{c}{Search} &
                                    &                            &
                                    & \multicolumn{1}{c}{(days)} &
\multicolumn{1}{c}{telescope}       &
\multicolumn{1}{c}{(s)}}
\startdata
SN~2001cw  & Subaru 01 & $15\ah23\am06\as.3$ & $+29\dd39\dm32\ds$ & 0.024 & 52056.6 & FOCAS/Subaru\tablenotemark{a}      &  4200          \\
SN~2001gn  & Spring 01 & $14\ah01\am59\as.9$ & $+05\dd05\dm00\ds$ & 0.028 & 52023.1 & ESI/Keck II\tablenotemark{b}       &  9700          \\
SN~2001go  & Spring 01 & $14\ah02\am00\as.9$ & $+05\dd00\dm59\ds$ & 0.027 & 52021.3 & \fors/Antu\tablenotemark{a}        &  2400          \\
SN~2001gq  & Spring 01 & $14\ah01\am51\as.4$ & $+04\dd53\dm12\ds$ & 0.027 & 52020.3 & LRIS/Keck\tablenotemark{a}         &  3600          \\
SN~2001gy  & Spring 01 & $13\ah57\am04\as.5$ & $+04\dd31\dm00\ds$ & 0.030 & 52021.3 & \fors/Antu\tablenotemark{a}        &  2400          \\
SN~2001hb  & Spring 01 & $13\ah57\am11\as.9$ & $+04\dd20\dm27\ds$ & 0.032 & 52024.3 & ESI/Keck II\tablenotemark{b}       &  3600          \\
\enddata
\tablenotetext{a}{\mbox{Long slit}}
\tablenotetext{b}{\mbox{Echellette}}
\end{deluxetable*}

Here, we describe the analysis of data that were taken with ESI and
LRIS. The analyses of the spectra taken with \fors and FOCAS
(SN~2001cw, SN~2001go and SN~2001gy) are described in
\citet{2005A&A...430..843L} and \citet{2010PASJ...62...19M}
respectively. The spectra of these SN are shown in these papers and
will not be repeated here.

\subsection{ESI}
The two highest redshift candidates, \sngn and \snhb, were observed
with the echelette mode of ESI \citep{2002PASP..114..851S}. A spectrum
taken with the echellette mode of ESI and the $20''$ slit covers the
0.39\,\micron\ to 1.09\,\micron\ wavelength range and is spread over
10 orders ranging in dispersion from 0.16\,\ang\ per pixel in the
bluest order (order 15) to 0.30\,\ang\ per pixel in the reddest (order
6). The detector is a MIT-Lincoln Labs 2048 x 4096 CCD with
15\,\micron\ pixels. The slit width was set according to the seeing
conditions and varied from $0.7''$ to $1.0''$, which corresponds to a
spectral resolution of $R\sim5000$. Compared to spectra obtained with
low-resolution spectrographs, such as FORS1, FORS2, LRIS and FOCAS,
the fraction of the ESI spectrum that is free of bright night sky
lines from the Earth's atmosphere is much greater. This allows one to
de-weight the low signal-to-noise regions that overlap these bright
lines when binning the spectra, a method that becomes inefficient with
low-resolution spectra as too much of the spectra are de-weighted.
Another advantage of ESI was that the MIT-LL CCD offered high quantum
efficiency at red wavelengths and significantly reduced fringing
compared to conventional backside-illuminated CCDs.

The data were reduced in a standard manner. The bias was removed by
subtracting the median of the pixel values in the overscan regions,
the relative gains of the two amplifiers were normalized by
multiplying one of the outputs with a constant and the data were flat
fielded with internal lamps. When extracting SN spectra, a bright star
was used to define the trace along each order, and the spectrum of the
\sn was used to define the center of the aperture. Once extracted, the
10 orders were wavelength calibrated (using internal arc lamps and
cross-checking the result with bright OH lines), flux calibrated and
stitched together to form a continuous spectrum.

To reduce the impact that residuals from bright OH lines have on
determining the redshift and classifying the candidate, the spectrum
was weighted according to the inverse square of the error spectrum and
then rebinned by a factor of 20, from 0.19\,\ang/pixel to
3.8\,\ang/pixel. The binning was chosen so that features from the host
were not lost.

The reduced spectra of \sngn and \snhb are presented in
Figures~\ref{fig:01gn} and~\ref{fig:01hb} respectively.

\subsection{LRIS}
The LRIS \citep{1995PASP..107..375O} data were taken with the 400/8500
grating and the GG495 order sorting filter and were reduced in a
standard manner. The bias was removed with a bias frame, the
pixel-to-pixel variations were normalized with flats that were taken
with internal lamps and the background was subtracted by fitting a low
order polynomial along detector columns. The fringes that were not
removed by the flat were removed with a fringe map, which was the
median of the sky subtracted data that was then smoothed with a 5x5
pixel box. The spectra were then combined, extracted, and calibrated
in wavelength and flux.

The reduced spectrum of \sngq is shown in Figure~\ref{fig:01gq}.

\subsection{Spectral fitting and supernova typing}
Light from the host and the \sn are often strongly blended in the
spectra of high redshift SN. To separate the two, we followed the
spectral fitting technique described in \citet{2005ApJ...634.1190H}.

To classify the \sne, we used the classification scheme described in
\citet{2005A&A...430..843L} and added to it the confidence index (CI)
described in \citet{2005ApJ...634.1190H}. In the
\citet{2005A&A...430..843L} scheme, an object is classified as a \snia
if the Si~II features at 4000\,\ang and/or 6150\,\ang or the S II W
feature can be clearly identified in the spectrum or if the spectrum
is best fit with the spectra of nearby \snia and other types do not
provide a good fit. We qualify the classification with the keys ``Si
II'' or ``SF'' in column 3 of Table~\ref{tab:SpecResults} depending on
whether the classification was done by identifying features or by
using the fit. In the scheme described in \citet{2005ApJ...634.1190H},
these \sne would have a CI of 5 and 4, respectively.

Less secure candidates are classified as Ia*. The asterisk indicates
some degree of uncertainty. Usually, this means that we can find an
acceptable match with nearby SNe Ia; however, other types, such as SNe Ibc,
also result in acceptable matches. These SNe have a confidence index
of 3.

Redshifts based on the host have an accuracy that is better than
0.001, and are, therefore, quoted to three decimal places. Redshifts
based on the fit are less accurate.  For completeness, the redshifts
and classifications reported in \citet{2005A&A...430..843L} and
\citet{2010PASJ...62...19M} are also included.

The agreement between the phase, $t_\mathrm{Spec}$, of the best fit
template and the corresponding phase, $t_\mathrm{LC}$, obtained from
the light curve fit is also shown in Table~\ref{tab:SpecResults}. The
weighted average difference for all six spectra is $\Delta t =
-0.4$~days with a dispersion of 2.0~days. The dispersion is similar in
magnitude to that found in other surveys
\citep{2005AJ....130.2788H,2008ApJ...684...68F}.

\begin{deluxetable*}{lccllclllr}
  \tabletypesize{\footnotesize}
  \tablewidth{0pc}
  \tablecaption{Classifications, redshifts,
    confidence index \citep{2005ApJ...634.1190H} and matches to nearby
    \sn spectra (or templates). Unless indicated otherwise, the
    uncertainty in the redshift is 0.001. Here $t_\mathrm{Spec}$ shows
    the spectroscopic phase relative to maximum light in the $B$-band
    of the template, $t_\mathrm{LC}$ shows the rest frame phase of the
    observation relative to the fitted $B$-band maximum
    ($\mathrm{MJD}_\mathrm{max}$ in Table~\ref{tb:salt2results}), and
    $\Delta t$ is the difference. The confidence index (CI) and
    spectroscopic classification are explained in detail in the text.
    We also specify whether the classification was done from the fit
    (``SF'') or if ``Si II'' could be identified.
    \label{tab:SpecResults}}
  \tablehead{
    \colhead{} & \colhead{} &
    \multicolumn{2}{c}{Spectroscopic}  & \colhead{} & 
    \multicolumn{1}{c}{Template}  & 
    \multicolumn{1}{c}{$t_\mathrm{Spec}$}  & 
    \multicolumn{1}{c}{$t_\mathrm{LC}$} & 
    \multicolumn{1}{c}{$\Delta t$} \\
    IAU name      &  CI  & 
    \multicolumn{2}{c}{classification} & 
    \multicolumn{1}{c}{Redshift} & 
    \multicolumn{1}{c}{match}     &
    \multicolumn{1}{c}{(days)}    &
    \multicolumn{1}{c}{(days)}    &
    \multicolumn{1}{c}{(days)}}
\startdata
%% automatically generated by 'specdates.py'
%%
\sncw & 3 & Ia* & SF    & $0.95\pm0.01$ & 1989B  & -5 & $-5.1\pm0.4$ & +0.1\\
\sngn & 3 & Ia* & SF    & $1.124$       & 1990N  & -7 & $-3.4\pm0.1$ & -3.6\\
\sngo & 5 & Ia  & Si~II & $0.552$       & 1992A  & +5 & $+6.5\pm0.2$ & -1.5\\
\sngq & 3 & Ia* & SF    & $0.672$       & 1999bp & -2 & $-5.1\pm0.2$ & +3.1\\
\sngy & 5 & Ia  & Si~II & $0.511$       & 1990N  & -7 & $-6.9\pm0.2$ & -0.1\\
\snhb & 3 & Ia* & SF    & $1.03\pm0.02$ & 1989B  & -7 & $-6.8\pm0.5$ & -0.2\\

\enddata
\end{deluxetable*}

%
%___________________________________________________________________________

\section{Photometric observations}
\label{sec:followup}
A total of nine different instruments, listed in
Table~\ref{tab:photinst}, were used for the photometric follow-up of
the \sne described in this work.
\begin{deluxetable*}{lllll}
  \tablewidth{0pc}
  \tablecaption{Instruments and telescopes used for photometric
    follow-up. For \wfpc we only list the properties of the PC camera
    which was always used for targeting the \sne.
    \label{tab:photinst}}
  \tablehead{
    \colhead{Tel./Instr.} & \colhead{Scale} & \colhead{FOV} &
    \colhead{Detectors} & \colhead{Search}\\
    \colhead{} & \colhead{($''$/px)} & \colhead{($'$)} & 
    \colhead{} & \colhead{}}
\startdata
\cfht/\cfhk       & 0.206       & \fov{42}{28}   & 
\detector{12 MIT \ccdkconf{2}{4} CCD} & Spring 2001\\[4ex]
\ctio/\mosaic     & 0.27        & \fov{36}{36}   &
\detector{8 SITe \ccdkconf{2}{2} CCD} & Spring 2001\\[4ex]
\vlt/\fors       & 0.20        & \fov{6.8}{6.8} &
\detector{1 Tektronix \ccdkconf{2}{2} CCD} & Spring 2001\\[4ex]
\vlt/\isaac      & 0.1484      & \fov{2.5}{2.5} &
\detector{Hawaii \ccdkconf{1}{1} HgCdTe array} & Spring 2001\\[4ex]
\ntt/\susi       & 0.08        & \fov{5.5}{5.5} &
\detector{2 EEV \ccdkconf{2}{4} CCD}  & Spring 2001\\[4ex]
\gemini/\niri       & 0.1171      & \fov{2.0}{2.0} & 
\detector{Aladdin \ccdkconf{1}{1} InSb array} & Spring 2001\\[4ex]
\hst/\acs        & 0.05        & \fov{2.4}{2.4} &
\detector{2 SITe \ccdkconf{2}{4} CCD} & Spring 2001\\[4ex]
\hst/\wfpc(PC)      & 0.046   & \fov{0.61}{0.61} &
\detector{1 Loral Aero\-space \ccdconf{800}{800} CCD} & Both\\[4ex]
\subaru/\suprimecam & 0.20        & \fov{34}{27}   &
\detector{10 MIT/LL \ccdkconf{2}{4} CCD} & Subaru 2001\\[4ex]
\enddata
\end{deluxetable*}

All observations are listed in Table~\ref{tb:photometry}. Here the
Modified Julian Date (MJD) is the weighted average of all images taken
during a given night except for the NIR data where data taken over
several nights were combined. We do not report the MJD for combined
reference images that were taken over several months.

\subsection{Ground-based optical observations and reductions}

\begin{figure*}[tbhp]
  \centering
  \includegraphics[height=\textwidth,angle=270]{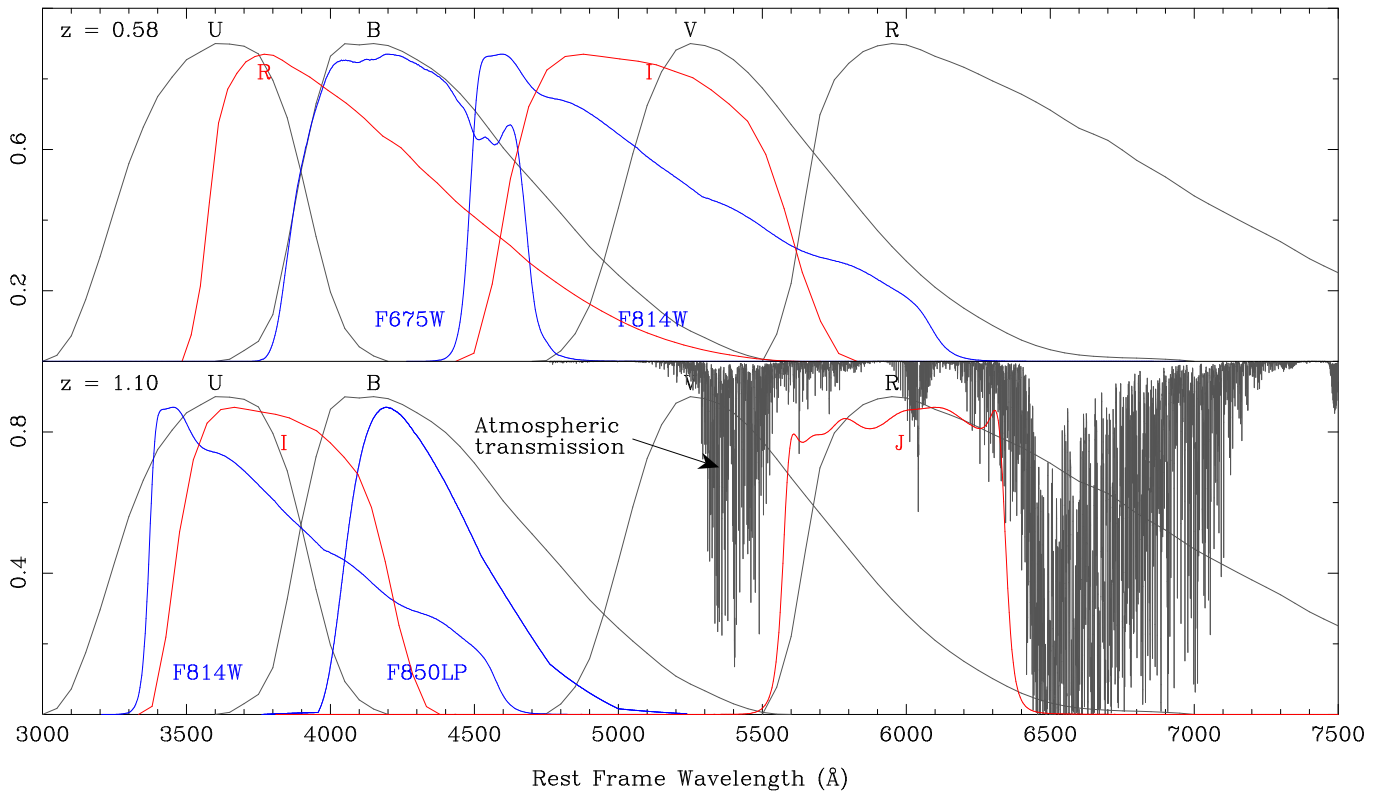}
  \caption{Illustration of filters (colored) used in the \sn campaign
    together with the rest frame Bessel filters (gray). The red
    (ground-based) and blue (\hst) filters were used for the
    observations of the three \sne below ({\bf upper panel}) and above
    ({\bf lower panel}) $z=0.7$ respectively.
    The filters have been blueshifted to the mean of the two sub
    samples.
    Note that the Bessel filters shown here differ slightly from the
    corresponding filters used at the different telescopes.
    Not shown in the figure is the \suprimecam $i$ filter, which is close
    to the $I$ filter and was only used for \sncw.
    We show the $J$ band filter curve for \niri which is very similar
    to the \isaac {\it Js} filter. The atmospheric transmission for 2~mm
    of precipitable water vapor, typical for Paranal and Mauna Kea,
    is also plotted between $10000-16000$\ang\ in the observer frame. The
    transmission curve was provided by Alain Smette (private
    communication). %
    \label{fig:filters}}
\end{figure*}

We obtained ground-based optical follow-up data of the \sne through
different combinations of passbands, shown in
Figure~\ref{fig:filters}, similar to Bessel $R$ and $I$
\citep{1990PASP..102.1181B}, and \sdss $i$
\citep{1996AJ....111.1748F}. Here we used \fors
\citep{1998Msngr..94....1A} at \vlt and \susi
\citep{1998SPIE.3355..507D} at \ntt in addition to the search
instruments. The \susi, \mosaic, \cfhk and \suprimecam data were
obtained in visitor mode, while the \fors observations were carried
out in service mode.

%% I think this was done in IRAF, hopefully someone in the SCP can
%% verify this.
%%
All optical ground data were reduced \citep{raux:2003} in a standard
manner including bias subtraction, flat fielding and fringe map
subtraction using the \iraf\footnote{\iraf is distributed by the
  National Optical Astronomy Observatories, which are operated by the
  Association of Universities for Research in Astronomy, Inc., under
  the cooperative agreement with the National Science Foundation.}
software.
%% According to Reynald the reduction was carried out by Julien Raux,
%% (which is cited in the paper) but some people have also said that
%% Rob Knop may have done it.
%%

\subsection{Ground-based IR observations and reduction}

The two most distant \sne in the sample, \snhb and \sngn, were also
observed from the ground in the near-IR.  Both \niri
\citep{2003PASP..115.1388H} and \isaac \citep{moorwood:1999} were used
to observe \snhb, while \sngn was observed with \isaac only.

The \isaac observations were done with the {\it Js} filter and the \niri
observations were carried out with the $J$ filter. The transmission
curves of the filters are similar to each other, and the transmission
curve of the latter is shown in Figure~\ref{fig:filters}. The red edges
of the filters are defined by the filters and not by the broad
telluric absorption band that lies between the $J$ and $H$ windows,
and the central wavelength is slightly redder than the central
wavelength of the $J$ filter of \citet{1998AJ....116.2475P}. Compared
to traditional $J$ band filters, photometry with the \isaac {\it Js} and
\niri $J$ band filters is less affected by water vapor and is
therefore more stable.

The \isaac observations were done in service mode and the data were
taken on 14 separate nights, starting on 2001 May 7 and ending on 2003
May 30. Individual exposures lasted 30 to 40~s, and three to four of
these were averaged to form a single image. Between images, the
telescope was offset by $10''$ to $30''$ in a semi-random manner, and
typically 20 to 25 images were taken in this way in a single observing
block. The observing block was repeated several times until sufficient
depth was reached.

The data, including the calibrations, were first processed to remove
two electronic artifacts. In about 10\% of the data, a difference in
the relative level of odd and even columns can be seen. The relative
difference is a function of the average count level and it evolves
with time, so it cannot be removed with flat fields. In those cases
%where the effect is present, the data is processed with the \texttt{eclipse}
where the effect is present, the data are processed with the
\texttt{eclipse}\footnote{\url{http://www.eso.org/projects/aot/eclipse/}}
odd-even routine. The second artifact, an electronic ghost, which is
most easily seen when there are bright stars in the field of view, is
removed with the \texttt{eclipse} ghost routine.

The \niri observations were done in queue mode and the data were taken
on 4 separate nights, starting on 2001 May 25 and ending 2002 Aug 5.
Individual exposures lasted 60~s. Between images, the telescope was
offset by $10''$ to $30''$ in a semi-random manner, and typically 60
images were taken in this way in a single observing sequence. The
sequence was repeated several times until sufficient depth was
reached.

Both the \isaac and \niri data were then reduced in a standard way
with the \iraf XDIMSUM package and our own \iraf scripts. From each
image, the zero-level offset was removed, a flatfield correction was
applied, and an estimate of the sky from other images in the sequence
was subtracted. Images were then combined with individual weights that
depend on the median sky background and the image quality.

\subsection{HST observations and reduction}

Observing \snia at high-$z$ from space has an enormous advantage for
accurately following their light curves. The absence of the
atmosphere and the high spatial resolution allows high signal-to-noise
measurements.
%background and the high spatial resolution. The latter which helps
%both by minimizing the host contamination and through focusing the
The high spatial resolution also helps minimize host contamination
through focusing the light over a smaller area. Space also permits
observations at longer
%wavelengths where the limited atmospheric transmission degrades
wavelengths where the limited atmospheric transmission and the high background
degrade ground-based data.

High quality follow-up data were obtained for all six \sne using the
Wide Field Planetary Camera~2 (\wfpc) on the Hubble Space Telescope
(\hst) during Cycle~9. All but two (\sncw and \sngq) of the objects
also have \sn-free reference images taken during Cycle~10, with the
Advanced Camera for Survey (\acs). \wfpc consists of four 800x800
pixel chips of which one, the Planetary Camera (\pc), has twice the
resolution of the others. The \sne were always targeted with the lower
left corner of the \pc in order to place them closer to the readout
amplifier so that the effects of charge-transfer inefficiency would be
reduced. Images of the same target were obtained with roughly the same
rotation for all epochs.

Each \sn was followed in two bands. All \sne were observed in the
F814W filter. Additionally, the F675W filter was used for the three
\sne at $z<0.7$ and the F850LP filter was used for the high redshift
targets. These filters were chosen to match the filters used on the
ground and correspond approximately to rest frame {\it UBV} bands as
illustrated in Figure~\ref{fig:filters}.

The data were reduced with software provided by the Space Telescope
Science Institute (\stsci). \wfpc images were processed through the
\stsci pipeline and then combined for each epoch to reject cosmic rays
using the \crrej task which is part of \stsdas\footnote{The Space
  Telescope Science Data Analysis System (\stsdas) is a software
  package for reducing and analyzing astronomical data. It provides
  general-purpose tools for astronomical data analysis as well as
  routines specifically designed for \hst data.} \iraf package.

The \acs images were processed using the multi-drizzle
\citep{2002PASP..114..144F} software, which also corrects for the
severe geometric distortion of the instrument. For this we used the
updated distortion coefficients from the \acs Data Handbook in
November 2006, and drizzled the images to the resolution of the \wfpc
images $0.046''$. Note that the \wfpc images were \emph{not} corrected
for geometric distortion at this stage.

\section{Photometry of the ground-based data}
\label{sec:groundphoto}
The photometry technique applied to the optical ground-based data is
the same as the one applied in \citet{2008A&A...486..375A}, except for the
\suprimecam $i$ band data. This is also very similar to the method
\citep{fabbro:2001} used in \citet{2006AA...447...31A}, and is briefly
summarized here:
\begin{enumerate}
\item Each exposure of a given \sn in a given passband was aligned to
  the best seeing \emph{photometric reference} image.
\item In order to properly compare images of different image quality
  we fitted convolution kernels, $K_i$, modelled by a linear
  decomposition of Gaussian and polynomial basis functions
  \citep{1998ApJ...503..325A,2000A&AS..144..363A}, between the
  photometric reference and each of the remaining images, $i$, for the
  given passband. The kernels were fitted by using image patches
  centered on fiducial objects across the field.
% The quality of both the geometric transformations as well as the
% fitted kernels was investigated by carrying out image subtractions and
% searching for residual artifacts. Such artifacts were in general
% absent and had minimal impact on the \sn photometry.

\item The background sky level for each image, $i$, is not
  expected to have any spatial variation
%\item The sky level is not expected to have any spatial variation for
%  a small patch centered on the \sn, and we assume that the patch can
  for a small patch, $I_i(x,y)$, centered on the \sn with a radius of
  the worst seeing FWHM. We assume that the patch
  can be modelled by a point spread function (\psf) at the location of
%  the \sn added to a model of the host galaxy and a constant sky
%  level. For a time-series of such patches we simultaneously fit the
  the \sn, a model of the host galaxy and a constant offset for
  the sky,
  \begin{eqnarray*}
    I_i(x,y) & = & f_i\cdot\left[K_i\otimes\mathrm{\psf}\right](x -
    x_0, y - y_0 ) + \\
    & + & \left[K_i\otimes G\right](x,y)\, + S_i\, .
  \end{eqnarray*}
  For a time-series of such patches we simultaneously fit the \sn
  position, $(x_0,y_0)$, and brightnesses, $f_i$, host model, $G(x,y)$, and
  background sky levels, $S_i$. We use a non-analytic host model with
  one parameter per pixel. This means that the model will be
  degenerate with the \sn and the sky background. We break these
  degeneracies by fixing the \sn flux to zero for all reference images
  and fix the sky level to zero in one of the images.
  Figure~\ref{fig:patchseries} shows an example of image patches,
  galaxy model residuals and resulting residuals when the full model
  has been subtracted, for increasing epochs.
  \begin{figure*}[p]
    \centering
    \resizebox{\hsize}{!}{\includegraphics{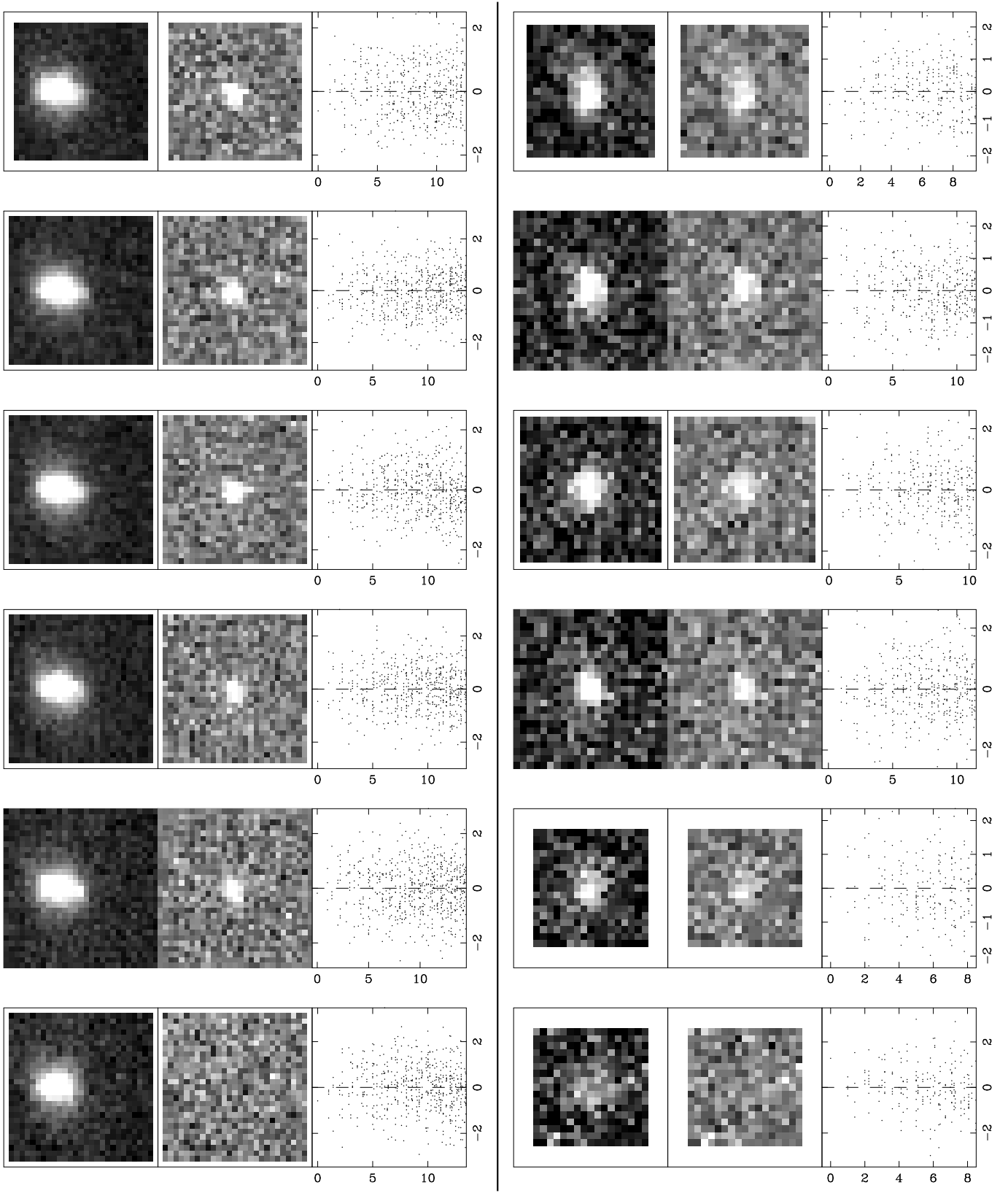}}
    \caption{Ground-based $I$-band images (patches) of \sngq (\emph{three left}
      columns; width $6.0''$) and \sngo (\emph{three right} columns; width
      $4.7''$). Each triplet represents from left to right the fully
      reduced data, the galaxy model subtracted data, and a profile plot
      where the full galaxy + \psf model has been subtracted from the
      data. The profile plot shows the deviation from zero in standard
      deviations of the noise versus the distance from the \sn position in
      pixels. Epoch increases from top to bottom. The images in the last
      row are the reference image, obtained approximately one year after
      discovery. The different image sizes (patches) reflect differences
      in image quality. A small image represents good seeing conditions. %
      \label{fig:patchseries}}
  \end{figure*}

\end{enumerate}
The \sn light curves were obtained in this manner for one filter and
instrument at a time where \sn-free reference images were available.
In the cases where \sn-free references did not exist for a given
telescope, reference images obtained with other telescopes were used
instead.

%%
%% ISSUE: I should do a more thorough comparison of how the measured
%% magnitude changes with aperture. I think the problem with the PSF
%% approach was that I got very inconsistent magnitude measurements
%% of the tertiary standards, i.e. the PSF and aperture photometry did
%% not agree, but I think this was because the field was crowded or
%% the PSF model was not correct?
%%
% For the \isaac and \niri data we could not isolate enough stars to fit
% an accurate \psf model.
%The brightness of the two high-$z$ \sne in the near infrared was
%obtained through aperture photometry. For \snhb, we performed aperture

\begin{figure*}[tbhp]
\centering\includegraphics[width=10cm,angle=0]{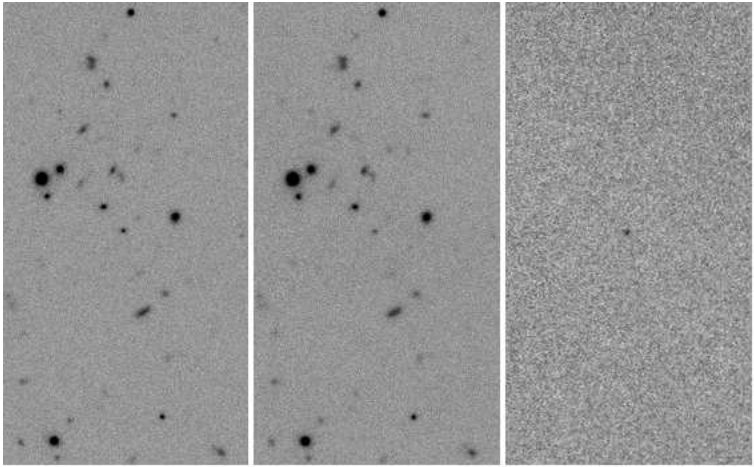}
\caption{\vlt/\isaac $J$ band observations of the distant \snia, \sngn
  (North is up and East is to the left). On the left, we show a $27''$
  wide and $51''$ high image with the \sn; in the middle we show the
  reference image, which was taken 2 years later. The images on the
  right is the subtraction between the two and shows the \sn with a
  signal-to-noise ratio of 10. Each image is the result of 10 hours of
  integration in good conditions. The image quality in the left hand
  image is FWHM=$0.39''$.%
  \label{fig:SN2001gn}}
\end{figure*}

For the IR image of \snhb, we performed aperture photometry directly on
the images, assuming the \sn to be hostless for the purpose of $J$
band photometry, since no host could be detected at the limit of the
\acs references (see below). For the IR image of \sngn, we carried out
steps~1--2 above, and then subtracted the reference image
(Figure~\ref{fig:SN2001gn}) and did aperture photometry on the
resulting image. In both cases the \sn positions from the \hst images
were used for centroiding the aperture and the diameter was chosen to
maximize the signal-to-noise ratio. The fluxes are corrected to larger
apertures by analyzing bright stars in the same image.

The $i$ band data for \sncw was analyzed together with a larger sample
of \sne discovered at Subaru. The details of this analysis are given
in Yasuda \etal (in preparation).

%
%___________________________________________________________________________

%\subsection{Calibration}
\subsection{Calibration of the ground-based data}
\begin{figure*}
\centering
\resizebox{\hsize}{!}{\includegraphics{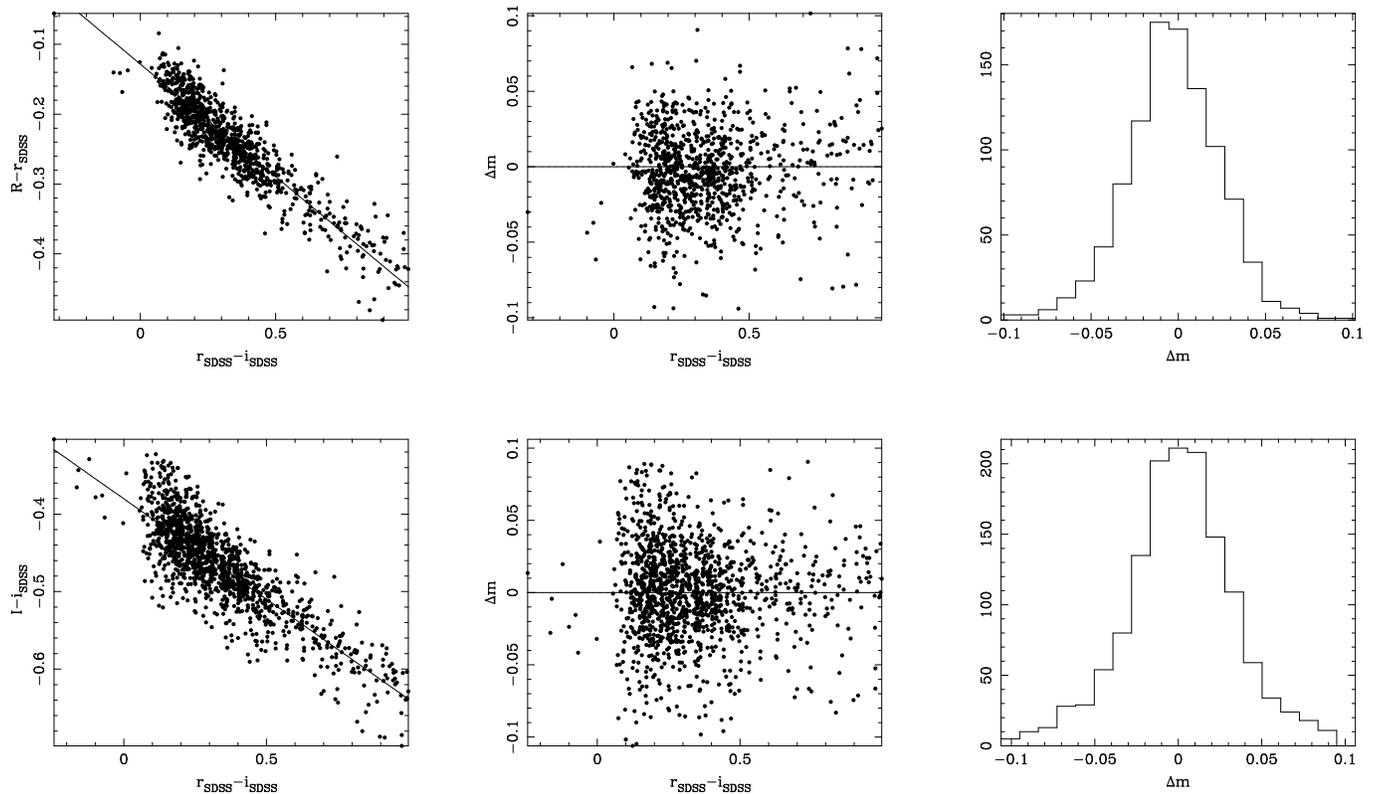}}
\caption{Stars observed both by Peter Stetson
  \citep{2000PASP..112..925S} and \sdss for which $-0.5 <
  r_\mathrm{SDSS}-i_\mathrm{SDSS} < 1.0$ together with fitted linear
  relations between the \sdss $\rsdss,\isdss$ and the Landolt $R,
  I$ magnitudes. The \emph{left} column shows the relation as a
  function of color, while the \emph{middle} column shows the
  residuals once the fitted relation has been subtracted, and a
  histogram of the residuals are presented in the \emph{right}
  column. %
  \label{fig:lupton}}
\end{figure*}

Nightly observations of standard stars were not available for all
optical instruments and filters. Instead the recipe from
\citet{2008A&A...486..375A}, using \sdss \citep{2007ApJS..172..634A}
measurements of the field stars was applied. However, the \sdss filter
system \citep{1996AJ....111.1748F} differs significantly from the
%filters used in this search except for the Subaru $i$ band. In order
%to cope with this we fitted relations between the \sdss filter system
filters used in this search except for the \suprimecam $i$ band. To overcome
this difference, we fitted relations between the \sdss filter system
and the Landolt system \citep{1992AJ....104..340L} in a similar
%manner to \citet{lupton2005}, using stars where \sdss photometry have
%been matched to Peter Stetson's published photometry
manner to \citet{lupton2005}, using stars with \sdss and Stetson
photometry \citep{2000PASP..112..925S}. Stetson has been publishing
photometry of a growing list of faint stars that is tied to the
Landolt system within $0.01$~mag \citep{2000PASP..112..925S}. The most
up to date version can be obtained from the Canadian Astronomy Data
Center\footnote{\texttt{
    \url{http://www.cadc.hia.nrc.gc.ca/community/STETSON/archive/}}\hfil}.
%    \url{http://www1.cadc-ccda.hia-iha.nrc-cnrc.gc.ca/community/STETSON/archive/}}\hfil}.
In contrast to \citet{lupton2005}, we applied magnitude, uncertainty
and color cuts ($15<\rsdss<18.5$, $\sigma_m<0.03$~mag in $\rsdss$ and
$\isdss$ and $-0.5 < \rsdss-\isdss < 1.0$) for the stars that went
into the fit. We also applied a $5\sigma$ outlier cut after our
initial fit (and lost about $\sim6\,\%$ of the sample). After
refitting, the following relations were derived for the Landolt $R$
and $I$ filters
\begin{eqnarray*}
  R-\rsdss & = -0.13 - 0.32\cdot(\rsdss-\isdss)\\
  I-\isdss & = -0.38 - 0.26\cdot(\rsdss-\isdss),
\end{eqnarray*}
which are also shown in Figure~\ref{fig:lupton}. The results for
$R-\rsdss$ are close to the ones derived by \citet{lupton2005} as well
as to \citet{2003ApJ...594....1T} who performed a similar operation.
Neither \citet{2003ApJ...594....1T} nor \citet{lupton2005} present fits
for $I-\isdss$ vs $\rsdss-\isdss$.

By forcing $\chi^2/\textrm{dof}=1$, we determined that there is a scatter of $0.03$~mag coming from intrinsic spectral distributions. This is a systematic uncertainty for individual stars,
but will average out when a big sample is used, assuming that the
color distribution of the sample is similar to the stars used to
derive the relations.

%% ISSUE: It should also be pointed out that the distribution plots in
%% Figure~\ref{fig:lupton} are not symmetric which could give rise to
%% systematic effects on the calibrated magnitudes??

%% ISSUE: May be worth to mention that including  higher order color terms
%% did not improve the fits.
%%

The transformations were applied to the \sdss stellar photometry of
our \sn field, and these were then used as tertiary standard stars in
order to tie the \sn photometry to the Landolt system. The flux of the
stars was determined on the photometric reference for each light
curve build using the same method as for the \sn and with the same
\psf model that was used for fitting the \sn fluxes. A zero-point
relation of the form
\begin{equation}
  \label{eq:ZP}
  m + 2.5\log_{10}f = \mathrm{ZP} + c_X\cdot(R-I)
\end{equation}
could then be fitted between the measured stellar fluxes, $f$, and
their Landolt magnitudes, $m$. Here $\mathrm{ZP}$ is the zero point
and $c_X$ is the color term for the filter. We applied the same color
cuts ($-0.5 < \rsdss-\isdss < 1.0$) to the stars that went into the
fit.
%\psf model that was used for fitting the \sn fluxes. A zero-point
%relation, $m + 2.5\log_{10}f = ZP + c_X\cdot(R-I)$, could then be
%fitted between the measured stellar fluxes, $f$, and their standard
%magnitudes, $m$. Here $ZP$ is the zero point and $c_X$ is the color
%term for the filter. We applied the same color cuts ($-0.5 <
%\rsdss-\isdss < 1.0$) for the stars that went into the fit.

Unfortunately we did not have enough stars over a wide enough color
range to accurately fit the color term. Instead, we used values from
the literature or from the observatories, which are summarized in
Table~\ref{tb:colorterms}. The zero-points could then be derived from
equation~\eqref{eq:ZP}. The values obtained this way are the sums of
three components; the instrumental zero-point, the aperture correction
to the \psf normalization radius and the atmospheric extinction for
the given airmass, and are shown along with the \sn fluxes in
Table~\ref{tb:photometry}.

We also calculated color terms synthetically by using Landolt standard
stars that have extensive spectrophotometry from
\citet{2005PASP..117..810S}. The synthetic Vega magnitudes of the stars
were calculated by multiplying the spectra and the Vega spectrum
\citep{2007ASPC..364..315B} with the filter and instrument throughputs
provided by the different observatories. The color terms were then
fitted by assuming a linear relation for the deviation between the
synthetic and the Landolt magnitudes as a function of the Landolt
color. The resulting fitted synthetic terms are also presented in
Table~\ref{tb:colorterms}.
\begin{deluxetable*}{lrrrr}
  \tablecolumns{5}
  \tablewidth{0pc}
  \tablecaption{Measured and synthetic color terms for the three
    instruments that were used as photometric references. 
%    The measured values are from
    \label{tb:colorterms}}
  \tablehead{
    \colhead{}
    & \multicolumn{2}{c}{Measured} &
    \multicolumn{2}{c}{Synthetic}\\
    \multicolumn{1}{l}{Inst./Tel.} & \multicolumn{1}{c}{$c_R$} &
    \multicolumn{1}{c}{$c_I$} & \multicolumn{1}{c}{$c_R$} &
    \multicolumn{1}{c}{$c_I$}}
  \startdata
  CFH12k/CFHT &
  $0.031$\tablenotemark{a} & $0.107$\tablenotemark{a} % ELIXIR
  & $0.066$ & $-0.023$\\
  \mosaic/CTIO & 
  \nodata & $0.030$\tablenotemark{b} % ESSENCE, Miknaitis et al. (2007)
  & \nodata &  $0.007$\\
  FORS1/VLT &
  $0.034$\tablenotemark{c} & $-0.050$\tablenotemark{c} % VLT/FORS1
  & $0.040$ & $-0.076$\\
  \enddata\\[1ex]
  \tablenotetext{a}{\footnotesize{\texttt{\url{http://www.cfht.hawaii.edu/Instruments/Elixir/filters.html}}}}
  \tablenotetext{b}{\footnotesize{\citet{2007ApJ...666..674M}}}
  \tablenotetext{c}{\footnotesize{\texttt{\url{http://www.eso.org/observing/dfo/quality/FORS1/qc/photcoeff/photcoeffs\_fors1.html}}}}  

%  { \raggedright%
%    $^a$\footnotesize{%
%      \texttt{http://www.cfht.hawaii.edu/Instruments/Elixir/filters.html}}\\
%    \raggedright%
%    $^b$\footnotesize{%
%      \citet{2007ApJ...666..674M}}\\
%    \raggedright%
%    $^c$\footnotesize{%
%      \texttt{http://www.eso.org/observing/dfo/quality/FORS1/qc/photcoeff/photcoeffs\_fors1.html}}\\}
\end{deluxetable*}

The difference is $\lesssim0.03$~mag for all but the \cfhk $I$-band,
where there is a significant discrepancy. A mismatch between the
effective filter transmission curves we use and the true photometric
system is most likely the origin of this. We have investigated if the
deviation could be explained by the differences in quantum
efficiencies between the different chips of the detector which are
presented on the \cfhk webpage. However these differences propagated
to the color term results in relatively low scatter and deviates
significantly from the measured value.

%% SYNTHETIC AND MEASURED COLOUR TERMS
%%
%%            R_R-I    I_R-I
%%            -----    -----
%% MEASURED   0.031    0.107
%%
%% NONE       0.068   -0.011
%%
%% HiRhoCCD   0.073    0.013
%% EPICCD     0.081   -0.016
%%
%% CHIP04     0.060   -0.041   2001gq
%% CHIP10     0.058   -0.055   2001gn, 2001go

For the purpose of fitting the zero-points, we can use the measured
color terms, but for light curve fitting erroneous filter
transmissions could introduce systematic effects. In order to study
the potential impact on the light curve fits we modified the filters
to match the measured color terms. The modifications were implemented
by either shifting or clipping the filters until the synthetic color
terms matched all measured values for a given filter (\eg $R$ or $I$
vs $R-I$, $V-I$ and $B-I$ for the ELIXIR measurements for the
$I$-band). The light curves were then refitted using the modified
filter transmissions and the resulting values were compared. Since the
light curves are tightly constrained by the high precision \hst data,
the modified filter only leads to negligible differences (less than
$30\,\%$ of the statistical uncertainty) in the fitted parameters.

% We update these zero-points with the color independent constants
% $b_X$ to obtain the instrument zero-points. These are shown along
% with the \sn-fluxes in Table~\ref{tb:photometry}, and can be used to
% obtain the \emph{instrumental} \sn-magnitudes for the given epoch
% and filter.

The $J$-band data, on the other hand, were calibrated using G-type
standard stars from the Persson LCO standard star catalogue
\citep{1998AJ....116.2475P}. Since the ISAAC and NIRI $J$-band filters
are slightly redder and narrower than the Persson $J$-band filter, we
subtracted $0.012$~magnitudes from the ISAAC and NIRI zero points to
place the ISAAC and NIRI IR photometry onto the natural system. Since
the data were taken over many nights, the photometry was carefully
cross checked. Differences in the absolute photometry usually amount
to less than $0.02$~magnitudes, which we conservatively adopt as our
zero-point uncertainty.

%
%___________________________________________________________________________

\section{Photometry of the HST data}
\label{sec:spacephoto}
A modified version of the photometric technique from
\citet{2003ApJ...598..102K} was used for the \hst data presented here.
This is similar to the method used for the ground-based optical data
above, but instead of aligning and resampling all images to a
common frame, the host$+$\sn model is resampled to each individual
image. A procedure like this is preferred when the \psf FWHM is of the
same order as the pixel scale, and it also preserves the image noise properties.

Linear geometric transformations
%, $T^x_k,T^y_k$,
were first fitted from each image, $k$, in a given filter to the
deepest image of the field using field objects. Due to the similar
orientation of the \wfpc images, linear transformations were
sufficient for these, and the geometric distortion of \wfpc could be
ignored. This was however not the case for the transformations between
the \acs and \wfpc images and the distortion was then hardcoded into
the fitting procedure. Due to the sparse number of objects in the tiny
\pc field the accuracy of all transformations were only good
to~$\lesssim1$~pixel ($\sim0.5$ FWHM). Unfortunately, using objects
from the remaining three chips for the alignment did not lead to
improved accuracy, which is probably due to small movements of the
chips between exposures \citep{2003PASP..115..113A}.
% Don't have a reference to this though...
This alignment precision was not enough for \psf photometry, and we
therefore allowed the \sn position, $(x_{k0},y_{k0})$ to float for the
individual images which increased the alignment precision by a
factor of 10. Allowing this extra degree of freedom could bias the
results toward higher fluxes since the fit will favor positive noise
fluctuations. However, as in \citet{2003ApJ...598..102K}, this was
shown to be of minor importance by studying the covariance between the
fitted flux and \sn position.
%%
%% ISSUE: I should take a closer look at the actual correlations
%% between the SN-fluxes.
%%

The full model used to describe each image patch can be expressed as
\begin{equation}
  \begin{split}
      I_k(x_k,y_k) & = f_k\cdot\mathrm{PSF}_k(x_k-x_{k0},y_k-y_{k0}) + \\
      & + G(x_k,y_k,a_j)
    + S_k\, .
    \label{eq:pcltcv}
  \end{split}
\end{equation}
Here $I_k$ is the value in pixel $(x_k,y_k)$ on image $k$, $f_k$ is
the \sn flux, $\mathrm{PSF}_k$ the point spread function, $G$ the host
galaxy model that is parameterized by $a_j$ and $S_k$ the local sky
background. The fits were carried out using a $\chi^2$ minimization
approach using MINUIT \citep{james:1975}.

Four \sne in the sample had \acs reference images, and for these cases
we used field objects to fit non-analytic convolution kernels, $K$,
between the \pc chip and the drizzled \acs image. These accounted for
the difference in quantum efficiency and \psf shape between the two
instruments. When kernels were used, the uncertainties of individual
pixels were propagated and the correlation between pixels produced by
the convolution were taken into account. Also, in this case
equation~\eqref{eq:pcltcv} above was modified so with both the \pc
images and the \pc PSF being convolved with the fitted kernel.

%%
%% ISSUE: Cite work of Andrew Dolphin that shows that PSF photometry
%% and aperture photometry. Should probably include a small discussion
%% of this topic.
%%
The \psf, $\mathrm{PSF}_k$, of the \wfpc \pc chip was simulated for
each filter and pixel position using the Tiny Tim software
\citep{tinytim} and normalized to the \wfpc calibration radius
$0.5''$. We also did an extensive test where we iteratively updated
the simulated \psf based on the knowledge of the \sn epoch, and
therefore the approximate spectral energy distribution (\sed), but
this did not have any significant effect on the fitted fluxes. The
Tiny Tim \psf was generated to be subsampled by a factor of 10. For
each iteration in the fitting procedure, any shift of the \psf
position was first applied in the subsampled space. The \psf was then
re-binned to normal sampling and convolved with a \emph{charge
  diffusion kernel} \citep{tinytim} before it was added to the patch
model.

For three of the \sne, \sncw, \sngq and \snhb we used an analytical
model for the host galaxy. Both \sncw and \sngq are offset from the
core of their respective host galaxies and we therefore chose not to
obtain \sn-free references images for these. Instead the hosts could
be modeled by a second order polynomial and constrained by the galaxy
light in the vicinity of the \sne. For \snhb, we did obtain a deep
%\acs reference image but no host could be detected on this. For this
%\sn we used a simple plane to model the host which was further
\acs reference image but no host could be detected. For this
\sn we used a simple plane to model the host which was further
constrained by fixing the \sn flux to zero for the \acs reference. One
caveat with analytical host modelling in general is that the patch
size must be chosen with care. The model will only work if there is no
dramatic change in the background across the patch, which is an
assumption that is likely to fail if the patch is too large and
includes the host galaxy core. On the other hand the patch can not
be too small either in order to successfully break the degeneracy
between the \sn and the background.

% To make sure that the choice of host model does not bias the fitted
% \sn fluxes we required (1) that the total $\chi^2$ matched the number
% of degrees of freedom, (2) clean residuals once the \sn+host model was
% subtracted from the data, (3) that the fitted \sn fluxes were
% insensitive to small changes in patch size. 
To make sure that the choice of host model does not bias the fitted
\sn fluxes we required, in addition to clean residuals once the \sn +
host was subtracted from the data, that the fitted \sn fluxes were
insensitive to variations in the patch size of a few pixels. Further,
for \sngq, we tested the host model by putting fake \sne at different
positions around the core of the host. The distances to the core and
the fluxes of the fakes were always chosen to match the corresponding
values for the real \sn, and the retrieved photometry was always
within the expected statistical uncertainty.
%%
%% ISSUE: make sure that I'm not lying here.... I don't quite remember
%% the details of this test, it was ~2 years ago...
%%

The recipe described above could not be used for the remaining three
\sne in the sample, since they were located too close to the cores of
%their host galaxies. The hosts did in addition to this have small
their host galaxies. Additionally, the hosts have small
angular sizes and the light gradients in the vicinity of the \sne were
steep enough to lead to biased photometry due to the coarse geometric
%alignment. In order to cope with this we were forced to change the
alignment. To overcome this we changed the
photometric procedure slightly. Instead of fitting the \sn position on
each image, we chose to fit it only on the geometric reference image,
and then introduce a free shift for the whole model. That is, in this
case we used the galaxy model $+$ \sn for the patch alignment. However, this
procedure did force us to apply some constraints on the galaxy
%modelling. Using a non-analytic matrix model as with the ground-based
modelling. Using a non-analytic pixel model, the approach used for the
ground-based optical data, was not feasible. It slowed down the fits
considerably and rarely converged. Instead we chose to use the \acs
images of the host galaxies directly. The \acs references were much
deeper then the \wfpc data and choosing this procedure did not
increase the uncertainty of the fitted fluxes.

A general problem with doing photometry on \hst images is that \ccd
photometry of faint objects over a low background suffers from an
%imperfect Charge Transfer Efficiency (\cte) that will lead to an
imperfect charge transfer, which will lead to an
%underestimate of the flux. We used the \cte recipe for point sources
underestimate of the flux. We used the Charge Transfer Efficiency
(\cte) recipe for point sources from \citet{2009PASP..121..655D}. The
correction for our data is usually around $5\,\%$--$8\,\%$ but it can
be as large as $\sim17\,\%$. The uncertainties of the corrections were
propagated to the flux uncertainties. The corrected \sn fluxes are
given in Table~\ref{tb:photometry} together with the instrumental
zero-points,
%also these obtained from Andrew Dolphin's webpage
which were also obtained from Andrew Dolphin's webpage

%
%___________________________________________________________________________

\newcommand{\mlcs}{MLCS2k2\xspace}
\section{Light curve fitting}\label{sec:lcfitting}
\snia that have bluer colors or broader light curves tend to be
intrinsically brighter
\citep{1993ApJ...413L.105P,1998A&A...331..815T}. Several methods of
combining this information into an accurate measure of the relative
distance have been used %
\citep{1996ApJ...473...88R,2001ApJ...558..359G,2003ApJ...590..944W,%
  2005A&A...443..781G,2007A&A...466...11G,2007ApJ...659..122J,%
  2008ApJ...681..482C}.

K08 consistently fitted all light curves using the \salt
\citep{2005A&A...443..781G} fitter, which is built on the \snia \sed
from \citet{2002PASP..114..803N}. In this paper, we use \saltii
\citep{2007A&A...466...11G}, which is based on more data.

% (including data from distant \sneia from the \snls survey), and
% has been shown to more accurately describe the rest-frame UV of
% the \snia \sed than its predecessor.

\citet{2008ApJ...681..482C} compared the performances of different
light curve fitters while also introducing their own empirical fitter,
SiFTO, and concluded that \saltii along with SiFTO perform better than
both \salt (which is conceptually different from its successor
\saltii) and \mlcs \citep{2007ApJ...659..122J} when judged by the
scatter around the best-fit luminosity distance relationship.
Furthermore, \saltii and SiFTO produce consistent cosmological results
when both are trained on the same data. Recently KS09 made a thorough
comparison between \saltii and their modified version of \mlcs
\citep{2007ApJ...659..122J} for a compilation of public data sets,
including the one from the \sdss \sn survey. The two light curve
fitters result in an estimate of $w$ (for a flat $w$CDM cosmology)
that differs by $0.2$. The difference exceeds their statistical and
systematic (from other sources) error budgets. They determine that
this deviation originates almost exclusively from the difference
between the two fitters in the rest-frame $U$-band region, and the
color prior used in \mlcs. They also noted that \mlcs is less accurate
at predicting the rest-frame $U$-band using data from filters at
longer wavelengths.

% I wanted to include this sentence to stress, that this is 
% hopefully not as serious a problem as it seems.
% 
% Further, they conclude that the situation with the discrepancy is
% likely to be temporary, and will probably be resolved as soon as
% the analysis of already available data is completed.
  
This difference in $U$-band performance is not surprising:
observations carried out in the observer-frame $U$-band are in general
associated with a high level of uncertainty due to atmospheric
variations. While the training of \mlcs is exclusively based on
observations of nearby \sne, the SiFTO and \saltii training address
this difficulty by also including high redshift data where the
rest-frame $U$-band is observed at redder wavelengths. This approach
also allows these fitters to extend blueward of the rest-frame
$U$-band.

%% 
%% Perhaps add a sentence that it is proven to show better
%% performance in the UV as shown in KS09, if we believe this.
%% 

In addition, for this paper, we have conducted our own test validating
the performance of \saltii by carrying out the Monte-Carlo simulation described
in~\ref{sec:verifyinglcfitter}, where we compare the fitted \saltii
parameters to the corresponding real values for mock samples with poor
cadence and low signal-to-noise drawn from individual well-measured
nearby \sne.

Given these tests that have been carried out on \saltii, and its high
redshift source for rest-frame $U$-band, we have chosen to use \saltii
in this paper.

%  We also include a systematic error term representing the higher
%  level of uncertainty in the rest frame $U$ band, described
%  in~\ref{sec:restUband}. %

% We also include a systematic error term representing the higher
% level of uncertainty in the rest frame $U$ band, described
% in~\ref{sec:restUband}. %

\subsection{\saltii}
The \saltii \sed model has been derived through a pseudo-Principal
component analysis based on both photometric and spectroscopic data. Most
of these data come from nearby \snia data, but \snls supernovae are
also included. To summarize, the \saltii \sed,
$F(\mathrm{\sn},p,\lambda)$, is a function of both wavelength,
$\lambda$, and time since $B$-band maximum, $p$. It consists of three
components; a model of the time dependent average \snia \sed,
$M_0(p,\lambda)$, a model of the variation from the average,
$M_1(p,\lambda)$, and a wavelength dependent function that warps the
model, $CL(\lambda)$. The three components have been determined from
the training process \citep{2007A&A...466...11G} and are combined as
\begin{eqnarray*}
F(\mathrm{\sn},p,\lambda) & = & x_0\times\left[M_0(p,\lambda) + x_1\times
  M_1(p,\lambda)\right]\times\\
  & \times & \exp\left[c\times CL(\lambda)\right]\, ,
\end{eqnarray*}
where $x_0$, $x_1$ and $c$ are free parameters that are fit for each
individual \sn.

Here, $x_0$, describes the overall \sed normalization, $x_1$, the
deviation from the average decline rate ($x_1=0$) of a \snia, and $c$,
the deviation from the mean \snia $B-V$ color at the time of $B$-band
maximum. These parameters are determined for each observed \sn by
fitting the model to the available data. The fit is carried out in the
observer frame by redshifting the model, correcting for Milky Way
extinction (using the CCM-law from \citet{1989ApJ...345..245C} with
$R_V=3.1$), and multiplying by the effective filter transmission
functions provided by the different observatories. 
  All synthetic photometry is carried out in the Vega system using the
  spectrum from \citet{2007ASPC..364..315B}. Following
  \citet{2006AA...447...31A} we adopt the magnitudes $(U,B,V,R_C,I_C)
  = (0.020,0.030,0.030,0.030,0.024)$~mag \citep{1996AJ....111.1748F} for
  Vega. For the near-infrared we adopt the values $J=0$ and $H=0$.
  
In the fit to our data, we take into account the correlations
introduced between different light curve points from using the same
host galaxy model. We also chose to run \saltii in the mode where the
diagonal of the covariance matrix is updated iteratively in order to
take model and $K$-correction uncertainties into account.
See \citet{2007A&A...466...11G} for details on this. The Milky Way
reddening for our supernovae from the \citet{1998ApJ...500..525S} dust
maps is given in Table~\ref{tab:SpecObservations}. The results of the
fits are shown in Table~\ref{tb:salt2results} and plotted in
Figure~\ref{fig:lightcurves} together with the data.
\begin{figure*}[p]
\centering
\ifthenelse{\boolean{salttwo}}{%
  \includegraphics[height=0.9\textheight]{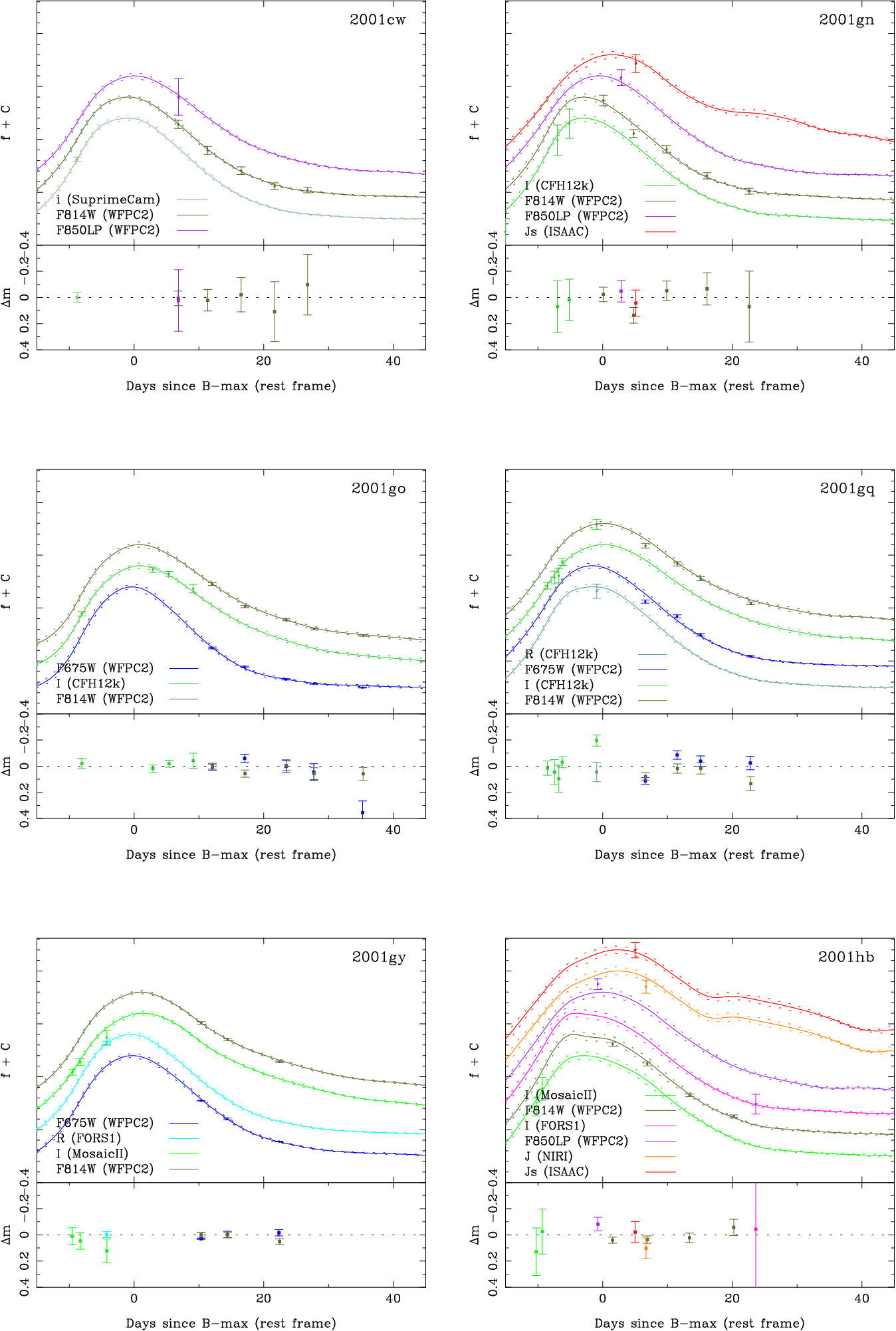}%
  \caption{\saltii light curve fits together with the data for the six
    \sne presented here. Light curves for different bands/instruments,
    indicated by the color coding, have been offset. The dotted lines
    following the curves represents the model errors. %
    \label{fig:lightcurves}}
}{%
  \includegraphics[height=0.9\textheight]{figures/current/lightcurves.eps}%
  \caption{\salt light curve fits together with the data for the six \sne
    presented here. Light curve builds for different bands/instruments,
    indicated by the color coding, have been offset.
    \label{fig:lightcurves}}
}
\end{figure*}

% The model errors are estimated from the results of the \salt
% training in \citet{2007astro.ph..1828G}, while the $K$-corrections
% uncertainties are determined by comparing the fitted peak
% magnitude for each filter with the predictions from the light
% curve fit of the other filter.
% 
The three parameters
\[
m_B^\mathrm{max} = -2.5\log_{10}\left[\int_B
  F(\mathrm{SN}, 0, \lambda)\, \lambda\,d\lambda\right]\,,\ x_1\
\mathrm{and}\ c
\]
can for each \sn be combined to form the distance modulus
\citep{2007A&A...466...11G},
\begin{equation}
    \mu_B = m_B^\mathrm{corr} -  M_B = m_B^\mathrm{max} + \alpha\cdot
  x_1 - \beta\cdot c - M_B\, ,\label{eq:magcor}
\end{equation}
where $M_B$ is the absolute $B$-band magnitude. The resulting color
and light curve shape corrected peak $B$-band magnitudes,
$m_B^\mathrm{corr}$, are presented in the third column of
Table~\ref{tb:salt2results}. The parameters $\alpha$, $\beta$ and
$M_B$ are nuisance parameters which are fitted simultaneously with the
cosmological parameters.

% SALT1/2 FIT RESULTS FOR ALL SNE
%
\ifthenelse{\boolean{salttwo}}{%
  \begin{deluxetable*}{lrrrrrrrrr}
    \tabletypesize{\footnotesize}
    \tablecaption{Light curve fit results using \saltii.
      The \saltii parameters are MJD$_\mathrm{max}$, $m_B$, $x_0$, $x_1$ and $c$.
      Here, the light curve shape, $x_1$, and color, $c$, corrected peak
      $B$-band magnitude is obtained from equation~\eqref{eq:magcor}
      as $m^\mathrm{corr}_B=m^\mathrm{max}_B + \alpha\cdot x_1 - \beta\cdot c$,
      where the values $\alpha=\salpha$ and $\beta=\cbeta$ have been used.
      For comparison we also present the \salt stretch, $s$, which has
      been derived from $x_1$ using the relation provided in
      \citet{2007A&A...466...11G}.
      \label{tb:salt2results}}
    \tablehead{
      \multicolumn{1}{c}{\sn} &
      \multicolumn{1}{c}{$\mathrm{MJD}_\mathrm{max}$} &
      \multicolumn{1}{c}{$m^{\mathrm{corr}}_B$} &       
      \multicolumn{1}{c}{$m^{\mathrm{max}}_B$} &               
      \multicolumn{1}{c}{$c$}   &  
      \multicolumn{1}{c}{$x_1$} &
      \multicolumn{1}{c}{$s$}}
    \startdata
    %% automatically generated by 'lightcurves.pl'
%%
\sncw & $52066.60\pm0.86$ & $24.88\pm0.42$ & $24.71\pm0.13$ & $-0.10\pm0.36$ & $0.140\pm0.844$ & $0.993\pm0.077$\\
\sngn & $52030.33\pm0.16$ & $25.16\pm0.23$ & $25.37\pm0.05$ & $0.07\pm0.05$ & $0.648\pm0.649$ & $1.040\pm0.061$\\
\sngo & $52010.99\pm0.35$ & $23.13\pm0.11$ & $23.08\pm0.07$ & $-0.11\pm0.07$ & $-1.149\pm0.259$ & $0.881\pm0.022$\\
\sngq & $52028.81\pm0.42$ & $23.62\pm0.19$ & $23.79\pm0.03$ & $0.06\pm0.04$ & $0.459\pm0.287$ & $1.022\pm0.027$\\
\sngy & $52031.66\pm0.25$ & $22.97\pm0.07$ & $23.05\pm0.03$ & $-0.03\pm0.03$ & $-0.312\pm0.265$ & $0.952\pm0.024$\\
\snhb & $52038.02\pm0.97$ & $24.84\pm0.13$ & $24.79\pm0.03$ & $0.00\pm0.04$ & $1.292\pm0.410$ & $1.101\pm0.039$\\

    \enddata
  \end{deluxetable*}}{%
  \input{tables/salt1}}

\newcommand{\sigmasys}{\sigma_\mathrm{sys}}
\newcommand{\sigmatot}{\sigma_\mathrm{ext}}
\section{The \compname Compilation}\label{sec:union}
K08 presented an analysis framework for combining different \snia data
sets in a consistent manner. Since then two other groups (H09 and
KS09) have made similar compilations, using different fitters. In this
work we carry out an improved analysis, using and refining the
approach of K08. We extend the sample with the six \sne presented
here, the \sne from \citet{2008A&A...486..375A}, the low-$z$ and
intermediate-$z$ data from \citet{2009ApJ...700..331H} and
\citet{2008AJ....136.2306H} respectively%
\footnote{The \ifthenelse{\boolean{salttwo}}{\saltii}{\salt} fit
  results for these samples are presented along with the entire
  \compname compilation fits at
  \texttt{\url{http://supernova.lbl.gov/Union/}}.}.

First, all light curves are fitted using a single light curve fitter
(the \saltii method) in order to eliminate differences that arise from
using different fitters. For all \sne going into the analysis we
require:
\begin{enumerate}
  \item data from at least two bands with rest-frame central wavelengths
    between $2900$\ang\ and $7000$\ang, the default wavelength range of
    SALT2
  \item that there is at least one point between $-15$~days and
    $6$~rest frame days relative to the $B$-band maximum.
  \item that there are in total at least five valid data points
    available.
  \item that the fitted $x_1$ values, including the fitted
    uncertainties, lie between $-5 < x_1 < 5$. This is a more
    conservative cut than that used in K08 and results in several poorly
    measured \sne being excluded. Part of the discrepancy observed by
    KS09 when using different light curve models could be traced to
    poorly measured \sne.
  \item that the CMB-centric redshift is greater than $z > 0.015$.
\end{enumerate}
We also exclude one \sn from the Union compilation that is 1991bg-like, which
neither the \salt nor the \saltii models are trained to handle. Note
that another 1991bg-like \sn from the Union compilation was removed by the
outlier rejection. 
% K08 investigated removing 1991bg-like \sne, but did not do so before
% the analysis was unblinded.
  All \sneia considered in this compilation are listed in
  Table~\ref{tb:unionsne}. For each \sn, the redshift and fitted light
  curve parameters are presented as well as the failed cuts, if any.

It should be pointed out that the choice of light curve model also has
an impact on the sample size. Using \saltii will allow more \sne to
pass the cuts above, since the \saltii model covers a broader
wavelength range than \salt. This is particularly important for
high-$z$ data that heavily rely on rest-frame $UV$ data.
For example, \saltmodelriessdrops net \sne would have been cut
from the \citet{2007ApJ...659...98R} sample with the \salt model.

\subsection{Revised HST zero-points and filter curves}\label{sec:nicmosphot}
Since \citet{2007ApJ...659...98R}, the reported zero-points of both
NICMOS and ACS were revised. For the F110W and F160W filters of
NICMOS, the revision is substantial. Using the latest calibration
\citep[and references within]{nicmos2009}, the revised zero-points
are, for both filters, approximately $5\,\%$ fainter than those
reported in \citet{2007ApJ...659...98R} and subsequently used by
K08.

For \sneia at $z>1.1$, observations with 
NICMOS cover the rest frame optical, so the fitted peak $B$-band
magnitudes and colors and the corrected $B$-band magnitudes of these
\sneia depend directly on the accuracy of the NICMOS photometry. With
the new zero-points, \sneia at $z>1.1$ are measured to be fainter and bluer.
Our current analysis also corrects an error in
the NICMOS filter curves that were used in K08, which also
acts in the same direction.

In the introduction, we had noted that almost all \sne at $z>1.1$ were
redder than the average \sn color over the redshift interval $0.3$ to
$1.1$.  This is surprising as redder \sne are also fainter and should
therefore be the harder to detect in magnitude limited surveys. K08
noted that these \sne, after light curve shape and color corrections,
are also on average $\sim0.1$~mag brighter than the line tracing the
best fit $\Lambda$CDM cosmology. They also noted that this was the
reason for the relatively high value for the binned value of $w$ in
the $[0.5,1]$ redshift bin.

After taking the NICMOS zero-point and filter updates just discussed
into account, we repeated the original K08 analysis. This made the
NICMOS observed \sne up to $\sim0.1$~mag fainter, and there no longer
is a significant offset from the best fit cosmology. Nor are these
\sne unusually red when compared to \sne over the redshift interval
$0.3$ to $1.1$. For \saltii the \sne at $z>1.1$ have an average color
of $c=\wmeancolourNICMOShiz$, compared to
$c=\wmeancolourptthreetooneptone$ for $0.3<z<1.1$, and no significant
offset in the Hubble diagram.
%
% However, the \saltii colors of \sne at $z>1.1$ are slightly redder
% than the average \saltii color of \sne over the redshift interval
% $0.3$ to $1.1$ ($\wmeancolourNICMOShiz$ and
% $\wmeancolourptthreetooneptone$, respectively), even after applying
% the revised NICMOS zero points. The offset may also be related to
% NICMOS calibration issues.

There could however still be unresolved NICMOS issues. For example the
NICMOS \sn photometry depends on extrapolating the non-linearity
correction to low flux levels. We have a program (HST GO-11799) to
obtain a calibration of NICMOS at low flux levels. The photometry of
the \sne observed with NICMOS will be revised once this program is
completed. The new data presented in this paper also allow us another
route to check this color discrepancy with IR data independent of the
NICMOS calibration.

\subsection{Fitting Cosmology}
Following \citet{2006ApJ...644....1C} and K08, we adopt a blind
analysis approach for cosmology fitting where the true fitted values
are not revealed until the complete analysis framework has been
settled. The blind technique is implemented by adjusting the
magnitudes of the \sne until they match a fiducial cosmology ($\om =
0.25$, $w = -1$). This procedure leaves the residuals only slightly
changed, so that the performance of the analysis framework can be
studied. The best fitted cosmology with statistical errors is
obtained through an iterative $\chi^2$-minimization of
\begin{equation}
  \chi_{\mathrm{stat}}^2 =
  \sum_\mathrm{\sne}\frac{\left[\mu_B(\alpha, \beta,M) - 
      \mu(z;\om,\Omega_w,w)\right]^2}{%
    \sigmatot^2 + \sigmasys^2 + \sigma_{\mathrm{lc}}^2}\,,\label{eq:cosmochi2}
\end{equation}
where,
\begin{equation}
  \sigma_{\mathrm{lc}}^2 = V_{m_B} + \alpha^2 V_{x_1} + \beta^2
  V_{c} + 2\alpha V_{m_B,x_1} - 2\beta V_{m_B,c} - 
  2\alpha\beta V_{x_1,c}\, \label{eq:lcchi2}
\end{equation}
is the propagated error from the covariance matrix, $V$, of the light
curve fits, with $\alpha$ and $\beta$ being the $x_1$ and color
correction coefficients of equation~\eqref{eq:magcor}. Uncertainties
due to host galaxy peculiar velocities of $300$~km/s and uncertainties
from Galactic extinction corrections and gravitational lensing as
described in~\ref{sec:systematics} are included in $\sigmatot$. A
floating dispersion term, $\sigmasys$, which contains potential
sample-dependent systematic errors that have not been accounted for
and the observed intrinsic \snia dispersion, is also added. The value
of $\sigmasys$ is obtained by setting the reduced $\chi^2$ to unity
for each sample. Computing a separate $\sigmasys$ for each sample
prevents samples with poorer-quality data from increasing the errors
of the whole sample. This approach does however still assume that all
\sne within a sample are measured with roughly the same accuracy. If
this is not the case there is a risk in degrading the constraints from
the sample by down weighting the best measured \sne. It should also be
pointed out that the fitted values of $\sigmasys$ will be less certain
for small samples and can therefore deviate significantly from the
average established by the larger samples (in particular, the six
high-$z$ \sne presented in this work are consistent with
$\sigmasys=0$), as are three other samples.

A number of systematic errors are also being considered for the full
cosmology analysis. These are taken into account by constructing a
covariance matrix for the entire sample which will be described below
in~\ref{sec:systematics}. The terms in the denominator of
equation~\eqref{eq:cosmochi2} are then added along the diagonal of
this covariance matrix.
  
Following K08, we carry out an iterative $\chi^2$ minimization
with $3\sigma$ outlier rejection. Each sample is fit for its own
absolute magnitude by minimizing the sum of the absolute residuals
from its Hubble line (rather than the sum of the squared residuals).
The line is then used for outlier rejection. This approach was
investigated in detail in K08, and it was shown with simulations that
the technique is robust and that the results are unaltered from
the Gaussian case in the absence of contamination and that in the
presence of a contaminating contribution, its impact is reduced.
Table~\ref{tb:outliercuts} summarizes the effect of the outlier cut on
each sample.  We also note that the residuals have a similar
distribution to a Gaussian in that $\sim5\%$ of the sample is outside
of $2\sigma$.
  
\begin{deluxetable*}{lcccccc}
  \tabletypesize{\small} \tablecaption{Statistics of each sample with
    no outlier rejection or $3\sigma$ outlier rejection (used in this
    paper). Here, $\sigmasys$ has the same meaning as in
    equation~\eqref{eq:cosmochi2}. Both $\sigmasys$ and the RMS are
    also plotted for each sample in Figure~\ref{fig:diagnostics}.
    Although each sample is independently fit for its $\sigmasys$ and
    RMS, a global $\alpha$ and $\beta$ are always used. This explains
    the minor shifts in parameters for samples where no supernovae are
    cut. A $2\sigma$ cut removes 34 more supernovae, so going from
    3 to $2\sigma$ is consistent with Gaussian residuals.
    \label{tb:outliercuts}}
\tablehead{
  \multicolumn{1}{c}{} &
  \multicolumn{3}{c}{No Outlier Cut} &
  \multicolumn{3}{c}{$\sigma_{\mathrm{cut}} = 3$}\\
  \colhead{Set} & \colhead{SNe} & \colhead{$\sigmasys (68\%)$} & 
  \colhead{RMS $(68\%)$} & \colhead{SNe} & \colhead{$\sigmasys
    (68\%)$} & 
  \colhead{RMS $(68\%)$}} 
\startdata
Hamuy et al. (1996) & $18$ & $ 0.15_{-0.03}^{+0.05} $ & $ 0.17_{-0.03}^{+0.03} $ &$18$ & $ 0.15_{-0.03}^{+0.05} $ & $ 0.17_{-0.03}^{+0.03} $ \\ 
Krisciunas et al. (2005) & $6$ & $ 0.01_{-0.01}^{+0.14} $ & $ 0.11_{-0.03}^{+0.02} $ &$6$ & $ 0.04_{-0.04}^{+0.13} $ & $ 0.11_{-0.03}^{+0.03} $ \\ 
Riess et al. (1999) & $11$ & $ 0.16_{-0.03}^{+0.07} $ & $ 0.17_{-0.04}^{+0.03} $ &$11$ & $ 0.15_{-0.03}^{+0.07} $ & $ 0.17_{-0.04}^{+0.03} $ \\ 
Jha et al. (2006) & $15$ & $ 0.20_{-0.04}^{+0.07} $ & $ 0.22_{-0.04}^{+0.04} $ &$15$ & $ 0.21_{-0.04}^{+0.07} $ & $ 0.22_{-0.04}^{+0.04} $ \\ 
Kowalski et al. (2008) & $8$ & $ 0.04_{-0.04}^{+0.08} $ & $ 0.14_{-0.04}^{+0.03} $ &$8$ & $ 0.07_{-0.06}^{+0.09} $ & $ 0.15_{-0.04}^{+0.03} $ \\ 
Hicken et al. (2009) & $104$ & $ 0.18_{-0.02}^{+0.02} $ & $ 0.21_{-0.01}^{+0.01} $ &$102$ & $ 0.15_{-0.01}^{+0.02} $ & $ 0.19_{-0.01}^{+0.01} $ \\ 
Holtzman et al. (2009) & $133$ & $ 0.19_{-0.01}^{+0.02} $ & $ 0.23_{-0.01}^{+0.01} $ &$129$ & $ 0.10_{-0.01}^{+0.01} $ & $ 0.15_{-0.01}^{+0.01} $ \\ 
Riess et al. (1998) + HZT & $11$ & $ 0.31_{-0.09}^{+0.19} $ & $ 0.53_{-0.12}^{+0.10} $ &$11$ & $ 0.31_{-0.09}^{+0.19} $ & $ 0.52_{-0.12}^{+0.10} $ \\ 
Perlmutter et al. (1999) & $33$ & $ 0.41_{-0.09}^{+0.12} $ & $ 0.64_{-0.08}^{+0.07} $ &$33$ & $ 0.41_{-0.09}^{+0.12} $ & $ 0.64_{-0.08}^{+0.07} $ \\ 
Barris et al. (2004) & $19$ & $ 0.19_{-0.10}^{+0.13} $ & $ 0.39_{-0.07}^{+0.06} $ &$19$ & $ 0.18_{-0.10}^{+0.13} $ & $ 0.38_{-0.07}^{+0.06} $ \\ 
Amanullah et al. (2008) & $5$ & $ 0.18_{-0.06}^{+0.21} $ & $ 0.20_{-0.07}^{+0.05} $ &$5$ & $ 0.19_{-0.06}^{+0.21} $ & $ 0.21_{-0.07}^{+0.05} $ \\ 
Knop et al. (2003) & $11$ & $ 0.04_{-0.04}^{+0.10} $ & $ 0.15_{-0.03}^{+0.03} $ &$11$ & $ 0.05_{-0.05}^{+0.10} $ & $ 0.15_{-0.03}^{+0.03} $ \\ 
Astier et al. (2006) & $73$ & $ 0.18_{-0.03}^{+0.03} $ & $ 0.25_{-0.02}^{+0.02} $ &$72$ & $ 0.13_{-0.02}^{+0.03} $ & $ 0.21_{-0.02}^{+0.02} $ \\ 
Miknaitis et al. (2007) & $77$ & $ 0.25_{-0.03}^{+0.04} $ & $ 0.34_{-0.03}^{+0.03} $ &$74$ & $ 0.19_{-0.03}^{+0.04} $ & $ 0.29_{-0.02}^{+0.02} $ \\ 
Tonry et al. (2003) & $6$ & $ 0.17_{-0.10}^{+0.21} $ & $ 0.24_{-0.07}^{+0.06} $ &$6$ & $ 0.15_{-0.12}^{+0.21} $ & $ 0.23_{-0.07}^{+0.05} $ \\ 
Riess et al. (2007) & $33$ & $ 0.27_{-0.04}^{+0.07} $ & $ 0.49_{-0.06}^{+0.06} $ &$31$ & $ 0.16_{-0.05}^{+0.06} $ & $ 0.45_{-0.06}^{+0.05} $ \\ 
This Paper & $6$ & $ 0.00_{0.00}^{+0.00} $ & $ 0.12_{-0.04}^{+0.03} $ &$6$ & $ 0.00_{0.00}^{+0.00} $ & $ 0.12_{-0.04}^{+0.03} $ \\ 
\hline
Total & 569 & & & 557 & &\\

\enddata
\end{deluxetable*}

\begin{figure*}[tbp]
%\begin{sidewaysfigure*}[p]
%% An update of Figure~8 from~K08 where the data from this work
%% has been added, together with other data sets that has recently
%% been published.
%%
\centering
\includegraphics[width=0.7\textwidth,angle=90]{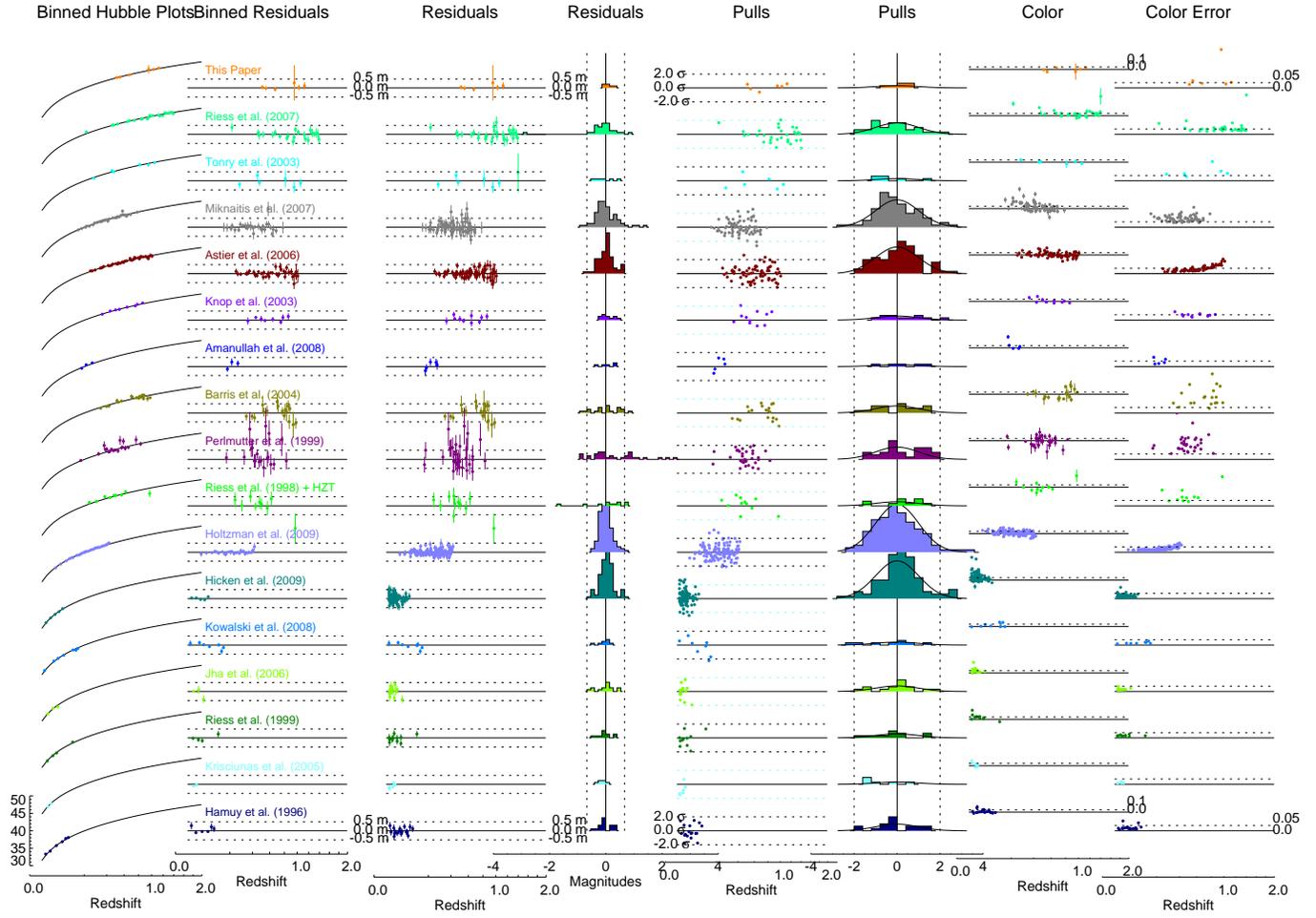}
\caption{Individual diagrams and distributions for the different data
  sets. From left to right: a) Hubble diagrams for the
  various samples; b) binned magnitude residuals from the best fit
  cosmology (bin-width: $\Delta z = 0.01$); c) unbinned magnitude
  residuals from the best fit; d) histogram of the residuals from the
  best fit; e) pull of individual SNe as a function of redshift; f)
  histogram of pulls; g) \sn color as a function of redshift; h)
  uncertainty of the color measurement as an illustration of the
  photometric quality of the data. %
  \label{fig:hubbleindiv}}
\end{figure*}

Figure~\ref{fig:hubbleindiv} shows the individual residuals and pulls
from the best fit cosmology together with the fitted \saltii colors
for the different samples. The photometric quality is illustrated by
the last column in the figure showing the color uncertainty. It is
notable how the photometric quality on the high redshift end has
improved from the K08 analysis. This is due to the extended rest-frame
range of the \saltii model compared to \salt.

%\end{sidewaysfigure*}

\begin{figure*}
\centering
\includegraphics[height=\textwidth,angle=90]{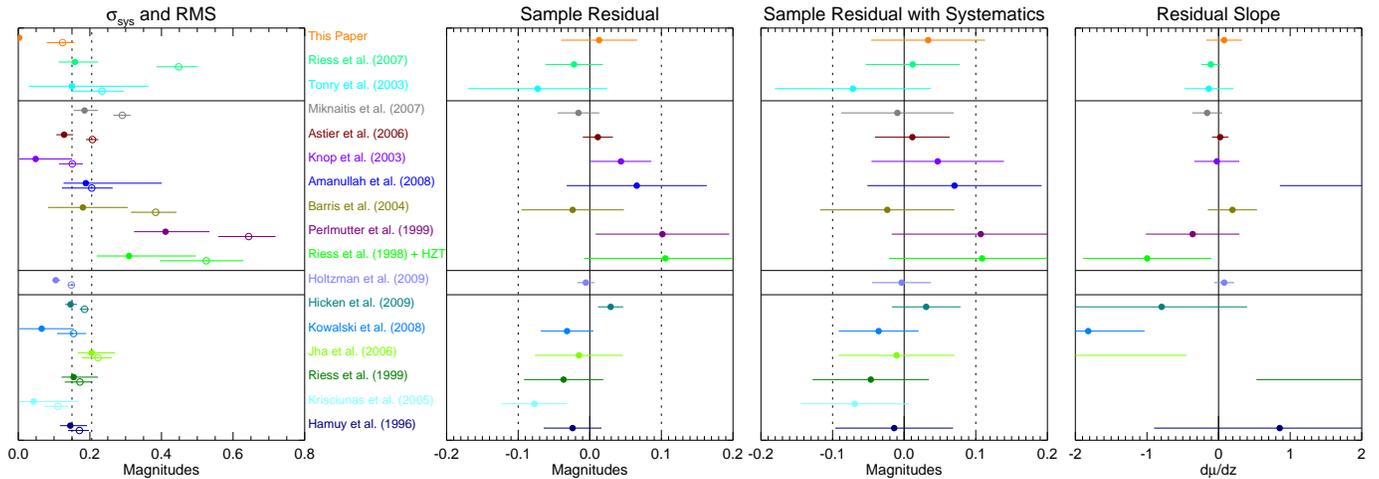}
\caption{Diagnostics plot for the individual data sets.
% The data in this work added to Figure~9 from~K08.
  From left to right: Systematic dispersion (filled circles) and RMS around the
  best fit model (open circles); The mean, sample averaged, deviation
  from the best fit model; The slope of the Hubble-residual (in
  magnitudes) versus redshift, $d\mu_\mathrm{residual}/dz$.
  Note that the errors on the systematic dispersion are the
  statistical errorbars and do not include possible systematic 
  effects such as misestimating photometry errors.
  \label{fig:diagnostics}}
\end{figure*}

Figure~\ref{fig:diagnostics} shows the diagnostics used for
studying the consistency between the different samples. The left panel
shows the fitted $\sigmasys$ values for each sample together with the
RMS around the best fitted cosmology. The intrinsic dispersion
associated with all \sne can be determined as the median of
$\sigmasys$ as long as the majority of the samples are not dominated
by observer-dependent uncertainties that have not been accounted for.
The median $\sigmasys$ for this analysis is $\intdisp$~mag, indicated
by the leftmost dashed vertical line in the figure. The two mid-panels
show the tensions for the individual samples, by comparing the average
residuals from the best-fit cosmology. The two panels show the
tensions without and with systematic errors (described
in~\ref{sec:systematics}) being considered. Most samples fall within
$1\sigma$ and no sample exceeds $2\sigma$. The right panel shows the
tension of the slopes of the residuals as a function of redshift. This
test may not be very meaningful for sparsely sampled data sets, but
could reveal possible Malmquist bias for large data sets.

\begin{figure}
  \centering
  \includegraphics[width=0.5\textwidth]{%
    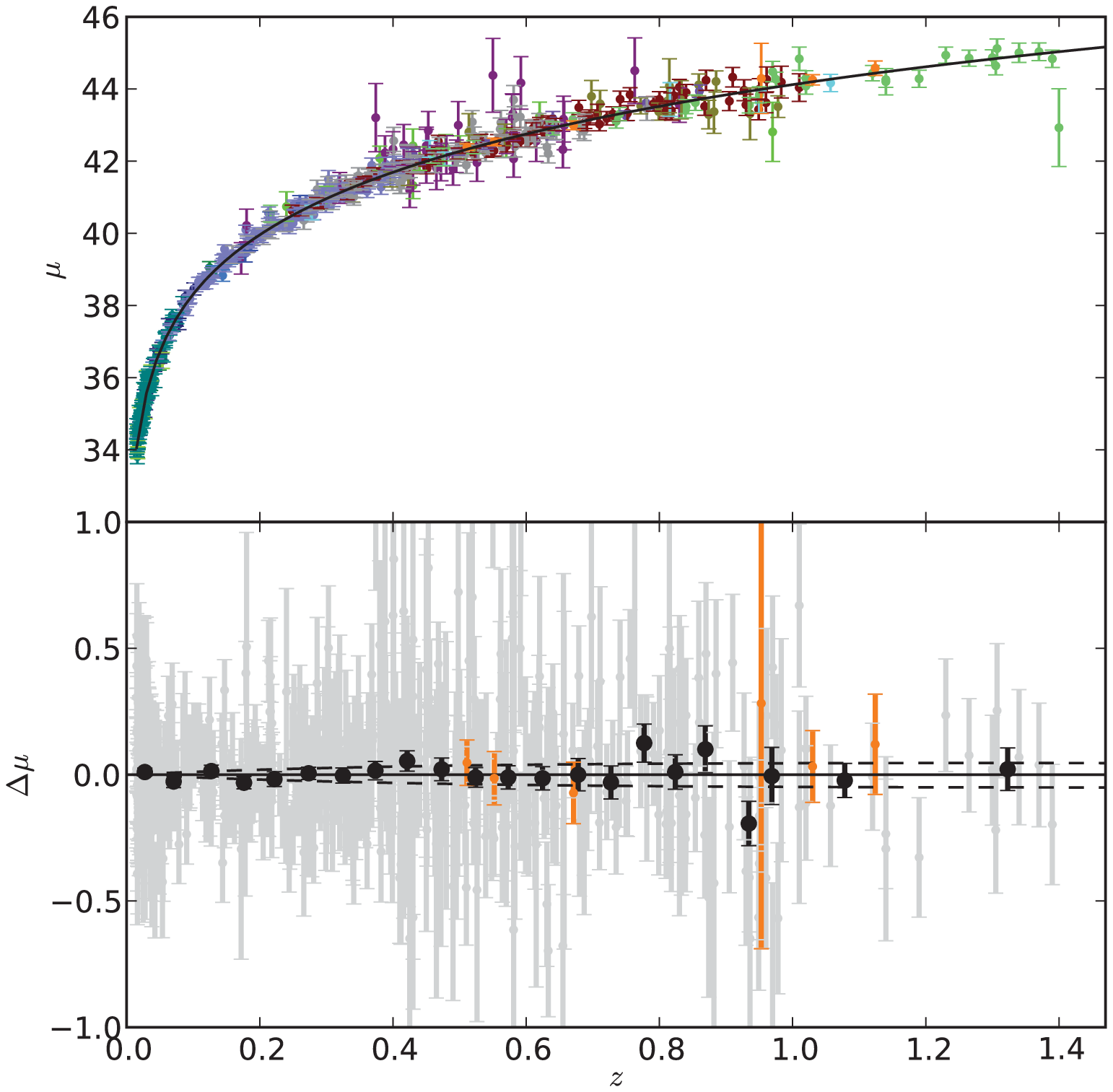}
  \caption{{\em Upper panel:} Hubble diagram for the \compname
    compilation.  The solid line
    represents the best fitted cosmology for a flat Universe including
    the CMB and BAO constraints discussed in the text. The different
    colors have the same interpretation as in
    Figures~\ref{fig:hubbleindiv} and~\ref{fig:diagnostics}.
    {\em Lower panel:} Hubble diagram residuals where the best fitted
    cosmology has been subtracted from the light curve shape and color
    corrected peak magnitudes. The gray points show the residuals
    for individual \sne, while the black points show the binned values in
    redshifts bins of $0.05$ for $z < 1.0$ and $0.2$ for $z > 1.0$.
    The orange points show the previously unpublished \sne introduced
    in this work.
    The dashed lines show the expected Hubble diagram residuals
    for cosmological models with $w\pm0.1$ from the best fitted value.
    \label{fig:hubble}}
\end{figure}

The \sne introduced in this work show no significant tension in any of
the panels. The Hubble residuals for these are also presented in
Figure~\ref{fig:hubble}. Here, the individual \sne are consistent with
the best fit cosmology.

All tables and figures, including the complete covariance matrix for
the sample, are available in electronic format on the Union
webpage\footnote{\texttt{\url{http://supernova.lbl.gov/Union/}}}. We also
provide a CosmoMC module for including this supernova compilation with
other datasets.

\subsection{Systematic errors}\label{sec:systematics}
The K08 analysis split systematic errors into two categories: the
first type affects each \sn sample independently, the second type
affects \sne at similar redshifts. Malmquist bias and uncertainty in
the colors of Vega are examples of the first and second type,
respectively. Typical numbers were derived for both of these types of
systematics, and they were included as covariances\footnote{Note that
  adding a covariance is equivalent to minimizing over a nuisance
  parameter that has a Gaussian prior around zero; the discussion in
  K08 is in terms of these nuisance parameters. This is further
  discussed in Appendix \ref{sec:statistics}.} between \sne. Each
sample received a common covariance, and all of the high-redshift ($z
> 0.2$) \sne shared an additional common covariance.
% This covariance matrix was made available for cosmology fitting.

Other analyses (\cite{2006AA...447...31A}, \cite{2007ApJ...666..694W},
KS09) have estimated the effect on $w$ for each systematic error and
summed these in quadrature. However, \cite{Kim2006451} show that
parameterizing systematic errors (such as uncertain zeropoints) with
nuisance parameters is a more apropriate approach and gives better
cosmological constraints. For this analysis, all contributing factors,
described below, were translated to nuisance parameters, which were
then incorporated into a covariance matrix for the distances of the
individual \sne. Appendix \ref{sec:statistics} contains the details of
converting nuisance parameters to a covariance matrix.

\subsubsection{Zero-point Uncertainties}
In order to correctly propagate calibration uncertainties, we computed
numerically the effect of each photometric passband on the distance
modulus.  For each \sn, the photometry from each band was shifted in
turn by $0.01$~magnitudes and then refit for $\mu$. We then computed the
change in distance modulus,
giving $\frac{d\mu(\alpha, \beta)}{d\mathrm{(ZP})}$ for each band. A list of
zero-point uncertainties is given in Table~\ref{tb:zperrs}.
% Each zero-point uncertainty was thus propagated into the covariance
% matrix.
For two \sne, $i$ and $j$, with calibrated photometry obtained in the
same photometric system, the zero-point uncertainty,
$\sigma_\mathrm{ZP}$ of that system was propagated into the covariance
matrix element $U_{ij}$ as $\frac{d\mu_i}{d\mathrm{(ZP)}}%
\frac{d\mu_j}{d\mathrm{(ZP)}} \sigma_\mathrm{ZP}^2\,$ according to
Appendix~\ref{sec:minsyserr}.

This procedure is a more efficient way of including zero-point
uncertainties than including a common magnitude covariance
(multiplicative in flux space) when performing all of the light curve
fits. In testing, both of these methods gave results that agreed at
the couple of a percent level. Our method has the advantage that the
zero-point errors can be adjusted without rerunning the light curve
fits.

Zero-point uncertainties are one of the largest systematic errors (see
Table~\ref{tb:werrsys}). However, we should note that this number is
based on a heterogeneous assessment of errors from different datasets
(Table~\ref{tb:zperrs}); the accuracy will vary.

%% TABLE WITH ZP UNCERTAINTIES USED
%%
\begin{deluxetable}{llll}
  \tablecaption{Assumed zero-point uncertainties for \sne in
    the \compname compilation. \label{tb:zperrs}}
  \tablehead{
    \colhead{Source} & \colhead{Band} & \colhead{$\sigma_\mathrm{ZP}$} & 
    \colhead{Reference}}
  \startdata
    HST     & WFPC2         & 0.02 & \citet{Heyer:2004}\\
            & ACS           & 0.03 & \citet{Bohlin:2007}\\
            & NICMOS        & 0.03 & \citet{nicmos2009}\\
    \hline
    SNLS    & $g$, $r$, $i$ & 0.01 & \citet{2006AA...447...31A}\\
            & $z$           & 0.03 &\\
    \hline
    ESSENCE & $R$, $I$      & 0.014 & \citet{2007ApJ...666..694W}\\
    \hline
    SDSS    & $u$           & 0.014 & \citet{2009ApJS..185...32K}\\
            & $g$, $r$, $i$ & 0.009 &\\
            & $z$           & 0.010 & \\
    \hline
    This paper & $R$, $I$    & 0.03 & \\
               & $J$     & 0.02 \\
    Other   & $U$-band      & 0.04 & \citet{2009ApJ...700..331H}\\
            & Other Band    & 0.02 & \citet{2009ApJ...700..331H}\\
  \enddata
\end{deluxetable}

\newcommand{\bd}{BD$+17^{\circ}4708$\xspace}

\subsubsection{Vega}
\citet{2006AA...447...31A} estimated the broadband Vega magnitude
system uncertainty to be within $1\,\%$ by comparing spectroscopy from
\citet{1985IAUS..111.....H} and \citet{2004AJ....127.3508B}. In their
analysis, only the uncertainties of Vega colors had implications for
cosmological measurements, which they chose to include by adopting a
flux uncertainty linear in wavelength that would offset the Vega $B-R$
color by 0.01. The uncertainty of Vega is the single largest source
of systematic error when estimating $w$, as shown in
Table~\ref{tb:werrsys}, suggesting that a better-understood reference
would allow for a significant reduction in systematic errors.
  
KS09, and recently SNLS \citep{2009A&A...506..999R}, chose \bd as
their primary reference star. This star has the advantage of having a
well-known SED, measured Landolt magnitudes (in contrast to Vega) and
colors that are close to the average colors of the Landolt standards
(in contrast to Vega which is much bluer). KS09 studied the
implications of switching between \bd and Vega and found zeropoints
consistent to $\sim 1\%$.

% KS09 derives offsets between magnitudes measured by Landolt and
% synthetic magnitudes from HST-measured spectra for BD+17 and three
% other standard stars. They find differences at the $1\%$ level,
% depending on the band.

Given this small difference between using \bd and Vega, we have
chosen, for this work, to continue using Vega as our primary reference
star. To account for the uncertainty of the magnitude of Vega on the
Landolt system, we have assumed a correlated uncertainty of $0.01$~mag
for all photometry with a rest-frame wavelength in each of six
wavelength intervals defined by the following wavelength boundaries:
2900\ang, 4000\ang, 5000\ang, 6000\ang, 7000\ang, 10000\ang,
16000\ang. %

\subsubsection{Rest-frame $U$-Band}\label{sec:restUband}
\sneia are known to show increasing spectroscopic and photometric
diversity for wavelengths shorter than the rest-frame $B$-band. Part
of this could perhaps be explained by differences in progenitor
metallicity \citep{1998ApJ...495..617H,2000ApJ...530..966L}, but the
spectral variations in the rest-frame UV \citep{2008ApJ...674...51E}
are larger than predicted by existing models.

As discussed in Section~\ref{sec:lcfitting}, KS09 studied how well the
\saltii model describes the rest-frame $U$-band by first running
\saltii with the rest frame $U$-band excluded. Using these fits, they
then generated a model for the rest frame $U$-band and binned the
magnitude residuals from the actual rest-frame $U$-band data as a
function of phase. For the \sdss and \snls datasets, the residuals
around the time of maximum are $\sim3\%$. For \snls, this is not
surprising, as the \snls data was used to train the \saltii model.  In
this analysis, we use the \sdss sample as a validation set, and
include a correlated $0.03$~magnitude uncertainty for all photometric
bands bluer than rest-frame 3500\ang.

We note that the HST and low-redshift datasets are less useful for
assessing the size of rest-frame $U$-band uncertainty.  For the \hst
data, the light curves are poorly constrained without the rest-frame
$U$-band. In the case of the nearby sample, the rest-frame $U$-band
overlaps with the observed $U$-band for which accurate photometry is
generally difficult to obtain (any potential problems with the nearby
$U$-band will not impact the light curve fits much, as the low-$z$
fits are typically very well constrained with the remaining bands).

\subsubsection{Malmquist bias}
K08 added a $0.02$~magnitude covariance for each sample representing
Malmquist bias uncertainty. More recently, KS09 completed a thorough
simulation of selection effects for each of the samples in their
analyses. They find a 0.024 change in $w$ when making a correction for
selection effects. As the $0.02$~magnitude covariance yields a quite
similar 0.026 error on $w$, and conducting a full simulation is beyond
the scope of this work, we reuse the covariance from K08.

\subsubsection{Gravitational lensing}
The effects from gravitational lensing on the Hubble diagram have been
discussed in detail in the literature
\citep{1987MNRAS.228..653S,1988A&A...206..190L,2000A&A...358...13B,2003A&A...397..819A,2005ApJ...631..678H}.
Gravitational lensing only affects the high redshift end of the data
that is currently available, and potential bias on the cosmological
parameters from the analysis carried out here due to the asymmetry of
the lensing probability density function is expected to be negligible.
We adopt the K08 approach of only treating gravitational lensing as a
statistical uncertainty by adding a value of $0.093z$
\citep{2005ApJ...631..678H} in quadrature to $\sigmatot$ in
equation~\eqref{eq:cosmochi2}. This is a conservative approach with
respect to the values presented by \citet{2006ApJ...639..991J}, where
they attempt to measure the lensing of individual \sne by determining
the mass distribution along the line of sight. A very important
conclusion of their work is that there is no evidence for selection
effects due to lensing of the high-redshift \sne.

\subsubsection{Light curve model}\label{sec:verifyinglcfitter}
We have studied any potential bias that could arise from poor light
curve sampling, by carrying out a similar analysis to K08, updated for
\saltii. We use nine {\it BVR} AQUAA templates \citep{2007ApJ...671.1084S}
constructed from observations of very well-observed nearby supernovae.
Each set of {\it BVR} templates is combined with a \saltii $U$-band
template generated for that supernova, as there were insufficient
observations in the $U$-band to construct an AQUAA template.

Mock data sets are then sampled from these templates with the same
rest frame dates and signal-to-noise ratios of the real \sne in our
sample. The mock sets are then fitted with \saltii and the offset
between AQUAA corrected magnitude and the corresponding \saltii fitted
value is investigated as a function of the fitted phase of each
supernova's first data point (the phase is with respect to $B$-band
maximum). K08 looked at other possible biases, but first phase was the
only significant one found. The test is carried out for each of the
nine template \sne.

For \sne with a first phase at $B$-band maximum, the average bias is
close to zero, with an RMS of about 0.03. For \sne with a first phase
at six days past $B$-band maximum, the average bias is still close to
zero, but the RMS has increased to about 0.08. Of course, our nine \sn
templates might not be a representative sample, but these results are
encouraging, since they both suggest that there is no significant bias
and indirectly validate the \saltii performance with respect to the
AQUAA templates.

We use a first phase cut of six days, but we conservatively give each
\sn that has first phase greater than zero a $0.03$~magnitude
covariance. Note that \saltii does not stretch its definition of phase
with light curve width.

\subsubsection{Contamination}
As already mentioned, we perform an iterative $\chi^2$ minimization
with a $3\sigma$ outlier rejection before fitting cosmology. In K08 we
showed that this technique greatly reduces the impact of potential
contamination, while maintaining roughly Gaussian statistics. We
carried out a Monte-Carlo study showing that the effect of
contamination on any individual sample is limited to less than
$0.015$~magnitudes. This is under the assumption that the dispersion
of the contaminating distribution is of the same order, or greater
than, the dispersion of \sneia and that the contamination is less than
$30\,\%$.
% For example, suppose that a contaminating distribution is added of
% $1/5$ the number of \sne. If the contaminating objects have a
% dispersion in brightness of $0.5$ magnitudes, then the effect on the
% mean is smaller than $0.013$ magnitudes, regardless of the mean of
% the contaminating distribution.
We include a $0.015$~magnitude uncertainty, correlated for each
sample, to account for possible contamination.

\subsubsection{Minimum redshift}
In order to test for possible effects from using a given minimum
redshift cut, we started by constructing a new sample with no minimum
redshift. Using this sample, we performed fits which allowed the
absolute magnitude to vary independently below and above a dividing
redshift in the range $0.01$ to $0.03$. This procedure should test for
a Hubble bubble or significantly correlated peculiar velocities. The
extra degree of freedom allowed by this step in $M_B$ improved the
$\chi^2$ by $\lesssim 1$ regardless of the dividing redshift and the
inclusion of systematic errors.  This confirms the results of
\citet{2007ApJ...664L..13C} for \saltii. We conclude that there is no
statistically significant difference between minimum redshifts and use
the value of 0.015, as was used in the K08 analysis.

\subsubsection{Galactic extinction}
All light curve photometry is corrected for Galactic extinction using
the extinction law from \citet{1989ApJ...345..245C}, assuming
$R_V=3.1$, together with the dust maps from
\citet{1998ApJ...500..525S}.
      
In the same procedure as with calibration uncertainties, we increased
the Galactic $E(B-V)$ by 0.01 for each supernova and repeated the fit,
giving $\frac{d\mu(\alpha, \beta)}{d (E(B-V))}$. A $10\%$ statistical and $16\%$
systematic error was assumed for the Galactic extinction of each
supernova \citep{1998ApJ...500..525S}.

\subsubsection{Intergalactic extinction}
Dimming of \sneia by hypothetical intergalactic gray dust has been
suggested by \citet{1999ApJ...525..583A} as an alternative to dark
energy to explain the \sn results \citep{2002A&A...384....1G}. This
potential dimming was however constrained by studying the colors of
high-$z$ quasars \citep{2003JCAP...09..009M,2005JCAP...02..005O} and
by observations \sneia in the rest frame I-band
\citep{2005A&A...437..789N,2009ApJ...700.1415N}.
  
Another possible extinction systematic comes from the dust in galaxy
halos that are along the line of sight.  \citet{2009arXiv0902.4240M}
used distant quasars to detect and measure extinction in galactic
halos at $z\sim0.3$. They find an average $R_V$ for their galaxies
of $3.9\pm2.6$. Using their observed $A_V(r)$, we find an average rest
frame $V$-band extinction of $0.004$~magnitudes per intersected halo,
assuming $R_V = 3.1$. At redshift 0.5, an average of three halos have
been intercepted. At redshift 1.0, the average is seven.

There are three mitigating factors. One is that expansion redshifts
photons between the supernova and the intervening galaxy. The CCM law
decreases with wavelength (in the relevant wavelength range), so less
light is absorbed.  \citet{2008MNRAS.385.1053M} finds that
$\rho_{\mathrm{dust}} \propto (1 + z)^{-1.1}$, which we use to scale
the extinction. Finally, most of the extinction is corrected by color
correction. The exact amount corrected depends on the redshift and the
filters used in the observations, but is around two thirds.

We find an error on $w$ of 0.008 due to this extinction, significantly
lower than the value of 0.024 derived by \citet{2009arXiv0903.4199M}.
However, they used $R_V = 3.9$, rather than $3.1$; the fraction of
extinction that is corrected by the color correction will decrease
with $R_V$.  We also numerically sum the CCM laws, rather than using
an analytic approximation. Since we know the exact redshift and
filters used in each observation, we can exactly calculate the amount
of extinction already handled by the color correction (using our
$\frac{d\mu}{d\mathrm{zp}}$ values), without approximation.

\subsubsection{Shape and Color Correction}
The most uncertain contribution to the dimming of \sneia is host
galaxy extinction. Several studies of \snia colors
%\citep[and references
%therein]{2004MNRAS.349.1344A,2005A&A...443..781G,2006AJ....131.1639K,2007AJ....133...58K,2006MNRAS.369.1880E,2007ApJ...664L..13C,
\citep[and references
therein]{2005A&A...443..781G,2007A&A...466...11G,2008ApJ...675..626W,2008A&A...487...19N}
indicate that the observed \snia reddening does not match the
Galactic CCM extinction law with $R_V=3.1$. A stronger wavelength
dependence has been found in the optical in most cases, and it remains
unclear if CCM models with any value of $R_V$ can be used to describe
the data accurately. It is possible that the observed steep reddening
originates from a mixture of local effects and host galaxy extinction.
Local effects could be intrinsic \sn variations, but also multiple
scattering on dust in the circumstellar environment has been suggested
\citep{2005ApJ...635L..33W,2008ApJ...686L.103G}. This model is
potentially supported by detection
\citep{2007Sci...317..924P,2009ApJ...702.1157S,2009ApJ...693..207B} of
circumstellar material but also by color excess measurements for two
of the best observed reddened \snia \citep{2010AJ....139..120F} being
consistent with the expected extinction from circumstellar material.

% While \mlcs attempts to handle intrinsic color variation and host
% galaxy extinction separately, 
The \saltii method approaches the lack of a consistent understanding
of \snia reddening by adopting a purely empirical approach. For
\saltii, the \snia luminosity is standardized by assuming that the
standardization is linear in both $x_1$ and $c$ as described in
equation~\eqref{eq:magcor}, where $\beta$ is the empirically
determined correction coeffecient that accounts for all linear
relations between color and observed peak magnitude. For example if
the only source for such \snia reddening originated from CCM
extinction then $\beta$ is identically equal to $R_V+1$. We test this approach and
propagate relevant systematic uncertainties by dividing the full
sample into smaller sets and carrying out independent fits for the
$x_1$ and $c$ correction coefficients, $\alpha$ and $\beta$, as shown
in Table~\ref{tb:albtredshift}.

When subdividing into redshift bins, we find that the values of
$\alpha$ and $\beta$ for the full sample are consistent with values
for the three first redshift bins. It is encouraging to see
consistency between the global values fit for the full dataset and the
values in the best-understood redshift range. Beyond this important
test, we also note that the value of $\beta$ in the redshift range 0.5
to 1 is significantly lower than the other values, while the value for
$z>1$ is higher than the global value, but is poorly measured. The
trend is similar to what was seen in KS09, but we use different
binning. This behavior is inconsistent with a monotonic drift in
redshift, so we consider other explanations for these results. That
conclusion is also supported by the observation that samples at
similar redshifts (e.g. \citet{2007ApJ...666..674M} and
\citet{2006AA...447...31A}) can have very different values of $\beta$
when fitted independently. The value of $\alpha$ is consistent across
redshift ranges, except at $z>1$, where many light curves are so
poorly sampled that it may not be possible to assign reasonable $x_1$
errors.

When subdividing into the four largest data sources (the lower half of
Table~\ref{tb:albtredshift}), we find values of $\alpha$ and $\beta$
generally consistent with the global values, with the exception of a lower value of $\beta$ for the
SNLS SNe \citep{2006AA...447...31A}. In general, ignoring or underestimating the errors in
$c$ or $x_1$ will decrease the associated correction coefficients, $\beta$ and $\alpha$, as
investigated in K08 and KS09 and this may be relevant here. Specifically, two potential sources of problems are an
incomplete understanding of calibration and underestimated \sn model
variations, either of which could affect these fits. If the SNLS \sne
are physically different, and they are allowed their own $\beta$, then
$w$ shifts by 0.02. Alternatively, as one $\beta$ is used for the global sample, the
possibility that that $\beta$ is biased from the true global value
must be considered. Selecting the global value of $\beta$ from any of
the other large samples shifts $w$ by less than 0.02. We have
accounted for this systematic by assigning each sample a
$0.02$~magnitude covariance (giving a 0.03 error on $w$), which avoids
the problem of handling an error on elements of the covariance matrix.
To study these details further, we look forward to more data for
$z>0.5$, with improved calibration and light curve models.
%For example, KS09
%found that $\beta$ was underestimated by 0.5 in their extensive
%Monte-Carlo study if simulated intrinsic color dispersion was ignored
%in the fit. \cite{2008A&A...487...19N} determined that the intrinsic
%\snia color dispersion is time-dependent, with the light curve shape
%independent $B-V$ dispersion reaching its minimum of $0.03$ at
%$B$-band maximum. Since \saltii is trained on a wide range of \sn
%data, it is expected that the intrinsic \sn variation is absorbed into
%the \saltii model uncertainties. However, if the resulting
%uncertainties of the light curve parameters are incorrect, it could
%bias the fitted cosmological parameters. For example, adding a $0.05$
%magnitude color dispersion to the \saltii $c$ errorbars of
%\citet{2006AA...447...31A} increases $\beta$ for that sample to within
%the errors of the value for the full sample. Adding this dispersion to
%all samples changes $w$ by 0.03, and causes several samples to yield
%$\sigmasys = 0$ in their fits. We thus believe that this 0.03 is a
%likely upper limit for effect of this systematic on $w$.

We also perform one additional sanity check by subdividing the data by
$x_1$ and $c$. There is evidence for two populations of normal \snia,
divided by light curve width (see K08, and references therein).
Star-forming galaxies tend to host the population with broader light
curves, while \sn hosted by passive galaxies tend to have narrower
light curves. As described below, we derive consistent cosmology for
these subdivisions as well.

We subdivide\footnote{Subdividing by $x_1$ or $c$ must be done
  carefully. When there are errors in both the dependent and
  independent variables (in this case, magnitude and $x_1$ or $c$),
  the true values of the independent variables must be explicitly
  solved for as part of the fit. Otherwise, the subdividing will be
  biased. For example, suppose that a supernova has a color that is
  poorly measured, and an uncorrected magnitude that is well-measured.
  If this supernova is faint and blue, then a fit for the true color
  will give a redder color. A color cut will place this supernova in
  the blue category, when the supernova is actually more likely to be
  red. As mentioned in K08, whenever one fits for $\alpha$ and
  $\beta$, the true values of $x_1$ and $c$ are only implicitly solved
  for; equation~\eqref{eq:lcchi2} is derived by analytically
  minimizing over the true $x_1$ and $c$. K08 provides the equation
  with the true values made explicit, we also include a discussion in
  Appendix \ref{sec:statistics}.} the full sample into two roughly
equal subsamples, split first by color and then by $x_1$. In total,
this makes four subsamples. We find that the cosmology is close for
all subsamples (as can be seen in Table~\ref{tb:albtredshift}) so the
difference from these subdivisions does not contribute significantly to the
systematic error on $w$.

It is interesting to note that $\alpha$ is substantially different for
the two samples split by light curve width. Likewise, $\beta$ is
substantially different for the two samples split by color. This might
suggest that the relationships between color and brightness and light
curve width and brightness are more complex than a simple linear
relationship, or it could be that the errors on $x_1$ and $c$ are not
perfectly understood. 
% A non-linear behavior is not unexpected. For example, the color
% brightness relation could result from the sum of two effects: an
% intrinsic redder-dimmer relation and dust extinction from the host
% galaxy.
We also find that $\beta$ is higher for the redder \sneia which is
similar to the results from \citet{2008ApJ...681..482C} based on
comparisons of $U-B$ colors to $B-V$, after correcting for the effect
of stretch on the $U$-band. At the same time it should be pointed out
that evidence of low $R_V$ values have also been found for a few
well-studied, and significantly reddened, nearby \sneia
\citet{2010AJ....139..120F}.

\subsubsection{Summary of systematic errors}
The effect of these systematic errors on $w$ is given in
Table~\ref{tb:werrsys}. The improvement in cosmology constraints over
the simple quadrature sum is also shown. Zeropoint and Vega
calibration dominate the systematics budget, but understanding the
color variations of \sne is also important. The benefit from making a
Malmquist bias correction can be seen; by doing so, KS09 reduce this
systematic error by a factor of two.

\begin{deluxetable}{lr}
  \tablewidth{0pc}
  \tablecaption{Effect on $w$ errorbar (including BAO and CMB
    constraints) for each of the systematic errors included.
    The proper way to sum systematic errors is to include each
    error in a covariance matrix. \label{tb:werrsys}}
  \tablehead{
    \colhead{Source} & \colhead{Error on $w$}}
  \startdata
  Zero point                        & 0.037\\
  Vega                              & 0.042\\
  Galactic Extinction Normalization & 0.012\\
  Rest-Frame $U$-Band               & 0.010\\
  Contamination                     & 0.021\\
  Malmquist Bias                    & 0.026\\
  Intergalactic Extinction          & 0.012\\
  Light curve Shape                 & 0.009\\
  Color Correction                  & 0.026 \\
  \hline
  \textit{Quadrature Sum (not used)}& \textit{0.073}\\
  Summed in Covariance Matrix       & 0.063\\
  \enddata
\end{deluxetable}

\section{Results and Discussion\label{sec:resultsandiscussion}}
\begin{figure*}[tb]
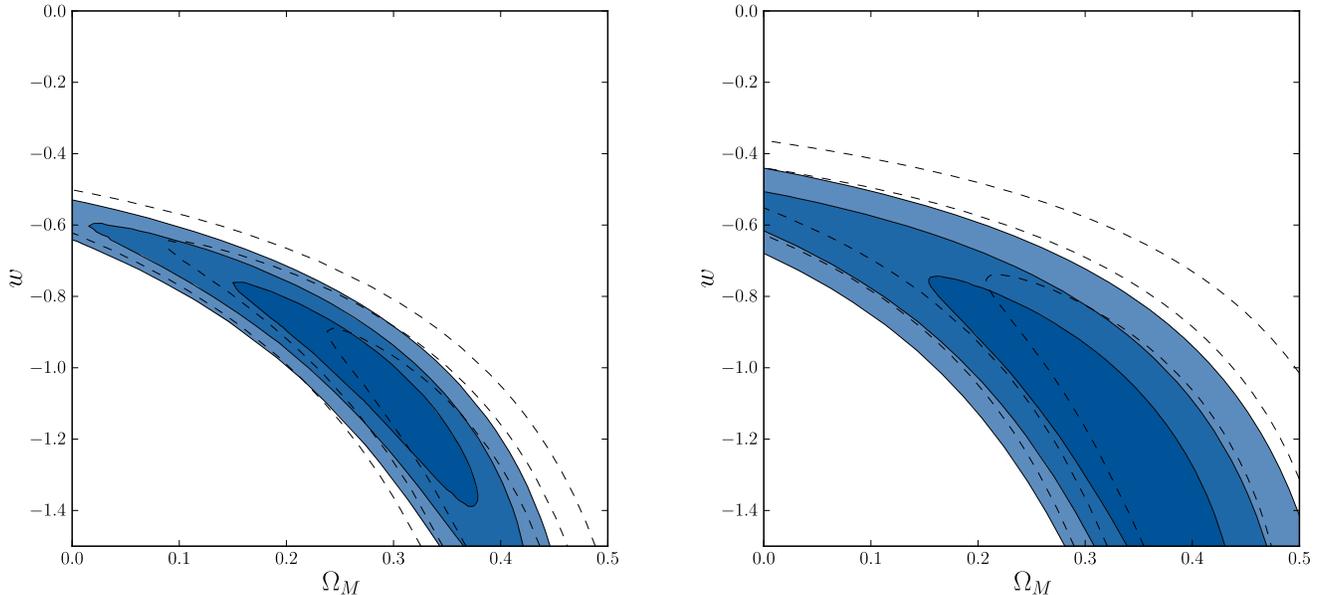

\includegraphics*[width=0.45\textwidth]{%
  figures/current/colorOm-w_comparisoninterp.epsi}%
\hspace{0.06\textwidth}%
\includegraphics*[width=0.45\textwidth]{%
  figures/current/colorOm-w_sys_comparisoninterp.epsi}%
\caption{$68.3\%$, $95.4\%$, and $99.7\%$ confidence regions of the
  $(\om,w)$ plane from \sne alone from K08 (dashed contours) and this
  compilation (shaded contours). Systematic errors are included in the
  right panel. Zero curvature has been assumed.
  \label{fig:omw_compare}}
\end{figure*}

In the cosmology analysis presented here, the statistical errors on
$\om$ have decreased by a significant $24\%$ over the K08 Union
analysis, while the estimated systematic errors have only improved by
$13\%$. When combining the \sn results with BAO and CMB constraints,
statistical errors on $w$ have improved by $16\%$ over K08, though the
quoted systematic errors have increased $7\%$.
Figure~\ref{fig:omw_compare} shows a comparison between the
constraints from K08 and this compilation in the $(\om-w)$ plane. Even
with some improvement on the understanding of systematic errors, it is
clear that the dataset is dominated by systematic error (at least at
low to mid-$z$).

\begin{figure*}[tb]
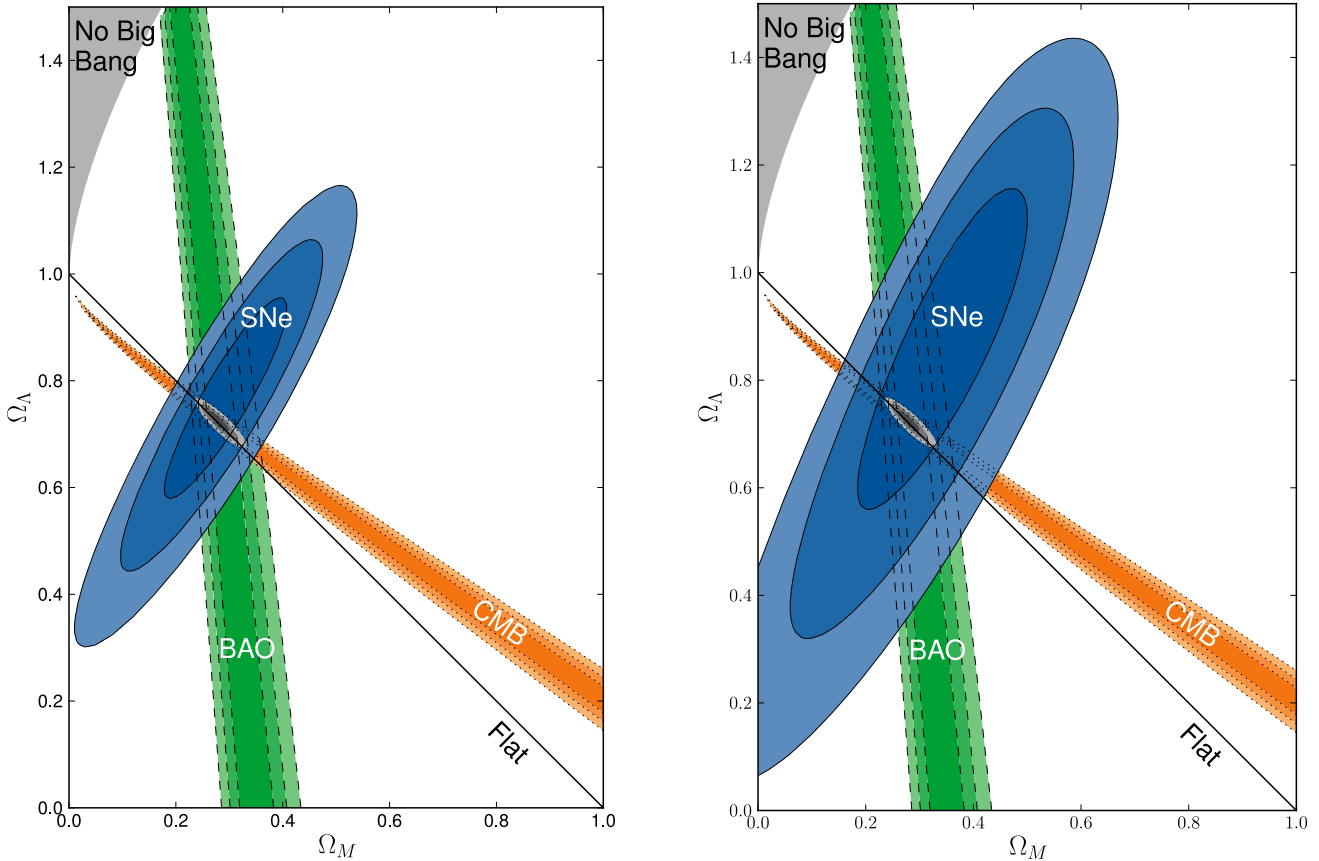

\includegraphics*[width=0.45\textwidth]{%
  figures/current/colorOm-Ol_BAO_CMBinterp.epsi}%
\hspace{0.06\textwidth}%
\includegraphics*[width=0.45\textwidth]{%
  figures/current/colorOm-Ol_BAO_CMB_sysinterp.epsi}%
\caption{$68.3\%$, $95.4\%$, and $99.7\%$ confidence regions in the
  $(\om,\ol)$ plane from \sne combined with the constraints from BAO
  and CMB both without (left panel) and with (right panel) systematic
  errors. Cosmological constant dark energy ($w=-1$) has been assumed.
  \label{fig:omol}}
\end{figure*}

\begin{figure*}[tb]
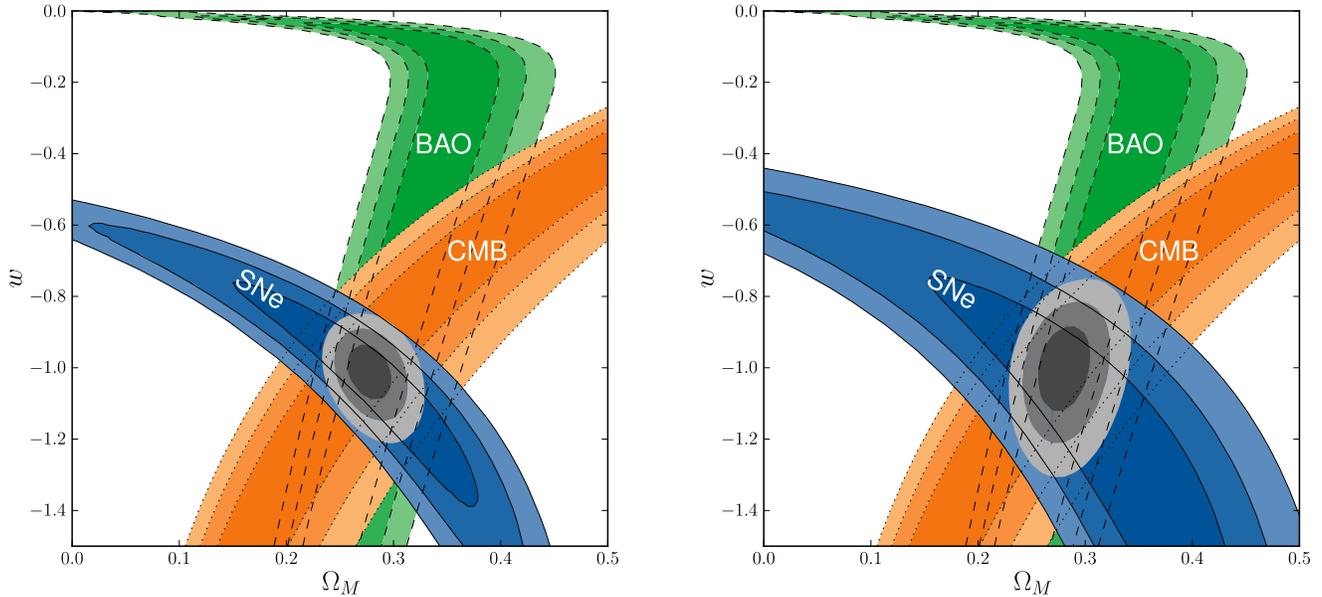

\includegraphics*[width=0.45\textwidth]{%
  figures/current/colorOm-w_interp.epsi}%
\hspace{0.06\textwidth}%
\includegraphics*[width=0.45\textwidth]{%
  figures/current/colorOm-w_sysinterp.epsi}%
\caption{$68.3\%$, $95.4\%$, and $99.7\%$ confidence regions of the
  $(\om,w)$ plane from \sne combined
  with the constraints from BAO and CMB both without (left panel)
  and with (right panel) systematic errors. Zero curvature and
  constant $w$ have been assumed.
  \label{fig:omw}}
\end{figure*}

The best fit cosmological parameters for the compilation are presented
in Table~\ref{tb:cosmoresults} with constraints from CMB and BAO. The
confidence regions in the $(\om,\ol)$ and $(\om,w)$ planes for the
last fit in the table are shown in Figures~\ref{fig:omol}
and~\ref{fig:omw} respectively.

\begin{deluxetable*}{ccccccc}
  \tabletypesize{\footnotesize}
  \tablecaption{Fit results on cosmological parameters $\om$, $w$ and
    $\Omega_k$. The parameter values are followed by their statistical
    (first column)
    and statistical and systematic (second column) uncertainties.
    \label{tb:cosmoresults}}
  \tablehead{
    \colhead{Fit} & \colhead{$\Omega_M$} & \colhead{$\Omega_M$ w/ Sys}
    & \colhead{$\Omega_k$}  & \colhead{$\Omega_k$ w/ Sys} &
    \colhead{$w$} & \colhead{$w$ w/ Sys}}
  \startdata
  SNe & $ 0.270^{+0.021}_{-0.021} $ & $ 0.274^{+0.040}_{-0.037} $ & $0$ (fixed) & $0$ (fixed) & $-1$ (fixed) & $-1$ (fixed) \\
SNe+BAO+$H_0$ & $ 0.309^{+0.032}_{-0.032} $ & $ 0.316^{+0.036}_{-0.035} $ & $0$ (fixed) & $0$ (fixed) & $ -1.114^{+0.098}_{-0.112} $ & $ -1.154^{+0.131}_{-0.150} $\\
SNe+CMB & $ 0.268^{+0.019}_{-0.017} $ & $ 0.269^{+0.023}_{-0.022} $ & $0$ (fixed) & $0$ (fixed) & $ -0.997^{+0.050}_{-0.055} $ & $ -0.999^{+0.074}_{-0.079} $\\
SNe+BAO+CMB & $ 0.277^{+0.014}_{-0.014} $ & $ 0.279^{+0.017}_{-0.016} $ & $0$ (fixed) & $0$ (fixed) & $ -1.009^{+0.050}_{-0.054} $ & $ -0.997^{+0.077}_{-0.082} $\\
SNe+BAO+CMB & $ 0.278^{+0.014}_{-0.014} $ & $ 0.281^{+0.018}_{-0.016} $ & $ -0.004^{+0.006}_{-0.006} $ & $ -0.004^{+0.006}_{-0.007} $ & $-1$ (fixed) & $-1$ (fixed) \\
SNe+BAO+CMB & $ 0.281^{+0.016}_{-0.015} $ & $ 0.281^{+0.018}_{-0.016} $ & $ -0.005^{+0.007}_{-0.007} $ & $ -0.006^{+0.008}_{-0.007} $ & $ -1.026^{+0.055}_{-0.059} $ & $ -1.035^{+0.093}_{-0.097} $\\
  \enddata
\end{deluxetable*}

For the CMB data we implement the constraints from the 7~year data
release of the Wilkinson Microwave Anisotropy Probe (WMAP) as outlined
in \citet{2010arXiv1001.4538K}. We take their results on $\zs$ (the
redshift of last scattering), $l_A(\zs)$, and $R(\zs)$, updating the central values for the cosmological model being considered. Here,
$l_A(\zs)$ is given by
\[
\l_A(\zs) \equiv (1+\zs)\frac{\pi D_A(\zs)}{r_s(\zs)}\,,
\]
where $D_A$ is the angular distance to $\zs$, while
\[
R(\zs) \equiv \frac{\sqrt{\om H_0^2}}{c} (1 + \zs)D_A(\zs)\,.
\]
  
\citet{2010MNRAS.401.2148P} measures the position of the BAO peak from the
SDSS DR7 and 2dFGRS data, constraining $d_z\equiv r_s(z_d)/D_V(0.275)$
to $0.1390\pm0.0037$, where $r_z(z_d)$ is the comoving sound horizon
and $D_V(z)\equiv\left[(1+z)^2D_A^2cz/H(z)\right]^{1/3}$.

% As the CMB temperature only constrains $\Omega_{\gamma}h^2$, the
% sound horizon is largely undetermined without a constraint on $h$.

For the \sne + BAO fit in Table~\ref{tb:cosmoresults}, we add an $H_0$
measurement of $74.2\pm3.6$~km/s/Mpc from \citet{2009ApJ...699..539R},
creating a constraint without the CMB that is therefore largely
independent of the high-redshift behavior of dark energy (as long as
the dark energy density contribution is negligible in the early
universe). Note that the $H_0$ constraint relies on most of the nearby supernovae used in this compilation. However, the effect on $w$ through $H_0$ from these supernovae is several times smaller than the effect through the Hubble diagram. Alternatively, adding a CMB constraint on $\Omega_m h^2$ of
$0.1338\pm0.0058$ from the WMAP7
webpage\footnote{\url{\texttt{http://lambda.gsfc.nasa.gov/product/map/current/params/wcdm\_sz\_lens\_wmap7.cfm}}}
allows us to create a constraint that is independent of $H_0$.  This
final result for \sne+BAO+CMB does not improve significantly if the
current $H_0$ constraint is added.

\subsection{Time variation of the dark energy equation of state}
\begin{figure}[t]
\centering
\includegraphics*[width=0.45\textwidth]{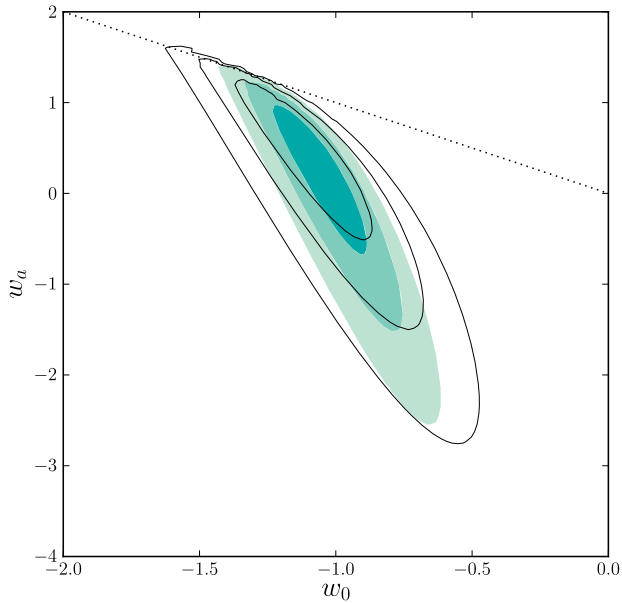}
\caption{$68.3\%$, $95.4\%$, and $99.7\%$ confidence regions of the
  $(w_0,w_a)$ plane from \sne combined with the constraints from BAO
  and CMB both with (solid contours) and without (shaded contours)
  systematic errors. Zero curvature has been assumed. Points above the
  dotted line ($w_0 + w_a = 0$) violate early matter domination and
  are implicitly disfavored in this analysis by the CMB and BAO data.
  \label{fig:w0wa}}
\end{figure}

The constraints shown in Figure~\ref{fig:omw} were obtained assuming
that the dark energy equation of state (EOS) is redshift independent.
\sneia are useful for constraining a redshift dependent $w(z)$ since,
unlike \eg CMB, their measured distances at a given redshift are
independent of the behavior of dark energy at higher redshifts. A
common method to parameterize $w(z)$ is
\[
  w(z) = w_0 + w_a\frac{z}{1+z}
\]
where a cosmological constant is described by $(w_0,w_a)=(-1,0)$. It
can be shown \citep{2003PhRvL..90i1301L} that this parameterization
provides an excellent approximation to a wide variety of dark energy
models. The constraints from the current \sn data together with the
CMB and BAO data are presented in Figure~\ref{fig:w0wa}. The
flattening of the contours in this diagram at $w_0 + w_a = 0$ comes
from the implicit constraint of matter domination in the early
Universe imposed by the CMB and BAO data. Only modest constraints can
currently be placed on $w_a$.

\begin{figure*}[tb]
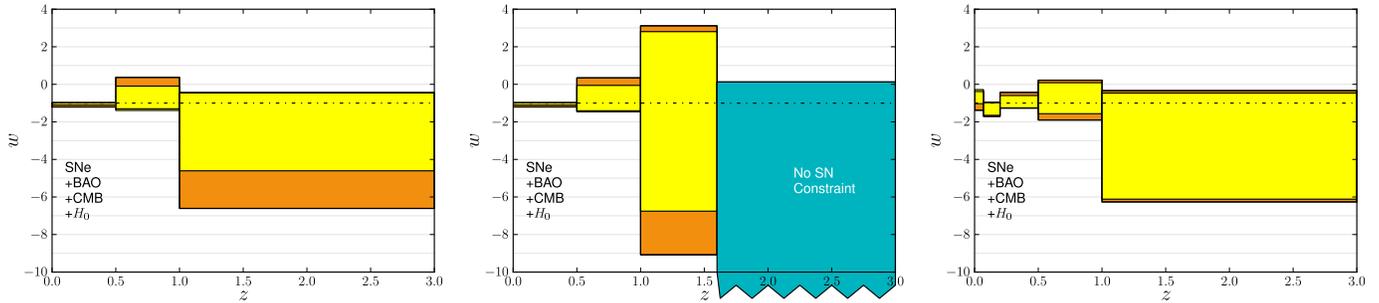

\includegraphics[width=0.32\textwidth]{%
  figures/current/wbins_3_BAO_CMB_H0.epsi}%
\hspace{0.02\textwidth}%
\includegraphics[width=0.32\textwidth]{%
  figures/current/wbins_3+1_BAO_CMB_H0.epsi}%
\hspace{0.02\textwidth}%
\includegraphics[width=0.32\textwidth]{%
  figures/current/wbins_eqerr5_BAO_CMB_H0.epsi}%
\caption{Constraints on $w(z)$, where $w(z)$ is assumed to be
  constant in each redshift bin, are plotted at the 68\% probability
  level. The results were obtained assuming a flat Universe for the
  joint data set of supernovae, BAO, CMB, and $H_0$, with
  (dark/orange) and without (light/yellow) SN systematics. The {\em
    left} panel shows three redshift bins, with the highest redshift
  bin keeping $w$ constant for all $z>1$. Dark energy is seen to exist
  at $z>1$ (at least at $68\,\%$ cl) since $w$ does not reach an
  infinitely negative value, indicating its density does not go to
  zero. The {\em middle} panel splits this last bin into two, showing
  that the seemingly tight constraints on dark energy at $z>1$ with
  current data depend on the combination of CMB with low-redshift
  data. No current probe alone can constrain the existence of dark
  energy at $z>1$. The {\em right} panel shows the effects of $w$
  binning at low redshift. The best fit values of $w$ go from less
  than $-1$ at $z=0.14$ to greater than $-1$ at $z=0.04$. While such a
  steep, late time transition in $w$ (corresponding to $dw/d\ln
  a\approx 7$) is unusual in physical models, it can easily appear due
  to offsets between heterogeneous data sets. We emphasize that the
  results are still consistent with the cosmological constant
  (dot-dashed line) at the $68\,\%$ confidence level.
%
%
%  $68.3\%$ confidence intervals with and without systematic errors for
%  three binned $w$ models. The left plot illustrates that no real
%  constraint can be placed on $w$ for $z > 1$. For three bin (middle)
%  and four bin (right) models, chosen to have roughly equal errors in
%  each bin, the confidence intervals are compatible with $\Lambda$CDM.
%  Note that almost all of the $w$ constraint above redshift 1 for these
%  models comes from supernovae at redshift $\sim 0.5$. Flatness is
%  assumed.
    \label{fig:wbins}}
\end{figure*}

It can be illuminating to study $w(z)$ in redshift bins, where $w$ is
assumed constant in each bin. This method has the advantage that
$w(z)$ can be studied without assuming a specific form for the
relation \citep{2003PhRvL..90c1301H}. We carry out the analysis
following \cite{2008ApJ...686..749K} and fit a constant $w$ in each
bin, while the remaining cosmological parameters are fit globally for
the entire redshift range. Figure~\ref{fig:wbins} shows three such
models for the combined constraints from \sne, BAO, CMB, and $H_0$
measurements where we assume a flat Universe. In these scenarios, the
$H_0$ measurement does not contribute much, but due to its small
improvement on the CMB constraints it gives a small ($\sim10\%$)
improvement on the errorbar of the highest redshift bin.

The left panel shows constraints on $w$ for three bins.  The first bin
($0<z<0.5$) shows a well-constrained $w$.  The middle bin ($0.5<z<1$)
shows a poorly-constrained $w$, though one that is distinct from
$-\infty$ (which would drive $\rho$ to $0$ above $z = 0.5$, resulting
in a matter-only universe) at high confidence, indicating the
detection of some kind of dark energy in this redshift range.  For $z
> 1$, there is little constraint on $w$, and only a weak constraint on
the existence of dark energy.
  
The middle panel shows the effect of dividing the highest redshift
bin. The constraints on $w$ for $z > 1$ get much weaker, showing that
most of the (weak) constraint on the highest bin in the left panel
comes from a combination of the CMB with the well-constrained
low-redshift supernova data. Current supernovae at $z > 1$ offer no
real constraint on $w(z>1)$. Providing a significant constraint at
these redshifts requires significantly better supernova measurements. As in the left panel, $w$ in the highest redshift
bin is constrained to be less than zero by the requirement from BAO
and CMB constraints that the early universe have a matter-dominated
epoch.
  
The right panel shows the effect of dividing the lowest redshift bin.
While no significant change in $w$ with redshift is detected, there is
still considerable room for evolution in $w$, even at low redshift.
  
\begin{figure*}[tb]
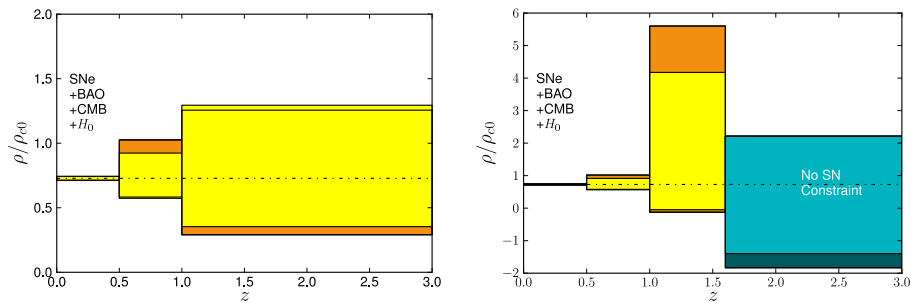

  \centering
  \includegraphics[width=0.32\textwidth]{%
    figures/current/rhobins_3_BAO_CMB_H0.epsi}
\hspace{0.02\textwidth}%
  \includegraphics[width=0.32\textwidth]{%
    figures/current/rhobins_3+1_BAO_CMB_H0.epsi}

  \caption{Here the dark energy density in units of the critical
    density today, $\rho/\rho_{c0}$, is assumed
      constant in each bin. The same binning as the left/center panels in
      Figure~\ref{fig:wbins} is chosen. As can be seen in the left
      panel, dark energy is detected
      between redshift $0.5$ and $1$ at high significance, but only
      hints of dark energy are seen above redshift $1$.  When the CMB
      and \sne are separated, neither one alone can provide any evidence for 
      dark energy at z>1.
      \label{fig:rhobins}}
\end{figure*}

Figure~\ref{fig:rhobins} shows dark energy \emph{density} constraints,
assuming the same redshift binning as in Figure~\ref{fig:wbins}. Note
that this is not equivalent to the left and center panels of
Figure~\ref{fig:wbins}; only in the limit of an infinite number of
bins do binned $\rho$ and binned $w$ give the same model. Dark energy
can be detected at high significance in the middle bin (redshift $0.5$
to $1$), but there is only weak evidence for dark energy above
redshift 1 (left panel). When the bin above redshift 1 is split at a
redshift greater than the supernova sample (right panel), it can be
seen that the current small sample of supernovae cannot constrain the
existence of dark energy above redshift 1.

%The left panel shows a binning of $w$, with one bin starting
%at $z=1$, two chosen to give roughly equal errors, and another
%starting at a redshift higher than the supernova data. There is no
%constraint on $w$ at redshift 1; we cannot even say that there was any
%component other than matter. The middle and right panel show models
%chosen to have roughly equal constraints on three and four bins,
%respectively. For these models, the high-redshift bin dips into the
%mid-redshift range, where constraints on $w$ can be placed.

\ifthenelse{\boolean{salttwo}}{}{
  \subsection{Rest frame UV and SALT2}
  \label{sec:salt2}

  As shown in Figure~\ref{fig:filters}, the bluest filters used for
  the observations partially overlap with the rest frame
  $U$-band. This is a wavelength region where the available \snia
  spectral templates are not well constrained, which is due to the
  fact that ground based observations of the UV-region of nearby
  \sneia are sparse, and because the diversity in \snia \sed is higher
  in the UV than for redder wavelengths
  \citep{2008ApJ...674...51E}. The \saltii \citep{2007A&A...466...11G}
  light curve fitter deals with these issues by deriving its \snia
  \sed
  % light curve fitter deals with this issue by deriving its \snia
  % \sed
  model from nearby \sne and from distant \sne from the \snls project,
  where the $UV$ is redshifted into the optical.  It also incorporates
  model uncertainties, resulting in light curve fits where each
  photometric point is weighted, not only by its measurement
  uncertainty, but also by how well the model is known for the given
  wavelength region and \sn epoch.
  
  \saltii is parameterized in a different manner than \salt, which
  originates from the pseudo-PCA nature of the method. The \saltii
  model consists of three components; a model of the time-dependent
  average \snia \sed, the departure of the \sed from average, and an
  average color correction law. The departure from the average \sed is
  parameterized by with $x_1$, which is closely related to stretch,
  $x_1<0$ and $x_1>0$ represents narrow and broad light curves,
  respectively. The phase-independent color correction is parameterized
  by the deviation, $c$, from the mean \snia $B-V$ color at the time
  of $B$-band maximum.
  
  The results of fitting the \sne in this work with \saltii are 
  presented in Table~\ref{tb:salt2results} and Figure~\ref{fig:hubble}.
  \input{tables/salt2}

  Note that we have not applied the K08 recipe of scaling the fit
  errors to obtain a reduced $\chi^2$ of unity to the \saltii result,
  with the motivation that \saltii already handles model errors by
  construction. There are in general small differences, $<0.04$~mag,
  between the fitted peak $B$-band magnitudes, and slightly larger
  color deviations between the two fitters. These do however propagate
  to differences of up to one standard deviation in the corrected peak
  $B$-band magnitude, but here the values used for $\alpha$ and
  $\beta$ become important. For \saltii the values $\alpha =
  0.13\pm0.01$ and $\beta = 1.77\pm0.16$ \citep{2007A&A...466...11G}
  were used. However, in order to make a fair comparison between
  differences in distance moduli, the full cosmology fit following the
  blinded analysis techniques described in K08 with \saltii must be
  carried out,
  including the fit to $\alpha$ and $\beta$ for the identical sample
  (Rubin \etal in preparation)
}

\subsection{SNe with ground-based near-IR data}\label{sec:zgt1}

Obtaining near-IR data of $z \gtrsim 1$ SNe Ia, whether from space or
from the ground, is critical for constraining the \saltii color
parameter, $c$. Without the near-IR data, the uncertainty in this
parameter for \snhb and \sngn, both beyond $z=1$, increases by a
factor of two. Precise measurements of $c$ are important, since
uncertainties in $c$ are inflated by $\beta\approx2.5$ and tend to
dominate the error budget when the corrected peak $B$-brightness of
\sneia are calculated.

Both \snhb and \sngn were observed with ground-based near-IR
instruments. The operational challenges associated in obtaining these
data are significant. Long exposure times (ten hours or more taken
within a few days) in excellent observing conditions are
necessary. Even with queue mode scheduling, these observations are
just feasible. Despite the challenges, the uncertainty in the \saltii
color of these two \sneia is comparable to the uncertainty in the
color of the {\it best} space-based measured \sneia at $z \gtrsim 1$.

The ground based near-IR data also allow us to search for systematic
offsets with near-IR data taken from space. For $z>1.1$ \sneia
observed with NICMOS, the average \saltii $c$ value is
$c=\wmeancolourNICMOShiz$~mag. By comparison, the weighted average
color of the three \sneia at $z\sim1.1$ with ground-based near-IR data
(\snhb and \sngn from this work, together with \snfk from
\citet{2003ApJ...594....1T}) that pass the light curve cuts is,
$\wmeancolourgroundnir$. Neither the ground-based or space-based
measurements show any Hubble diagram offset, ($\Delta\mu =
\wmeanoffsetgroundnir$ and $\wmeanoffsetNICMOShiz$, respectively),
from the best fit cosmology. These results include a $0.1$~magnitude
dispersion in color, and the fitted systematic dispersions in
magnitude.
%
%Clearly, the number of distant \sneia with ground based near-IR
%observations is currently too small to search for tension between
%space and ground based near-IR data. 
Additional \sneia at $z\sim1.1$ with ground based near-IR photometry
will be published in Suzuki \etal (in preparation) and we will revisit
this issue in that paper, where we will also incorporate the results
from the NICMOS calibration program referenced in section 
\ref{sec:nicmosphot}.

\subsection{Comparison of KeplerCam and SDSS Photometry}
H09 and KS09 share three normal supernovae in common: 2005hc, 2005hj,
and 2005ir. In comparing the data for these \sne, we noticed
  that the H09 KeplerCam $r$-band photometry is generally
  $0.05\pm0.02$ magnitudes fainter than the KS09 SDSS $r$-band
  photometry based on \saltii fits including both data sets. The
  quoted uncertainty is completely dominated by the zero-point
  uncertainties, as both sets of data have high S/N. The offset is
consistent for each supernova. This offset is the correct size and
direction to explain the tension in Hubble residuals seen between
these datasets in second panel of Figure~\ref{fig:diagnostics}.

\sn 2005hc and \sn 2005ir were also observed by CSP using the Swope
telescope. In these cases, the Swope photometry agrees with the SDSS
$r$-band photometry.

\begin{deluxetable*}{rrrrrrr}
  \tabletypesize{\footnotesize} %
  \tablecaption{Subdivisions of the \compname compilation. Values of
    absolute $B$-band magnitude, $M_B$ (assuming $H_0=70$~km/s/Mpc),
    as well as stretch and color correction coefficients, $\alpha$ and
    $\beta$, for several redshift ranges. $\om$ and $w$ are shown for
    properly conducted $x_1$ and $c$ cuts, chosen to give similar
    uncertainty of the fitted $w$. The outlier rejection is redone for
    each bin, so the totals may not add up to the whole sample. The
    constraints are computed including BAO and CMB data.
    \label{tb:albtredshift}}
  \tablehead{
    \colhead{Cut} & \colhead{Number} & \colhead{$M_B$} &
    \colhead{$\alpha$} & \colhead{$\beta$}  & \colhead{$\om$} &
    \colhead{$w$}}
  \startdata
  $ 0.015\leq z \leq  0.10 $ & 166 & $ -19.323^{+0.016}_{-0.016} $ & $ 0.112^{+0.011}_{-0.011} $ & $ 2.77^{+0.09}_{-0.09} $ & $ 0.270 $ (fixed)  & $ -1.000 $ (fixed) \\ 
$ 0.100\leq z \leq  0.25 $ & 74 & $ -19.326^{+0.020}_{-0.020} $ & $ 0.154^{+0.019}_{-0.018} $ & $ 2.49^{+0.15}_{-0.14} $ & $ 0.270 $ (fixed)  & $ -1.000 $ (fixed) \\ 
$ 0.250\leq z \leq  0.50 $ & 154 & $ -19.305^{+0.014}_{-0.014} $ & $ 0.110^{+0.013}_{-0.013} $ & $ 2.50^{+0.12}_{-0.11} $ & $ 0.270 $ (fixed)  & $ -1.000 $ (fixed) \\ 
$ 0.500\leq z \leq  1.00 $ & 133 & $ -19.309^{+0.016}_{-0.017} $ & $ 0.129^{+0.018}_{-0.018} $ & $ 1.45^{+0.19}_{-0.19} $ & $ 0.270 $ (fixed)  & $ -1.000 $ (fixed) \\ 
$ z \geq  1.000 $ & 16 & $ -19.450^{+0.083}_{-0.106} $ & $ -0.124^{+0.085}_{-0.118} $ & $ 3.84^{+1.20}_{-0.85} $ & $ 0.270 $ (fixed)  & $ -1.000 $ (fixed) \\ 
\hline
$c \geq 0.05 $ & 245 & $ -19.373^{+0.026}_{-0.026} $ & $ 0.112^{+0.011}_{-0.011} $ & $ 2.96^{+0.10}_{-0.10} $ & $ 0.283^{+0.017}_{-0.016} $ & $ -0.969^{+0.070}_{-0.074} $\\ 
$c \leq 0.05 $ & 308 & $ -19.305^{+0.019}_{-0.020} $ & $ 0.122^{+0.010}_{-0.010} $ & $ 1.13^{+0.30}_{-0.28} $ & $ 0.284^{+0.015}_{-0.015} $ & $ -0.959^{+0.058}_{-0.062} $\\ 
$x_1 \geq -0.25$ & 302 & $ -19.358^{+0.023}_{-0.023} $ & $ 0.026^{+0.021}_{-0.021} $ & $ 2.60^{+0.10}_{-0.09} $ & $ 0.278^{+0.015}_{-0.014} $ & $ -0.996^{+0.061}_{-0.065} $\\ 
$x_1 \leq -0.25$ & 254 & $ -19.335^{+0.031}_{-0.032} $ & $ 0.147^{+0.020}_{-0.019} $ & $ 2.42^{+0.10}_{-0.10} $ & $ 0.276^{+0.016}_{-0.015} $ & $ -1.016^{+0.069}_{-0.073} $\\ 
\hline
Holtzman et al. (2009) & 129 & $ -19.315^{+0.013}_{-0.013} $ & $ 0.147^{+0.014}_{-0.013} $ & $ 2.38^{+0.15}_{-0.14} $ & $ 0.270 $ (fixed)  & $ -1.000 $ (fixed) \\ 
Hicken et al. (2009) & 102 & $ -19.299^{+0.021}_{-0.022} $ & $ 0.113^{+0.013}_{-0.013} $ & $ 2.73^{+0.10}_{-0.10} $ & $ 0.270 $ (fixed)  & $ -1.000 $ (fixed) \\ 
Miknaitis et al. (2007) & 74 & $ -19.325^{+0.032}_{-0.033} $ & $ 0.112^{+0.037}_{-0.035} $ & $ 2.50^{+0.17}_{-0.16} $ & $ 0.270 $ (fixed)  & $ -1.000 $ (fixed) \\ 
Astier et al. (2006) & 71 & $ -19.287^{+0.016}_{-0.017} $ & $ 0.140^{+0.017}_{-0.017} $ & $ 1.72^{+0.18}_{-0.17} $ & $ 0.270 $ (fixed)  & $ -1.000 $ (fixed) \\ 
\hline
$ z \geq  0.015 $ & 557 & $ -19.310^{+0.014}_{-0.014} $ & $ 0.121^{+0.007}_{-0.007} $ & $ 2.51^{+0.07}_{-0.07} $ & $ 0.277^{+0.014}_{-0.014} $ & $ -1.009^{+0.050}_{-0.054} $

  \enddata
\end{deluxetable*}

\section{Summary and Conclusions}\label{sec:conclusions}
We have presented the light curves of \sneia in the redshift range
$0.511<z<1.12$. The \sne were discovered as part of a search conducted
by the Supernova Cosmology Project using the Subaru, CFHT and CTIO
Blanco telescopes in~2001. The fitted light curve shapes and colors at
maximum light are all consistent with the corresponding distributions
from previous \sn surveys.

Following K08, we add these six \sne and other \snia data sets to the
Union compilation. We have also improved the Union analysis in a
number of respects, creating the new \compname compilation.  The most
important improvements are: (1) Systematic errors are directly
computed using the effect they have on the distance modulus (2) All
\sn light curves are fitted with the \saltii light curve fitter.

We determine the best fit cosmology for the \compname compilation, and
the concordance $\Lambda$CDM model remains an excellent fit. The new
analysis results in a significant improvement over K08 in constraining
$w$ over the redshift interval $0 < z < 1$.  Above $z \gtrsim 1$,
evidence for dark energy is weak.  This will remain the case until
there is much more high redshift data, with better signal-to-noise and
wavelength coverage.

\acknowledgements
% \section{Acknowledgments}
Based, in part, on observations obtained at the ESO La Silla Paranal
Observatory (ESO programs 67.A-0361 and 169.A-0382). Based, in part,
on observations obtained at the Cerro Tololo Inter-American
Observatory, National Optical Astronomy Observatory, which are
operated by the Association of Universities for Research in Astronomy,
under contract with the National Science Foundation. Based, in part,
on observations obtained at the Gemini Observatory (Gemini programs
GN-2001A-SV-19 and GN-2002A-Q-31), which is operated by the
Association of Universities for Research in Astronomy, Inc., under a
cooperative agreement with the NSF on behalf of the Gemini
partnership: the National Science Foundation (United States), the
Particle Physics and Astronomy Research Council (United Kingdom), the
National Research Council (Canada), CONICYT (Chile), the Australian
Research Council (Australia), CNPq (Brazil) and CONICET (Argentina).
Based, in part on observations obtained at the Subaru Telescope, which
is operated by the National Observatory of Japan. Some of imaging data
taken with Suprime-Cam were obtained during the commissioning phase of
the instrument, and we would like to thank all members of Suprime-Cam
instrument team. Based, in part, on data that were obtained at the
W.M. Keck Observatory, which is operated as a scientific partnership
among the California Institute of Technology, the University of
California and the National Aeronautics and Space Administration. The
Observatory was made possible by the generous financial support of the
W.M. Keck Foundation. Based on observations obtained at the
Canada-France-Hawaii Telescope (CFHT) which is operated by the
National Research Council of Canada, the Institut National des
Sciences de l'Univers of the Centre National de la Recherche
Scientifique of France, and the University of Hawaii. The authors wish
to recognize and acknowledge the very significant cultural role and
reverence that the summit of Mauna Kea has always had within the
indigenous Hawaiian community. We are most fortunate to have the
opportunity to conduct observations from this mountain.

Based, in part, on observations made with the NASA/ESA Hubble Space
Telescope, obtained at the Space Telescope Science Institute, which is
operated by the Association of Universities for Research in Astronomy,
Inc., under NASA contract NAS 5-26555. These observations are
associated with programs HST-GO-08585 and HST-GO-09075.

Support for programs HST-GO-08585.14-A and HST-GO-09075.01-A was
provided by NASA through a grant from the Space Telescope Science
Institute, which is operated by the Association of Universities for
Research in Astronomy, Inc., under NASA contract NAS 5-26555.

This work is supported in part by a JSPS core-to-core program
``International Research Network for Dark Energy'' and by JSPS research
grants (20040002).

This work was supported by the Director, Office of Science, Office of
High Energy Physics, of the U.S.\ Department of Energy under Contract
No. DE- AC02-05CH11231.

T.M. is financially supported by the Japan Society for the Promotion
of Science (JSPS) through the JSPS Research Fellowship.

C.L. acknowledges the support provided by the Oskar Klein Centre at
Stockholm University.

The authors would like to thank the anonymous referee for its helpful
comments and suggestions.

{\it Facilities:} \facility{HST (WFPC2,ACS)}, \facility{Keck:I (LRIS)}, \facility{Keck:II (ESI)},
\facility{CFHT (\cfhk)}, \facility{Gemini:Gillett (NIRI)}, \facility{Subaru
  (SuprimeCam)}, \facility{Blanco (\mosaic)}, \facility{NTT (\susi)}, \facility{VLT:Antu (\fors,\isaac)}

\bibliography{rahman}

\begin{thebibliography}{120}
\expandafter\ifx\csname natexlab\endcsname\relax\def\natexlab#1{#1}\fi

\bibitem[{{Adelman-McCarthy} {et~al.}(2007){Adelman-McCarthy}, {Ag{\"u}eros},  {Allam}, {Anderson}, {Anderson}, {Annis}, {Bahcall}, {Bailer-Jones},  {Baldry}, {Barentine}, {Beers}, {Belokurov}, {Berlind}, {Bernardi},  {Blanton}, {Bochanski}, {Boroski}, {Bramich}, {Brewington}, {Brinchmann},  {Brinkmann}, {Brunner}, {Budav{\'a}ri}, {Carey}, {Carliles}, {Carr},  {Castander}, {Connolly}, {Cool}, {Cunha}, {Csabai}, {Dalcanton}, {Doi},  {Eisenstein}, {Evans}, {Evans}, {Fan}, {Finkbeiner}, {Friedman}, {Frieman},  {Fukugita}, {Gillespie}, {Gilmore}, {Glazebrook}, {Gray}, {Grebel}, {Gunn},  {de Haas}, {Hall}, {Harvanek}, {Hawley}, {Hayes}, {Heckman}, {Hendry},  {Hennessy}, {Hindsley}, {Hirata}, {Hogan}, {Hogg}, {Holtzman}, {Ichikawa},  {Ichikawa}, {Ivezi{\'c}}, {Jester}, {Johnston}, {Jorgensen}, {Juri{\'c}},  {Kauffmann}, {Kent}, {Kleinman}, {Knapp}, {Kniazev}, {Kron}, {Krzesinski},  {Kuropatkin}, {Lamb}, {Lampeitl}, {Lee}, {Leger}, {Lima}, {Lin}, {Long},  {Loveday}, {Lupton}, {Mandelbaum}, {Margon}, {Mart{\'{\i}}nez-Delgado},  {Matsubara}, {McGehee}, {McKay}, {Meiksin}, {Munn}, {Nakajima}, {Nash},  {Neilsen}, {Newberg}, {Nichol}, {Nieto-Santisteban}, {Nitta}, {Oyaizu},  {Okamura}, {Ostriker}, {Padmanabhan}, {Park}, {Peoples}, {Pier}, {Pope},  {Pourbaix}, {Quinn}, {Raddick}, {Re Fiorentin}, {Richards}, {Richmond},  {Rix}, {Rockosi}, {Schlegel}, {Schneider}, {Scranton}, {Seljak}, {Sheldon},  {Shimasaku}, {Silvestri}, {Smith}, {Smol{\v c}i{\'c}}, {Snedden}, {Stebbins},  {Stoughton}, {Strauss}, {SubbaRao}, {Suto}, {Szalay}, {Szapudi}, {Szkody},  {Tegmark}, {Thakar}, {Tremonti}, {Tucker}, {Uomoto}, {Vanden Berk},  {Vandenberg}, {Vidrih}, {Vogeley}, {Voges}, {Vogt}, {Weinberg}, {West},  {White}, {Wilhite}, {Yanny}, {Yocum}, {York}, {Zehavi}, {Zibetti}, \&  {Zucker}}]{2007ApJS..172..634A}
{Adelman-McCarthy}, J.~K., {et~al.} 2007, \apjs, 172, 634

\bibitem[{{Aguirre}(1999)}]{1999ApJ...525..583A}
{Aguirre}, A. 1999, \apj, 525, 583

\bibitem[{{Alard}(2000)}]{2000A&AS..144..363A}
{Alard}, C. 2000, \aaps, 144, 363

\bibitem[{{Alard} \& {Lupton}(1998)}]{1998ApJ...503..325A}
{Alard}, C., \& {Lupton}, R.~H. 1998, \apj, 503, 325

\bibitem[{{Amanullah} {et~al.}(2003){Amanullah}, {M{\"o}rtsell}, \&  {Goobar}}]{2003A&A...397..819A}
{Amanullah}, R., {M{\"o}rtsell}, E., \& {Goobar}, A. 2003, \aap, 397, 819

\bibitem[{{Amanullah} {et~al.}(2008){Amanullah}, {Stanishev}, {Goobar},  {Schahmaneche}, {Astier}, {Balland}, {Ellis}, {Fabbro}, {Hardin}, {Hook},  {Irwin}, {McMahon}, {Mendez}, {Mouchet}, {Pain}, {Ruiz-Lapuente}, \&  {Walton}}]{2008A&A...486..375A}
{Amanullah}, R., {et~al.} 2008, \aap, 486, 375

\bibitem[{{Anderson} \& {King}(2003)}]{2003PASP..115..113A}
{Anderson}, J., \& {King}, I.~R. 2003, \pasp, 115, 113

\bibitem[{{Appenzeller} {et~al.}(1998){Appenzeller}, {Fricke}, {F{\"u}rtig},  {G{\"a}ssler}, {H{\"a}fner}, {Harke}, {Hess}, {Hummel}, {J{\"u}rgens},  {Kudritzki}, {Mantel}, {Meisl}, {Muschielok}, {Nicklas}, {Rupprecht},  {Seifert}, {Stahl}, {Szeifert}, \& {Tarantik}}]{1998Msngr..94....1A}
{Appenzeller}, I., {et~al.} 1998, The Messenger, 94, 1

\bibitem[{{Astier} {et~al.}(2006){Astier}, {Guy}, {Regnault}, {Pain},  {Aubourg}, {Balam}, {Basa}, {Carlberg}, {Fabbro}, {Fouchez}, {Hook},  {Howell}, {Lafoux}, {Neill}, {Palanque-Delabrouille}, {Perrett}, {Pritchet},  {Rich}, {Sullivan}, {Taillet}, {Aldering}, {Antilogus}, {Arsenijevic},  {Balland}, {Baumont}, {Bronder}, {Courtois}, {Ellis}, {Filiol}, {Gon{\c  c}alves}, {Goobar}, {Guide}, {Hardin}, {Lusset}, {Lidman}, {McMahon},  {Mouchet}, {Mourao}, {Perlmutter}, {Ripoche}, {Tao}, \&  {Walton}}]{2006AA...447...31A}
{Astier}, P., {et~al.} 2006, \aap, 447, 31

\bibitem[{{Barris} {et~al.}(2004){Barris}, {Tonry}, {Blondin}, {Challis},  {Chornock}, {Clocchiatti}, {Filippenko}, {Garnavich}, {Holland}, {Jha},  {Kirshner}, {Krisciunas}, {Leibundgut}, {Li}, {Matheson}, {Miknaitis},  {Riess}, {Schmidt}, {Smith}, {Sollerman}, {Spyromilio}, {Stubbs}, {Suntzeff},  {Aussel}, {Chambers}, {Connelley}, {Donovan}, {Henry}, {Kaiser}, {Liu},  {Mart{\'{\i}}n}, \& {Wainscoat}}]{2004ApJ...602..571B}
{Barris}, B.~J., {et~al.} 2004, \apj, 602, 571

\bibitem[{{Bergstr{\"o}m} {et~al.}(2000){Bergstr{\"o}m}, {Goliath}, {Goobar},  \& {M{\"o}rtsell}}]{2000A&A...358...13B}
{Bergstr{\"o}m}, L., {Goliath}, M., {Goobar}, A., \& {M{\"o}rtsell}, E. 2000,  \aap, 358, 13

\bibitem[{{Bessell}(1990)}]{1990PASP..102.1181B}
{Bessell}, M.~S. 1990, \pasp, 102, 1181

\bibitem[{{Blondin} {et~al.}(2009){Blondin}, {Prieto}, {Patat}, {Challis},  {Hicken}, {Kirshner}, {Matheson}, \& {Modjaz}}]{2009ApJ...693..207B}
{Blondin}, S., {Prieto}, J.~L., {Patat}, F., {Challis}, P., {Hicken}, M.,  {Kirshner}, R.~P., {Matheson}, T., \& {Modjaz}, M. 2009, \apj, 693, 207

\bibitem[{{Bohlin}(2007)}]{2007ASPC..364..315B}
{Bohlin}, R.~C. 2007, in Astronomical Society of the Pacific Conference Series,  Vol. 364, The Future of Photometric, Spectrophotometric and Polarimetric  Standardization, ed. {C.~Sterken}, 315--+

\bibitem[{Bohlin(2007)}]{Bohlin:2007}
Bohlin, R.~C. 2007, Photometric Calibration of the ACS CCD Cameras

\bibitem[{{Bohlin} \& {Gilliland}(2004)}]{2004AJ....127.3508B}
{Bohlin}, R.~C., \& {Gilliland}, R.~L. 2004, \aj, 127, 3508

\bibitem[{{Cardelli} {et~al.}(1989){Cardelli}, {Clayton}, \&  {Mathis}}]{1989ApJ...345..245C}
{Cardelli}, J.~A., {Clayton}, G.~C., \& {Mathis}, J.~S. 1989, \apj, 345, 245

\bibitem[{{Conley} {et~al.}(2007){Conley}, {Carlberg}, {Guy}, {Howell}, {Jha},  {Riess}, \& {Sullivan}}]{2007ApJ...664L..13C}
{Conley}, A., {Carlberg}, R.~G., {Guy}, J., {Howell}, D.~A., {Jha}, S.,  {Riess}, A.~G., \& {Sullivan}, M. 2007, \apjl, 664, L13

\bibitem[{{Conley} {et~al.}(2006){Conley}, {Goldhaber}, {Wang}, {Aldering},  {Amanullah}, {Commins}, {Fadeyev}, {Folatelli}, {Garavini}, {Gibbons},  {Goobar}, {Groom}, {Hook}, {Howell}, {Kim}, {Knop}, {Kowalski}, {Kuznetsova},  {Lidman}, {Nobili}, {Nugent}, {Pain}, {Perlmutter}, {Smith}, {Spadafora},  {Stanishev}, {Strovink}, {Thomas}, \& {Wood-Vasey}}]{2006ApJ...644....1C}
{Conley}, A., {et~al.} 2006, \apj, 644, 1

\bibitem[{{Conley} {et~al.}(2008){Conley}, {Sullivan}, {Hsiao}, {Guy},  {Astier}, {Balam}, {Balland}, {Basa}, {Carlberg}, {Fouchez}, {Hardin},  {Howell}, {Hook}, {Pain}, {Perrett}, {Pritchet}, \&  {Regnault}}]{2008ApJ...681..482C}
{Conley}, A., {et~al.} 2008, \apj, 681, 482

\bibitem[{{Copin} {et~al.}(2006){Copin}, {Blanc}, {Bongard}, {Gangler},  {Saug{\'e}}, {Smadja}, {Antilogus}, {Garavini}, {Gilles}, {Pain}, {Aldering},  {Bailey}, {Lee}, {Loken}, {Nugent}, {Perlmutter}, {Scalzo}, {Thomas}, {Wang},  {Weaver}, {P{\'e}contal}, {Kessler}, {Baltay}, {Rabinowitz}, \&  {Bauer}}]{2006NewAR..50..436C}
{Copin}, Y., {et~al.} 2006, New Astronomy Review, 50, 436

\bibitem[{Cuillandre {et~al.}(2000)Cuillandre, Luppino, Starr, \&  Isani}]{cuillandre:2000}
Cuillandre, J.-C., Luppino, G., Starr, B., \& Isani, S. 2000, Astronomical  Telescopes and Instrumentation, 4008, 1010

\bibitem[{{D.~S.~Hayes, L.~E.~Pasinetti, \&  A.~G.~D.~Philip}(1985)}]{1985IAUS..111.....H}
{D.~S.~Hayes, L.~E.~Pasinetti, \& A.~G.~D.~Philip}, ed. 1985, IAU Symposium,  Vol. 111, {Calibration of fundamental stellar quantities; Proceedings of the  Symposium, Como, Italy, May 24-29, 1984}, ed. {D.~S.~Hayes, L.~E.~Pasinetti,  \& A.~G.~D.~Philip}

\bibitem[{{Dawson} {et~al.}(2009){Dawson}, {Aldering}, {Amanullah}, {Barbary},  {Barrientos}, {Brodwin}, {Connolly}, {Dey}, {Doi}, {Donahue}, {Eisenhardt},  {Ellingson}, {Faccioli}, {Fadeyev}, {Fakhouri}, {Fruchter}, {Gilbank},  {Gladders}, {Goldhaber}, {Gonzalez}, {Goobar}, {Gude}, {Hattori}, {Hoekstra},  {Huang}, {Ihara}, {Jannuzi}, {Johnston}, {Kashikawa}, {Koester}, {Konishi},  {Kowalski}, {Lidman}, {Linder}, {Lubin}, {Meyers}, {Morokuma}, {Munshi},  {Mullis}, {Oda}, {Panagia}, {Perlmutter}, {Postman}, {Pritchard}, {Rhodes},  {Rosati}, {Rubin}, {Schlegel}, {Spadafora}, {Stanford}, {Stanishev}, {Stern},  {Strovink}, {Suzuki}, {Takanashi}, {Tokita}, {Wagner}, {Wang}, {Yasuda},  {Yee}, \& {Supernova Cosmology Project}}]{dawson:2009}
{Dawson}, K.~S., {et~al.} 2009, \aj, 138, 1271

\bibitem[{{D'Odorico} {et~al.}(1998){D'Odorico}, {Beletic}, {Amico}, {Hook},  {Marconi}, \& {Pedichini}}]{1998SPIE.3355..507D}
{D'Odorico}, S., {Beletic}, J.~W., {Amico}, P., {Hook}, I., {Marconi}, G., \&  {Pedichini}, F. 1998, in Presented at the Society of Photo-Optical  Instrumentation Engineers (SPIE) Conference, Vol. 3355, Proc. SPIE Vol. 3355,  p. 507-511, Optical Astronomical Instrumentation, Sandro D'Odorico; Ed., ed.  S.~{D'Odorico}, 507--511

\bibitem[{{Dolphin}(2000)}]{2000PASP..112.1397D}
{Dolphin}, A.~E. 2000, \pasp, 112, 1397

\bibitem[{{Dolphin}(2009)}]{2009PASP..121..655D}
---. 2009, \pasp, 121, 655

\bibitem[{{Eisenstein} {et~al.}(2005){Eisenstein}, {Zehavi}, {Hogg},  {Scoccimarro}, {Blanton}, {Nichol}, {Scranton}, {Seo}, {Tegmark}, {Zheng},  {Anderson}, {Annis}, {Bahcall}, {Brinkmann}, {Burles}, {Castander},  {Connolly}, {Csabai}, {Doi}, {Fukugita}, {Frieman}, {Glazebrook}, {Gunn},  {Hendry}, {Hennessy}, {Ivezi{\'c}}, {Kent}, {Knapp}, {Lin}, {Loh}, {Lupton},  {Margon}, {McKay}, {Meiksin}, {Munn}, {Pope}, {Richmond}, {Schlegel},  {Schneider}, {Shimasaku}, {Stoughton}, {Strauss}, {SubbaRao}, {Szalay},  {Szapudi}, {Tucker}, {Yanny}, \& {York}}]{2005ApJ...633..560E}
{Eisenstein}, D.~J., {et~al.} 2005, \apj, 633, 560

\bibitem[{{Ellis} {et~al.}(2008){Ellis}, {Sullivan}, {Nugent}, {Howell},  {Gal-Yam}, {Astier}, {Balam}, {Balland}, {Basa}, {Carlberg}, {Conley},  {Fouchez}, {Guy}, {Hardin}, {Hook}, {Pain}, {Perrett}, {Pritchet}, \&  {Regnault}}]{2008ApJ...674...51E}
{Ellis}, R.~S., {et~al.} 2008, \apj, 674, 51

\bibitem[{Fabbro(2001)}]{fabbro:2001}
Fabbro, S. 2001, PhD thesis, l'Universit\'e Paris XI Orsay

\bibitem[{{Filippenko} {et~al.}(2001){Filippenko}, {Li}, {Treffers}, \&  {Modjaz}}]{2001ASPC..246..121F}
{Filippenko}, A.~V., {Li}, W.~D., {Treffers}, R.~R., \& {Modjaz}, M. 2001, in  Astronomical Society of the Pacific Conference Series, Vol. 246, IAU Colloq.  183: Small Telescope Astronomy on Global Scales, ed. B.~{Paczynski}, W.-P.  {Chen}, \& C.~{Lemme}, 121--+

\bibitem[{{Folatelli} {et~al.}(2010){Folatelli}, {Phillips}, {Burns},  {Contreras}, {Hamuy}, {Freedman}, {Persson}, {Stritzinger}, {Suntzeff},  {Krisciunas}, {Boldt}, {Gonz{\'a}lez}, {Krzeminski}, {Morrell}, {Roth},  {Salgado}, {Madore}, {Murphy}, {Wyatt}, {Li}, {Filippenko}, \&  {Miller}}]{2010AJ....139..120F}
{Folatelli}, G., {et~al.} 2010, \aj, 139,  120

\bibitem[{{Foley} {et~al.}(2008){Foley}, {Filippenko}, {Aguilera}, {Becker},  {Blondin}, {Challis}, {Clocchiatti}, {Covarrubias}, {Davis}, {Garnavich},  {Jha}, {Kirshner}, {Krisciunas}, {Leibundgut}, {Li}, {Matheson}, {Miceli},  {Miknaitis}, {Pignata}, {Rest}, {Riess}, {Schmidt}, {Smith}, {Sollerman},  {Spyromilio}, {Stubbs}, {Suntzeff}, {Tonry}, {Wood-Vasey}, \&  {Zenteno}}]{2008ApJ...684...68F}
{Foley}, R.~J., {et~al.} 2008, \apj, 684, 68

\bibitem[{{Fruchter} \& {Hook}(2002)}]{2002PASP..114..144F}
{Fruchter}, A.~S., \& {Hook}, R.~N. 2002, \pasp, 114, 144

\bibitem[{{Fukugita} {et~al.}(1996){Fukugita}, {Ichikawa}, {Gunn}, {Doi},  {Shimasaku}, \& {Schneider}}]{1996AJ....111.1748F}
{Fukugita}, M., {Ichikawa}, T., {Gunn}, J.~E., {Doi}, M., {Shimasaku}, K., \&  {Schneider}, D.~P. 1996, \aj, 111, 1748

\bibitem[{{Garnavich} {et~al.}(1998){Garnavich}, {Jha}, {Challis},  {Clocchiatti}, {Diercks}, {Filippenko}, {Gilliland}, {Hogan}, {Kirshner},  {Leibundgut}, {Phillips}, {Reiss}, {Riess}, {Schmidt}, {Schommer}, {Smith},  {Spyromilio}, {Stubbs}, {Suntzeff}, {Tonry}, \&  {Carroll}}]{1998ApJ...509...74G}
{Garnavich}, P.~M., {et~al.} 1998, \apj, 509, 74

\bibitem[{{Goldhaber} {et~al.}(2001){Goldhaber}, {Groom}, {Kim}, {Aldering},  {Astier}, {Conley}, {Deustua}, {Ellis}, {Fabbro}, {Fruchter}, {Goobar},  {Hook}, {Irwin}, {Kim}, {Knop}, {Lidman}, {McMahon}, {Nugent}, {Pain},  {Panagia}, {Pennypacker}, {Perlmutter}, {Ruiz-Lapuente}, {Schaefer},  {Walton}, \& {York}}]{2001ApJ...558..359G}
{Goldhaber}, G., {et~al.} 2001, \apj, 558, 359

\bibitem[{{Goobar}(2008)}]{2008ApJ...686L.103G}
{Goobar}, A. 2008, \apjl, 686, L103

\bibitem[{{Goobar} {et~al.}(2002){Goobar}, {Bergstr{\"o}m}, \&  {M{\"o}rtsell}}]{2002A&A...384....1G}
{Goobar}, A., {Bergstr{\"o}m}, L., \& {M{\"o}rtsell}, E. 2002, \aap, 384, 1

\bibitem[{{Guy} {et~al.}(2007){Guy}, {Astier}, {Baumont}, {Hardin}, {Pain},  {Regnault}, {Basa}, {Carlberg}, {Conley}, {Fabbro}, {Fouchez}, {Hook},  {Howell}, {Perrett}, {Pritchet}, {Rich}, {Sullivan}, {Antilogus}, {Aubourg},  {Bazin}, {Bronder}, {Filiol}, {Palanque-Delabrouille}, {Ripoche}, \&  {Ruhlmann-Kleider}}]{2007A&A...466...11G}
{Guy}, J., {et~al.} 2007, \aap, 466, 11

\bibitem[{{Guy} {et~al.}(2005){Guy}, {Astier}, {Nobili}, {Regnault}, \&  {Pain}}]{2005A&A...443..781G}
{Guy}, J., {Astier}, P., {Nobili}, S., {Regnault}, N., \& {Pain}, R. 2005,  \aap, 443, 781

\bibitem[{Hager(1989)}]{Hager75570}
Hager, W.~W. 1989, SIAM Rev., 31, 221

\bibitem[{{Hamuy} {et~al.}(2006){Hamuy}, {Folatelli}, {Morrell}, {Phillips},  {Suntzeff}, {Persson}, {Roth}, {Gonzalez}, {Krzeminski}, {Contreras},  {Freedman}, {Murphy}, {Madore}, {Wyatt}, {Maza}, {Filippenko}, {Li}, \&  {Pinto}}]{2006PASP..118....2H}
{Hamuy}, M., {et~al.} 2006, \pasp, 118, 2

\bibitem[{{Hamuy} {et~al.}(1996){Hamuy}, {Phillips}, {Suntzeff}, {Schommer},  {Maza}, \& {Aviles}}]{1996AJ....112.2398H}
{Hamuy}, M., {Phillips}, M.~M., {Suntzeff}, N.~B., {Schommer}, R.~A., {Maza},  J., \& {Aviles}, R. 1996, \aj, 112, 2398

\bibitem[{Heyer {et~al.}(2004)Heyer, Richardson, Whitmore, \&  Lubin}]{Heyer:2004}
Heyer, I., Richardson, M., Whitmore, B., \& Lubin, L. 2004, The Accuracy of  WFPC2 Photometric Zeropoints

\bibitem[{{Hicken} {et~al.}(2009a){Hicken}, {Challis}, {Jha},  {Kirshner}, {Matheson}, {Modjaz}, {Rest}, {Michael Wood-Vasey}, {Bakos},  {Barton}, {Berlind}, {Bragg}, {Brice{\~n}o}, {Brown}, {Caldwell}, {Calkins},  {Cho}, {Ciupik}, {Contreras}, {Dendy}, {Dosaj}, {Durham}, {Eriksen},  {Esquerdo}, {Everett}, {Falco}, {Fernandez}, {Gaba}, {Garnavich}, {Graves},  {Green}, {Groner}, {Hergenrother}, {Holman}, {Hradecky}, {Huchra},  {Hutchison}, {Jerius}, {Jordan}, {Kilgard}, {Krauss}, {Luhman}, {Macri},  {Marrone}, {McDowell}, {McIntosh}, {McNamara}, {Megeath}, {Mochejska},  {Munoz}, {Muzerolle}, {Naranjo}, {Narayan}, {Pahre}, {Peters}, {Peterson},  {Rines}, {Ripman}, {Roussanova}, {Schild}, {Sicilia-Aguilar}, {Sokoloski},  {Smalley}, {Smith}, {Spahr}, {Stanek}, {Barmby}, {Blondin}, {Stubbs},  {Szentgyorgyi}, {Torres}, {Vaz}, {Vikhlinin}, {Wang}, {Westover}, {Woods}, \&  {Zhao}}]{2009ApJ...700..331H}
{Hicken}, M., {et~al.} 2009a,  \apj, 700, 331

\bibitem[{{Hicken} {et~al.}(2009b){Hicken}, {Wood-Vasey},  {Blondin}, {Challis}, {Jha}, {Kelly}, {Rest}, \&  {Kirshner}}]{2009ApJ...700.1097H}
{Hicken}, M., {Wood-Vasey}, W.~M., {Blondin}, S., {Challis}, P., {Jha}, S.,  {Kelly}, P.~L., {Rest}, A., \& {Kirshner}, R.~P. 2009b, \apj,  700, 1097

\bibitem[{{Hodapp} {et~al.}(2003){Hodapp}, {Jensen}, {Irwin}, {Yamada},  {Chung}, {Fletcher}, {Robertson}, {Hora}, {Simons}, {Mays}, {Nolan}, {Bec},  {Merrill}, \& {Fowler}}]{2003PASP..115.1388H}
{Hodapp}, K.~W., {et~al.} 2003, \pasp, 115,  1388

\bibitem[{{Hoeflich} {et~al.}(1998){Hoeflich}, {Wheeler}, \&  {Thielemann}}]{1998ApJ...495..617H}
{Hoeflich}, P., {Wheeler}, J.~C., \& {Thielemann}, F.~K. 1998, \apj, 495, 617

\bibitem[{{Holtzman} {et~al.}(2008){Holtzman}, {Marriner}, {Kessler}, {Sako},  {Dilday}, {Frieman}, {Schneider}, {Bassett}, {Becker}, {Cinabro}, {DeJongh},  {Depoy}, {Doi}, {Garnavich}, {Hogan}, {Jha}, {Konishi}, {Lampeitl},  {Marshall}, {McGinnis}, {Miknaitis}, {Nichol}, {Prieto}, {Riess}, {Richmond},  {Romani}, {Smith}, {Takanashi}, {Tokita}, {van der Heyden}, {Yasuda}, \&  {Zheng}}]{2008AJ....136.2306H}
{Holtzman}, J.~A., {et~al.} 2008, \aj,  136, 2306

\bibitem[{{Holz} \& {Linder}(2005)}]{2005ApJ...631..678H}
{Holz}, D.~E., \& {Linder}, E.~V. 2005, \apj, 631, 678

\bibitem[{{Hook} {et~al.}(2005){Hook}, {Howell}, {Aldering}, {Amanullah},  {Burns}, {Conley}, {Deustua}, {Ellis}, {Fabbro}, {Fadeyev}, {Folatelli},  {Garavini}, {Gibbons}, {Goldhaber}, {Goobar}, {Groom}, {Kim}, {Knop},  {Kowalski}, {Lidman}, {Nobili}, {Nugent}, {Pain}, {Pennypacker},  {Perlmutter}, {Ruiz-Lapuente}, {Sainton}, {Schaefer}, {Smith}, {Spadafora},  {Stanishev}, {Thomas}, {Walton}, {Wang}, \&  {Wood-Vasey}}]{2005AJ....130.2788H}
{Hook}, I.~M., {et~al.} 2005, \aj, 130, 2788

\bibitem[{{Howell} {et~al.}(2007){Howell}, {Sullivan}, {Conley}, \&  {Carlberg}}]{2007ApJ...667L..37H}
{Howell}, D.~A., {Sullivan}, M., {Conley}, A., \& {Carlberg}, R. 2007, \apjl,  667, L37

\bibitem[{{Howell} {et~al.}(2005){Howell}, {Sullivan}, {Perrett}, {Bronder},  {Hook}, {Astier}, {Aubourg}, {Balam}, {Basa}, {Carlberg}, {Fabbro},  {Fouchez}, {Guy}, {Lafoux}, {Neill}, {Pain}, {Palanque-Delabrouille},  {Pritchet}, {Regnault}, {Rich}, {Taillet}, {Knop}, {McMahon}, {Perlmutter},  \& {Walton}}]{2005ApJ...634.1190H}
{Howell}, D.~A., {et~al.} 2005, \apj, 634, 1190

\bibitem[{{Huterer} \& {Starkman}(2003)}]{2003PhRvL..90c1301H}
{Huterer}, D., \& {Starkman}, G. 2003, Physical Review Letters, 90, 031301

\bibitem[{James \& Roos(1975)}]{james:1975}
James, F., \& Roos, M. 1975, Comput. Phys. Commun., 10, 343

\bibitem[{{Jha} {et~al.}(2006){Jha}, {Kirshner}, {Challis}, {Garnavich},  {Matheson}, {Soderberg}, {Graves}, {Hicken}, {Alves}, {Arce}, {Balog},  {Barmby}, {Barton}, {Berlind}, {Bragg}, {Brice{\~n}o}, {Brown}, {Buckley},  {Caldwell}, {Calkins}, {Carter}, {Concannon}, {Donnelly}, {Eriksen},  {Fabricant}, {Falco}, {Fiore}, {Garcia}, {G{\'o}mez}, {Grogin}, {Groner},  {Groot}, {Haisch}, {Hartmann}, {Hergenrother}, {Holman}, {Huchra},  {Jayawardhana}, {Jerius}, {Kannappan}, {Kim}, {Kleyna}, {Kochanek},  {Koranyi}, {Krockenberger}, {Lada}, {Luhman}, {Luu}, {Macri}, {Mader},  {Mahdavi}, {Marengo}, {Marsden}, {McLeod}, {McNamara}, {Megeath}, {Moraru},  {Mossman}, {Muench}, {Mu{\~n}oz}, {Muzerolle}, {Naranjo}, {Nelson-Patel},  {Pahre}, {Patten}, {Peters}, {Peters}, {Raymond}, {Rines}, {Schild},  {Sobczak}, {Spahr}, {Stauffer}, {Stefanik}, {Szentgyorgyi}, {Tollestrup},  {V{\"a}is{\"a}nen}, {Vikhlinin}, {Wang}, {Willner}, {Wolk}, {Zajac}, {Zhao},  \& {Stanek}}]{2006AJ....131..527J}
{Jha}, S., {et~al.} 2006, \aj, 131, 527

\bibitem[{{Jha} {et~al.}(2007){Jha}, {Riess}, \&  {Kirshner}}]{2007ApJ...659..122J}
{Jha}, S., {Riess}, A.~G., \& {Kirshner}, R.~P. 2007, \apj, 659, 122

\bibitem[{{J{\"o}nsson} {et~al.}(2006){J{\"o}nsson}, {Dahl{\'e}n}, {Goobar},  {Gunnarsson}, {M{\"o}rtsell}, \& {Lee}}]{2006ApJ...639..991J}
{J{\"o}nsson}, J., {Dahl{\'e}n}, T., {Goobar}, A., {Gunnarsson}, C.,  {M{\"o}rtsell}, E., \& {Lee}, K. 2006, \apj, 639, 991

\bibitem[{{Kessler} {et~al.}(2009){Kessler}, {Becker}, {Cinabro}, {Vanderplas},  {Frieman}, {Marriner}, {Davis}, {Dilday}, {Holtzman}, {Jha}, {Lampeitl},  {Sako}, {Smith}, {Zheng}, {Nichol}, {Bassett}, {Bender}, {Depoy}, {Doi},  {Elson}, {Filippenko}, {Foley}, {Garnavich}, {Hopp}, {Ihara}, {Ketzeback},  {Kollatschny}, {Konishi}, {Marshall}, {McMillan}, {Miknaitis}, {Morokuma},  {M{\"o}rtsell}, {Pan}, {Prieto}, {Richmond}, {Riess}, {Romani}, {Schneider},  {Sollerman}, {Takanashi}, {Tokita}, {van der Heyden}, {Wheeler}, {Yasuda}, \&  {York}}]{2009ApJS..185...32K}
{Kessler}, R., {et~al.} 2009, \apjs, 185,  32

\bibitem[{Kim \& Miquel(2006)}]{Kim2006451}
Kim, A.~G., \& Miquel, R. 2006, Astroparticle Physics, 24, 451

\bibitem[{{Knop} {et~al.}(2003){Knop}, {Aldering}, {Amanullah}, {Astier},  {Blanc}, {Burns}, {Conley}, {Deustua}, {Doi}, {Ellis}, {Fabbro}, {Folatelli},  {Fruchter}, {Garavini}, {Garmond}, {Garton}, {Gibbons}, {Goldhaber},  {Goobar}, {Groom}, {Hardin}, {Hook}, {Howell}, {Kim}, {Lee}, {Lidman},  {Mendez}, {Nobili}, {Nugent}, {Pain}, {Panagia}, {Pennypacker}, {Perlmutter},  {Quimby}, {Raux}, {Regnault}, {Ruiz-Lapuente}, {Sainton}, {Schaefer},  {Schahmaneche}, {Smith}, {Spadafora}, {Stanishev}, {Sullivan}, {Walton},  {Wang}, {Wood-Vasey}, \& {Yasuda}}]{2003ApJ...598..102K}
{Knop}, R.~A., {et~al.} 2003, \apj, 598, 102

\bibitem[{{Komatsu} {et~al.}(2010){Komatsu}, {Smith}, {Dunkley}, {Bennett},  {Gold}, {Hinshaw}, {Jarosik}, {Larson}, {Nolta}, {Page}, {Spergel},  {Halpern}, {Hill}, {Kogut}, {Limon}, {Meyer}, {Odegard}, {Tucker}, {Weiland},  {Wollack}, \& {Wright}}]{2010arXiv1001.4538K}
{Komatsu}, E., {et~al.} 2010, ArXiv e-prints

\bibitem[{{Kowalski} {et~al.}(2008){Kowalski}, {Rubin}, {Aldering},  {Agostinho}, {Amadon}, {Amanullah}, {Balland}, {Barbary}, {Blanc}, {Challis},  {Conley}, {Connolly}, {Covarrubias}, {Dawson}, {Deustua}, {Ellis}, {Fabbro},  {Fadeyev}, {Fan}, {Farris}, {Folatelli}, {Frye}, {Garavini}, {Gates},  {Germany}, {Goldhaber}, {Goldman}, {Goobar}, {Groom}, {Haissinski}, {Hardin},  {Hook}, {Kent}, {Kim}, {Knop}, {Lidman}, {Linder}, {Mendez}, {Meyers},  {Miller}, {Moniez}, {Mour{\~a}o}, {Newberg}, {Nobili}, {Nugent}, {Pain},  {Perdereau}, {Perlmutter}, {Phillips}, {Prasad}, {Quimby}, {Regnault},  {Rich}, {Rubenstein}, {Ruiz-Lapuente}, {Santos}, {Schaefer}, {Schommer},  {Smith}, {Soderberg}, {Spadafora}, {Strolger}, {Strovink}, {Suntzeff},  {Suzuki}, {Thomas}, {Walton}, {Wang}, {Wood-Vasey}, \&  {Yun}}]{2008ApJ...686..749K}
{Kowalski}, M., {et~al.} 2008, \apj,  686, 749

\bibitem[{{Krisciunas} {et~al.}(2005){Krisciunas}, {Garnavich}, {Challis},  {Prieto}, {Riess}, {Barris}, {Aguilera}, {Becker}, {Blondin}, {Chornock},  {Clocchiatti}, {Covarrubias}, {Filippenko}, {Foley}, {Hicken}, {Jha},  {Kirshner}, {Leibundgut}, {Li}, {Matheson}, {Miceli}, {Miknaitis}, {Rest},  {Salvo}, {Schmidt}, {Smith}, {Sollerman}, {Spyromilio}, {Stubbs}, {Suntzeff},  {Tonry}, \& {Wood-Vasey}}]{2005AJ....130.2453K}
{Krisciunas}, K., {et~al.} 2005, \aj, 130, 2453

\bibitem[{Krist(2001)}]{tinytim}
Krist, John E.;~Hook, R.~N. 2001, The Tiny Tim User's Guide, STScI

\bibitem[{{Kuznetsova} {et~al.}(2008){Kuznetsova}, {Barbary}, {Connolly},  {Kim}, {Pain}, {Roe}, {Aldering}, {Amanullah}, {Dawson}, {Doi}, {Fadeyev},  {Fruchter}, {Gibbons}, {Goldhaber}, {Goobar}, {Gude}, {Knop}, {Kowalski},  {Lidman}, {Morokuma}, {Meyers}, {Perlmutter}, {Rubin}, {Schlegel},  {Spadafora}, {Stanishev}, {Strovink}, {Suzuki}, {Wang}, \&  {Yasuda}}]{2008ApJ...673..981K}
{Kuznetsova}, N., {et~al.} 2008, \apj, 673, 981

\bibitem[{{Landolt}(1992)}]{1992AJ....104..340L}
{Landolt}, A.~U. 1992, \aj, 104, 340

\bibitem[{{Law} {et~al.}(2009){Law}, {Kulkarni}, {Dekany}, {Ofek}, {Quimby},  {Nugent}, {Surace}, {Grillmair}, {Bloom}, {Kasliwal}, {Bildsten}, {Brown},  {Cenko}, {Ciardi}, {Croner}, {Djorgovski}, {van Eyken}, {Filippenko}, {Fox},  {Gal-Yam}, {Hale}, {Hamam}, {Helou}, {Henning}, {Howell}, {Jacobsen},  {Laher}, {Mattingly}, {McKenna}, {Pickles}, {Poznanski}, {Rahmer}, {Rau},  {Rosing}, {Shara}, {Smith}, {Starr}, {Sullivan}, {Velur}, {Walters}, \&  {Zolkower}}]{2009PASP..121.1395L}
{Law}, N.~M., {et~al.} 2009, \pasp, 121, 1395

\bibitem[{Lentz {et~al.}(2000)Lentz, {Baron}, {Branch}, {Hauschildt}, \&  {Nugent}}]{2000ApJ...530..966L}
Lentz, E.~J., {Baron}, E., {Branch}, D., {Hauschildt}, P.~H., \& {Nugent},  P.~E. 2000, \apj, 530, 966

\bibitem[{{Lidman} {et~al.}(2005){Lidman}, {Howell}, {Folatelli}, {Garavini},  {Nobili}, {Aldering}, {Amanullah}, {Antilogus}, {Astier}, {Blanc}, {Burns},  {Conley}, {Deustua}, {Doi}, {Ellis}, {Fabbro}, {Fadeyev}, {Gibbons},  {Goldhaber}, {Goobar}, {Groom}, {Hook}, {Kashikawa}, {Kim}, {Knop}, {Lee},  {Mendez}, {Morokuma}, {Motohara}, {Nugent}, {Pain}, {Perlmutter}, {Prasad},  {Quimby}, {Raux}, {Regnault}, {Ruiz-Lapuente}, {Sainton}, {Schaefer},  {Schahmaneche}, {Smith}, {Spadafora}, {Stanishev}, {Walton}, {Wang},  {Wood-Vasey}, {Yasuda}, \& {The Supernova Cosmology  Project}}]{2005A&A...430..843L}
{Lidman}, C., {et~al.} 2005, \aap, 430, 843

\bibitem[{{Linder}(1988)}]{1988A&A...206..190L}
{Linder}, E.~V. 1988, \aap, 206, 190

\bibitem[{{Linder}(2003)}]{2003PhRvL..90i1301L}
---. 2003, Physical Review Letters, 90, 091301

\bibitem[{Lupton(2005)}]{lupton2005}
Lupton, R. 2005, Transformations between SDSS magnitudes and UBVRcIc, {\tt  http://www.sdss.org/dr6/algorithms/sdssUBVRITransform.html\#Lupton2005}

\bibitem[{{M{\'e}nard} {et~al.}(2009a){M{\'e}nard}, {Kilbinger},  \& {Scranton}}]{2009arXiv0903.4199M}
{M{\'e}nard}, B., {Kilbinger}, M., \& {Scranton}, R. 2009a, ArXiv  e-prints

\bibitem[{{M{\'e}nard} {et~al.}(2008){M{\'e}nard}, {Nestor}, {Turnshek},  {Quider}, {Richards}, {Chelouche}, \& {Rao}}]{2008MNRAS.385.1053M}
{M{\'e}nard}, B., {Nestor}, D., {Turnshek}, D., {Quider}, A., {Richards}, G.,  {Chelouche}, D., \& {Rao}, S. 2008, \mnras, 385, 1053

\bibitem[{{M{\'e}nard} {et~al.}(2009b){M{\'e}nard}, {Scranton},  {Fukugita}, \& {Richards}}]{2009arXiv0902.4240M}
{M{\'e}nard}, B., {Scranton}, R., {Fukugita}, M., \& {Richards}, G.  2009b, ArXiv e-prints

\bibitem[{{Miknaitis} {et~al.}(2007){Miknaitis}, {Pignata}, {Rest},  {Wood-Vasey}, {Blondin}, {Challis}, {Smith}, {Stubbs}, {Suntzeff}, {Foley},  {Matheson}, {Tonry}, {Aguilera}, {Blackman}, {Becker}, {Clocchiatti},  {Covarrubias}, {Davis}, {Filippenko}, {Garg}, {Garnavich}, {Hicken}, {Jha},  {Krisciunas}, {Kirshner}, {Leibundgut}, {Li}, {Miceli}, {Narayan}, {Prieto},  {Riess}, {Salvo}, {Schmidt}, {Sollerman}, {Spyromilio}, \&  {Zenteno}}]{2007ApJ...666..674M}
{Miknaitis}, G., {et~al.} 2007,  \apj, 666, 674

\bibitem[{{Miyazaki} {et~al.}(2002){Miyazaki}, {Komiyama}, {Sekiguchi},  {Okamura}, {Doi}, {Furusawa}, {Hamabe}, {Imi}, {Kimura}, {Nakata}, {Okada},  {Ouchi}, {Shimasaku}, {Yagi}, \& {Yasuda}}]{2002PASJ...54..833M}
{Miyazaki}, S., {et~al.} 2002, \pasj,  54, 833

\bibitem[{Moorwood {et~al.}(1999)Moorwood, Cuby, \& Lidman}]{moorwood:1999}
Moorwood, A., Cuby, J.~G., \& Lidman, C. 1999

\bibitem[{{Morokuma} {et~al.}(2010){Morokuma}, {Tokita}, {Lidman}, {Doi},  {Yasuda}, {Aldering}, {Amanullah}, {Barbary}, {Dawson}, {Fadeyev},  {Fakhouri}, {Goldhaber}, {Goobar}, {Hattori}, {Hayano}, {Hook}, {Howell},  {Furusawa}, {Ihara}, {Kashikawa}, {Knop}, {Konishi}, {Meyers}, {Oda}, {Pain},  {Perlmutter}, {Rubin}, {Spadafora}, {Suzuki}, {Takanashi}, {Totani},  {Utsunomiya}, \& {Wang}}]{2010PASJ...62...19M}
{Morokuma}, T., {et~al.} 2010, \pasj, 62, 19

\bibitem[{{M{\"o}rtsell} \& {Goobar}(2003)}]{2003JCAP...09..009M}
{M{\"o}rtsell}, E., \& {Goobar}, A. 2003, Journal of Cosmology and  Astro-Particle Physics, 9, 9

\bibitem[{{Muller} {et~al.}(1998){Muller}, {Reed}, {Armandroff}, {Boroson}, \&  {Jacoby}}]{1998SPIE.3355..577M}
{Muller}, G.~P., {Reed}, R., {Armandroff}, T., {Boroson}, T.~A., \& {Jacoby},  G.~H. 1998, in Society of Photo-Optical Instrumentation Engineers (SPIE)  Conference Series, Vol. 3355, Society of Photo-Optical Instrumentation  Engineers (SPIE) Conference Series, ed. {S.~D'Odorico}, 577--585

\bibitem[{{Nobili} {et~al.}(2005){Nobili}, {Amanullah}, {Garavini}, {Goobar},  {Lidman}, {Stanishev}, {Aldering}, {Antilogus}, {Astier}, {Burns}, {Conley},  {Deustua}, {Ellis}, {Fabbro}, {Fadeyev}, {Folatelli}, {Gibbons}, {Goldhaber},  {Groom}, {Hook}, {Howell}, {Kim}, {Knop}, {Nugent}, {Pain}, {Perlmutter},  {Quimby}, {Raux}, {Regnault}, {Ruiz-Lapuente}, {Sainton}, {Schahmaneche},  {Smith}, {Spadafora}, {Thomas}, {Wang}, \& {The Supernova Cosmology  Project}}]{2005A&A...437..789N}
{Nobili}, S., {et~al.} 2005, \aap, 437, 789

\bibitem[{{Nobili} {et~al.}(2009){Nobili}, {Fadeyev}, {Aldering}, {Amanullah},  {Barbary}, {Burns}, {Dawson}, {Deustua}, {Faccioli}, {Fruchter}, {Goldhaber},  {Goobar}, {Hook}, {Howell}, {Kim}, {Knop}, {Lidman}, {Meyers}, {Nugent},  {Pain}, {Panagia}, {Perlmutter}, {Rubin}, {Spadafora}, {Strovink}, {Suzuki},  \& {The Supernova Cosmology Project}}]{2009ApJ...700.1415N}
{Nobili}, S., {et~al.} 2009, \apj, 700, 1415

\bibitem[{{Nobili} \& {Goobar}(2008)}]{2008A&A...487...19N}
{Nobili}, S., \& {Goobar}, A. 2008, \aap, 487, 19

\bibitem[{{Nugent} {et~al.}(2002){Nugent}, {Kim}, \&  {Perlmutter}}]{2002PASP..114..803N}
{Nugent}, P., {Kim}, A., \& {Perlmutter}, S. 2002, \pasp, 114, 803

\bibitem[{{Oke} {et~al.}(1995){Oke}, {Cohen}, {Carr}, {Cromer}, {Dingizian},  {Harris}, {Labrecque}, {Lucinio}, {Schaal}, {Epps}, \&  {Miller}}]{1995PASP..107..375O}
{Oke}, J.~B., {et~al.} 1995, \pasp, 107, 375

\bibitem[{{{\"O}stman} \& {M{\"o}rtsell}(2005)}]{2005JCAP...02..005O}
{{\"O}stman}, L., \& {M{\"o}rtsell}, E. 2005, Journal of Cosmology and  Astro-Particle Physics, 2, 5

\bibitem[{{Patat} {et~al.}(2007){Patat}, {Chandra}, {Chevalier}, {Justham},  {Podsiadlowski}, {Wolf}, {Gal-Yam}, {Pasquini}, {Crawford}, {Mazzali},  {Pauldrach}, {Nomoto}, {Benetti}, {Cappellaro}, {Elias-Rosa}, {Hillebrandt},  {Leonard}, {Pastorello}, {Renzini}, {Sabbadin}, {Simon}, \&  {Turatto}}]{2007Sci...317..924P}
{Patat}, F., {et~al.} 2007,  Science, 317, 924

\bibitem[{{Percival} {et~al.}(2010){Percival}, {Reid}, {Eisenstein}, {Bahcall},  {Budavari}, {Frieman}, {Fukugita}, {Gunn}, {Ivezi{\'c}}, {Knapp}, {Kron},  {Loveday}, {Lupton}, {McKay}, {Meiksin}, {Nichol}, {Pope}, {Schlegel},  {Schneider}, {Spergel}, {Stoughton}, {Strauss}, {Szalay}, {Tegmark},  {Vogeley}, {Weinberg}, {York}, \& {Zehavi}}]{2010MNRAS.401.2148P}
{Percival}, W.~J., {et~al.} 2010, \mnras, 401, 2148

\bibitem[{{Perlmutter} {et~al.}(1998){Perlmutter}, {Aldering}, {della Valle},  {Deustua}, {Ellis}, {Fabbro}, {Fruchter}, {Goldhaber}, {Groom}, {Hook},  {Kim}, {Kim}, {Knop}, {Lidman}, {McMahon}, {Nugent}, {Pain}, {Panagia},  {Pennypacker}, {Ruiz-Lapuente}, {Schaefer}, \&  {Walton}}]{1998Natur.391...51P}
{Perlmutter}, S., {et~al.} 1998, \nat, 391, 51

\bibitem[{{Perlmutter} {et~al.}(1999){Perlmutter}, {Aldering}, {Goldhaber},  {Knop}, {Nugent}, {Castro}, {Deustua}, {Fabbro}, {Goobar}, {Groom}, {Hook},  {Kim}, {Kim}, {Lee}, {Nunes}, {Pain}, {Pennypacker}, {Quimby}, {Lidman},  {Ellis}, {Irwin}, {McMahon}, {Ruiz-Lapuente}, {Walton}, {Schaefer}, {Boyle},  {Filippenko}, {Matheson}, {Fruchter}, {Panagia}, {Newberg}, {Couch}, \& {The  Supernova Cosmology Project}}]{1999ApJ...517..565P}
{Perlmutter}, S., {et~al.} 1999, \apj, 517, 565

\bibitem[{{Perlmutter} {et~al.}(1997){Perlmutter}, {Gabi}, {Goldhaber},  {Goobar}, {Groom}, {Hook}, {Kim}, {Kim}, {Lee}, {Pain}, {Pennypacker},  {Small}, {Ellis}, {McMahon}, {Boyle}, {Bunclark}, {Carter}, {Irwin},  {Glazebrook}, {Newberg}, {Filippenko}, {Matheson}, {Dopita}, {Couch}, \& {The  Supernova Cosmology Project}}]{1997ApJ...483..565P}
{Perlmutter}, S., {et~al.} 1997,  \apj, 483, 565

\bibitem[{{Perlmutter} {et~al.}(1995){Perlmutter}, {Pennypacker}, {Goldhaber},  {Goobar}, {Muller}, {Newberg}, {Desai}, {Kim}, {Kim}, {Small}, {Boyle},  {Crawford}, {McMahon}, {Bunclark}, {Carter}, {Irwin}, {Terlevich}, {Ellis},  {Glazebrook}, {Couch}, {Mould}, {Small}, \& {Abraham}}]{1995ApJ...440L..41P}
{Perlmutter}, S., {et~al.} 1995, \apjl, 440, L41

\bibitem[{{Persson} {et~al.}(1998){Persson}, {Murphy}, {Krzeminski}, {Roth}, \&  {Rieke}}]{1998AJ....116.2475P}
{Persson}, S.~E., {Murphy}, D.~C., {Krzeminski}, W., {Roth}, M., \& {Rieke},  M.~J. 1998, \aj, 116, 2475

\bibitem[{{Phillips}(1993)}]{1993ApJ...413L.105P}
{Phillips}, M.~M. 1993, \apjl, 413, L105

\bibitem[{Raux(2003)}]{raux:2003}
Raux, J. 2003, PhD thesis, l'Universit\'e Paris XI Orsay

\bibitem[{{Regnault} {et~al.}(2009){Regnault}, {Conley}, {Guy}, {Sullivan},  {Cuillandre}, {Astier}, {Balland}, {Basa}, {Carlberg}, {Fouchez}, {Hardin},  {Hook}, {Howell}, {Pain}, {Perrett}, \& {Pritchet}}]{2009A&A...506..999R}
{Regnault}, N., {et~al.} 2009, \aap, 506, 999

\bibitem[{{Riess} {et~al.}(1998){Riess}, {Filippenko}, {Challis},  {Clocchiatti}, {Diercks}, {Garnavich}, {Gilliland}, {Hogan}, {Jha},  {Kirshner}, {Leibundgut}, {Phillips}, {Reiss}, {Schmidt}, {Schommer},  {Smith}, {Spyromilio}, {Stubbs}, {Suntzeff}, \&  {Tonry}}]{1998AJ....116.1009R}
{Riess}, A.~G., {et~al.} 1998, \aj, 116, 1009

\bibitem[{{Riess} {et~al.}(1999){Riess}, {Kirshner}, {Schmidt}, {Jha},  {Challis}, {Garnavich}, {Esin}, {Carpenter}, {Grashius}, {Schild}, {Berlind},  {Huchra}, {Prosser}, {Falco}, {Benson}, {Brice{\~n}o}, {Brown}, {Caldwell},  {dell'Antonio}, {Filippenko}, {Goodman}, {Grogin}, {Groner}, {Hughes},  {Green}, {Jansen}, {Kleyna}, {Luu}, {Macri}, {McLeod}, {McLeod}, {McNamara},  {McLean}, {Milone}, {Mohr}, {Moraru}, {Peng}, {Peters}, {Prestwich},  {Stanek}, {Szentgyorgyi}, \& {Zhao}}]{1999AJ....117..707R}
{Riess}, A.~G., {et~al.} 1999, \aj, 117, 707

\bibitem[{{Riess} {et~al.}(2009){Riess}, {Macri}, {Casertano}, {Sosey},  {Lampeitl}, {Ferguson}, {Filippenko}, {Jha}, {Li}, {Chornock}, \&  {Sarkar}}]{2009ApJ...699..539R}
{Riess}, A.~G., {et~al.} 2009, \apj, 699, 539

\bibitem[{{Riess} {et~al.}(1996){Riess}, {Press}, \&  {Kirshner}}]{1996ApJ...473...88R}
{Riess}, A.~G., {Press}, W.~H., \& {Kirshner}, R.~P. 1996, \apj, 473, 88

\bibitem[{{Riess} {et~al.}(2007){Riess}, {Strolger}, {Casertano}, {Ferguson},  {Mobasher}, {Gold}, {Challis}, {Filippenko}, {Jha}, {Li}, {Tonry}, {Foley},  {Kirshner}, {Dickinson}, {MacDonald}, {Eisenstein}, {Livio}, {Younger}, {Xu},  {Dahl{\'e}n}, \& {Stern}}]{2007ApJ...659...98R}
{Riess}, A.~G., {et~al.} 2007, \apj, 659, 98

\bibitem[{{Riess} {et~al.}(2004){Riess}, {Strolger}, {Tonry}, {Casertano},  {Ferguson}, {Mobasher}, {Challis}, {Filippenko}, {Jha}, {Li}, {Chornock},  {Kirshner}, {Leibundgut}, {Dickinson}, {Livio}, {Giavalisco}, {Steidel},  {Ben{\'{\i}}tez}, \& {Tsvetanov}}]{2004ApJ...607..665R}
{Riess}, A.~G., {et~al.} 2004, \apj, 607, 665

\bibitem[{{Sasaki}(1987)}]{1987MNRAS.228..653S}
{Sasaki}, M. 1987, \mnras, 228, 653

\bibitem[{{Schlegel} {et~al.}(1998){Schlegel}, {Finkbeiner}, \&  {Davis}}]{1998ApJ...500..525S}
{Schlegel}, D.~J., {Finkbeiner}, D.~P., \& {Davis}, M. 1998, \apj, 500, 525

\bibitem[{{Schmidt} {et~al.}(1998){Schmidt}, {Suntzeff}, {Phillips},  {Schommer}, {Clocchiatti}, {Kirshner}, {Garnavich}, {Challis}, {Leibundgut},  {Spyromilio}, {Riess}, {Filippenko}, {Hamuy}, {Smith}, {Hogan}, {Stubbs},  {Diercks}, {Reiss}, {Gilliland}, {Tonry}, {Maza}, {Dressler}, {Walsh}, \&  {Ciardullo}}]{1998ApJ...507...46S}
{Schmidt}, B.~P., {et~al.} 1998, \apj, 507, 46

\bibitem[{{Sheinis} {et~al.}(2002){Sheinis}, {Bolte}, {Epps}, {Kibrick},  {Miller}, {Radovan}, {Bigelow}, \& {Sutin}}]{2002PASP..114..851S}
{Sheinis}, A.~I., {Bolte}, M., {Epps}, H.~W., {Kibrick}, R.~I., {Miller},  J.~S., {Radovan}, M.~V., {Bigelow}, B.~C., \& {Sutin}, B.~M. 2002, \pasp,  114, 851

\bibitem[{{Simon} {et~al.}(2009){Simon}, {Gal-Yam}, {Gnat}, {Quimby},  {Ganeshalingam}, {Silverman}, {Blondin}, {Li}, {Filippenko}, {Wheeler},  {Kirshner}, {Patat}, {Nugent}, {Foley}, {Vogt}, {Butler}, {Peek},  {Rosolowsky}, {Herczeg}, {Sauer}, \& {Mazzali}}]{2009ApJ...702.1157S}
{Simon}, J.~D., {et~al.} 2009, \apj, 702, 1157

\bibitem[{{Stetson}(2000)}]{2000PASP..112..925S}
{Stetson}, P.~B. 2000, \pasp, 112, 925

\bibitem[{{Stritzinger} {et~al.}(2005){Stritzinger}, {Suntzeff}, {Hamuy},  {Challis}, {Demarco}, {Germany}, \& {Soderberg}}]{2005PASP..117..810S}
{Stritzinger}, M., {Suntzeff}, N.~B., {Hamuy}, M., {Challis}, P., {Demarco},  R., {Germany}, L., \& {Soderberg}, A.~M. 2005, \pasp, 117, 810

\bibitem[{{Strovink}(2007)}]{2007ApJ...671.1084S}
{Strovink}, M. 2007, \apj, 671, 1084

\bibitem[{Thatte {et~al.}(2009)Thatte, Dahl\'en, {et~al.}}]{nicmos2009}
Thatte, D., Dahl\'en, T., {et~al.} 2009, NICMOS Data Handbook

\bibitem[{{Tonry} {et~al.}(2003){Tonry}, {Schmidt}, {Barris}, {Candia},  {Challis}, {Clocchiatti}, {Coil}, {Filippenko}, {Garnavich}, {Hogan},  {Holland}, {Jha}, {Kirshner}, {Krisciunas}, {Leibundgut}, {Li}, {Matheson},  {Phillips}, {Riess}, {Schommer}, {Smith}, {Sollerman}, {Spyromilio},  {Stubbs}, \& {Suntzeff}}]{2003ApJ...594....1T}
{Tonry}, J.~L., {et~al.} 2003, \apj, 594, 1

\bibitem[{{Tripp}(1998)}]{1998A&A...331..815T}
{Tripp}, R. 1998, \aap, 331, 815

\bibitem[{{Wang}(2005)}]{2005ApJ...635L..33W}
{Wang}, L. 2005, \apjl, 635, L33

\bibitem[{{Wang} {et~al.}(2003){Wang}, {Goldhaber}, {Aldering}, \&  {Perlmutter}}]{2003ApJ...590..944W}
{Wang}, L., {Goldhaber}, G., {Aldering}, G., \& {Perlmutter}, S. 2003, \apj,  590, 944

\bibitem[{{Wang} {et~al.}(2008){Wang}, {Li}, {Filippenko}, {Krisciunas},  {Suntzeff}, {Li}, {Zhang}, {Deng}, {Foley}, {Ganeshalingam}, {Li}, {Lou},  {Qiu}, {Shang}, {Silverman}, {Zhang}, \& {Zhang}}]{2008ApJ...675..626W}
{Wang}, X., {et~al.} 2008, \apj, 675, 626

\bibitem[{{Wood-Vasey} {et~al.}(2007){Wood-Vasey}, {Miknaitis}, {Stubbs},  {Jha}, {Riess}, {Garnavich}, {Kirshner}, {Aguilera}, {Becker}, {Blackman},  {Blondin}, {Challis}, {Clocchiatti}, {Conley}, {Covarrubias}, {Davis},  {Filippenko}, {Foley}, {Garg}, {Hicken}, {Krisciunas}, {Leibundgut}, {Li},  {Matheson}, {Miceli}, {Narayan}, {Pignata}, {Prieto}, {Rest}, {Salvo},  {Schmidt}, {Smith}, {Sollerman}, {Spyromilio}, {Tonry}, {Suntzeff}, \&  {Zenteno}}]{2007ApJ...666..694W}
{Wood-Vasey}, W.~M., {et~al.} 2007, \apj, 666, 694

\end{thebibliography}

%\begin{figure}
%  \centering
%  \ifthenelse{\boolean{salttwo}}{%
%    \includegraphics[width=\textwidth]{figures/hubbleresid-SALT12.eps}
%    \caption{Residuals from the Hubble diagram where a
%      $(\Omega_\Lambda,w,M_B(h_{0.7})) = (0.713,-0.969,-19.31)$
%      cosmology (K08) has been subtracted from
%      the light curve shape and color corrected peak magnitudes. The
%      results from using \salt (blue squares) and \saltii (red circles
%      shifted $\Delta z=0.01$) for the \sne presented in this paper are
%      shown together with the \sne from~K08
%      (grey).}}{%
%    \includegraphics[width=\textwidth]{figures/hubbleresid.eps}
%    \caption{Residuals from the Hubble diagram where the best fit
%      cosmology has been subtracted from the light curve shape and
%      color corrected peak magnitudes The \sne presented in this paper
%      (black) is shown together with the \sne from~K08 (grey).}}
%  \label{fig:hubble}
%\end{figure}

\appendix

\section{Spectra and notes on individual candidates}\label{sec:spectra}
The spectra of the six \sne in this paper. Details of the spectra for
\sngo and \sngy can be found in \citet{2005A&A...430..843L}, while the
spectrum for \sncw is described in \citet{2010PASJ...62...19M}.

The spectrum of each candidate is plotted twice. In the upper panel,
the spectrum of the candidate is plotted in the observer frame and is
uncorrected for host galaxy light. Telluric absorption features are
marked with the symbol $\oplus$. In the lower panel, contamination
from the host (if any) is removed, and the resulting spectrum is
rescaled and re-binned. This spectrum is plotted in black as a
histogram and it is plotted in both the rest frame (lower axis) and
the observer frame (upper axis). For comparison, low redshift \sne are
plotted as a blue continuous line.

%\begin{figure}
%  \centering
%  \resizebox{\hsize}{!}{\includegraphics{figures/SN01cw.eps}}
%  \caption{{\bf Upper panel:} The unbinned FOCAS spectrum of \sncw, a
%    probable \snia at $z=0.95$. {\bf Lower Panel:} A rebinned version
%    of the spectrum in the upper panel and SN~1989B at
%    $-5$ days. The redshift of \sncw is derived by fitting the SN to
%    local SN templates as no features from the host galaxy were
%    detected. %
%    \label{fig:01cw}}
%\end{figure}

\begin{figure}[p]
\centering
\includegraphics[width=0.5\textwidth]{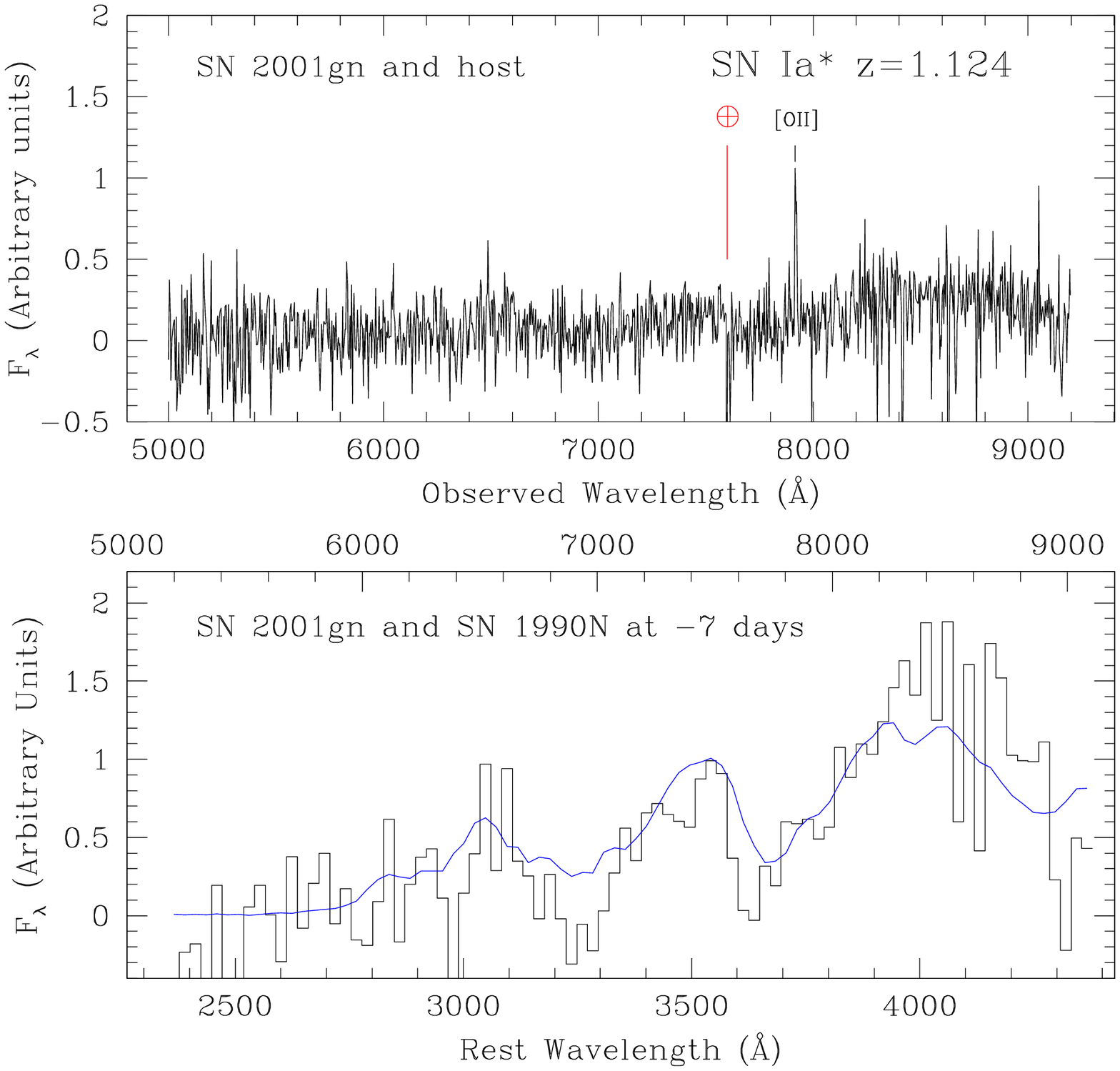}
\caption{{\bf Upper panel:} The error weighted and rebinned ESI
  spectrum of \sngn, a probable SN~Ia at $z=1.124$. The redshift is
  derived from the [OII] $\lambda\lambda 3727,3729$ doublet,
  which is clearly resolved in the ESI spectrum. {\bf Lower Panel:}
  The rebinned host galaxy subtracted spectrum and SN~1990N at $-7$ days. %
  \label{fig:01gn}}
\end{figure}

\begin{figure}[p]
\centering
\includegraphics[width=0.5\textwidth]{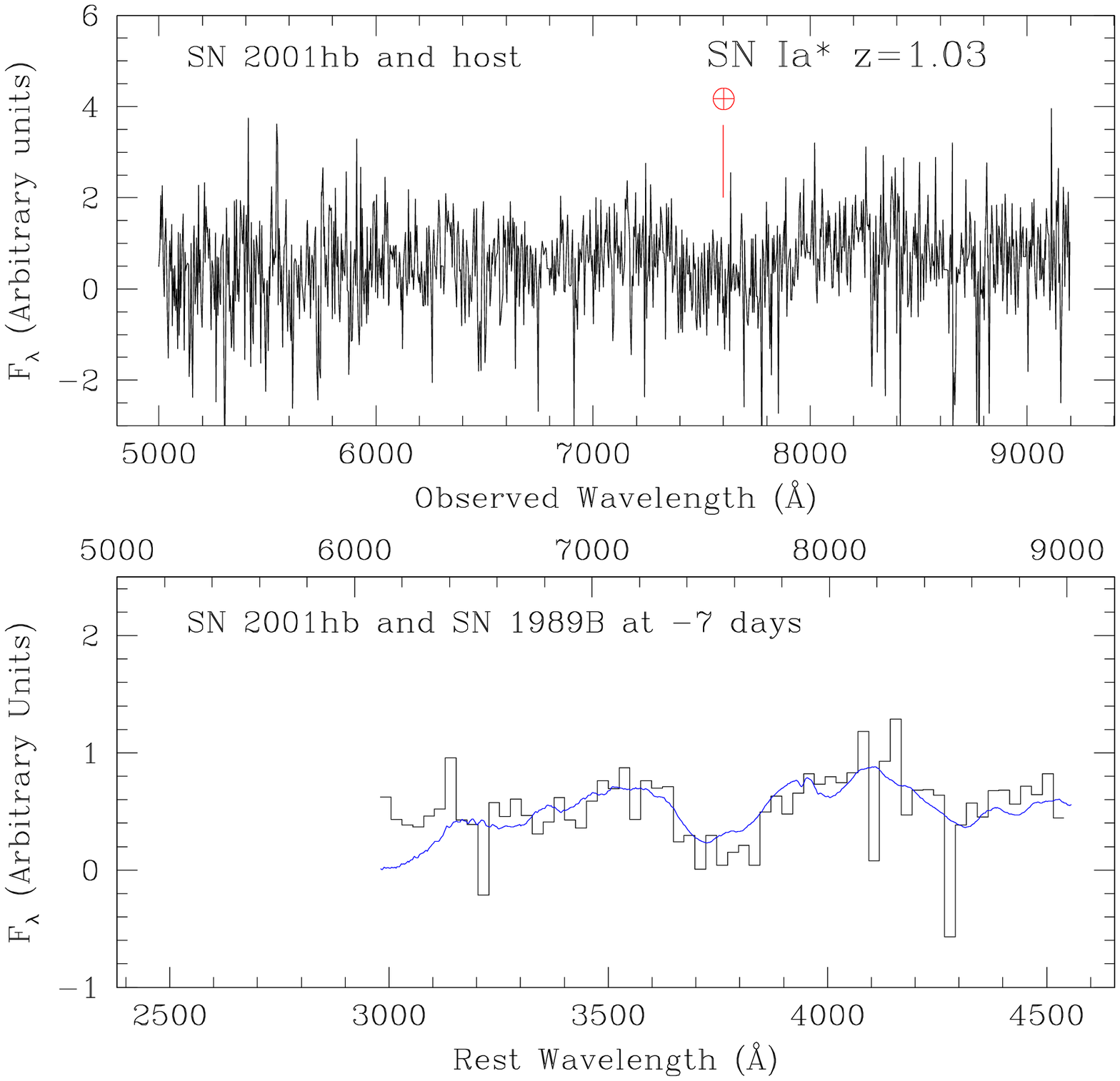}
\caption{{\bf Upper panel:} The error weighted and rebinned ESI
  spectrum of \snhb, a probable \snia at $z=1.03$. {\bf Lower Panel:}
  A rebinned version of the spectrum in the upper panel and SN~1989B
  at $-7$ days. The redshift of \snhb is derived by fitting the SN to
  local SN templates as no features from the host galaxy are
  detected. %
  \label{fig:01hb}}
\end{figure}

\begin{figure*}[tb]
\centering
\includegraphics[width=0.5\textwidth]{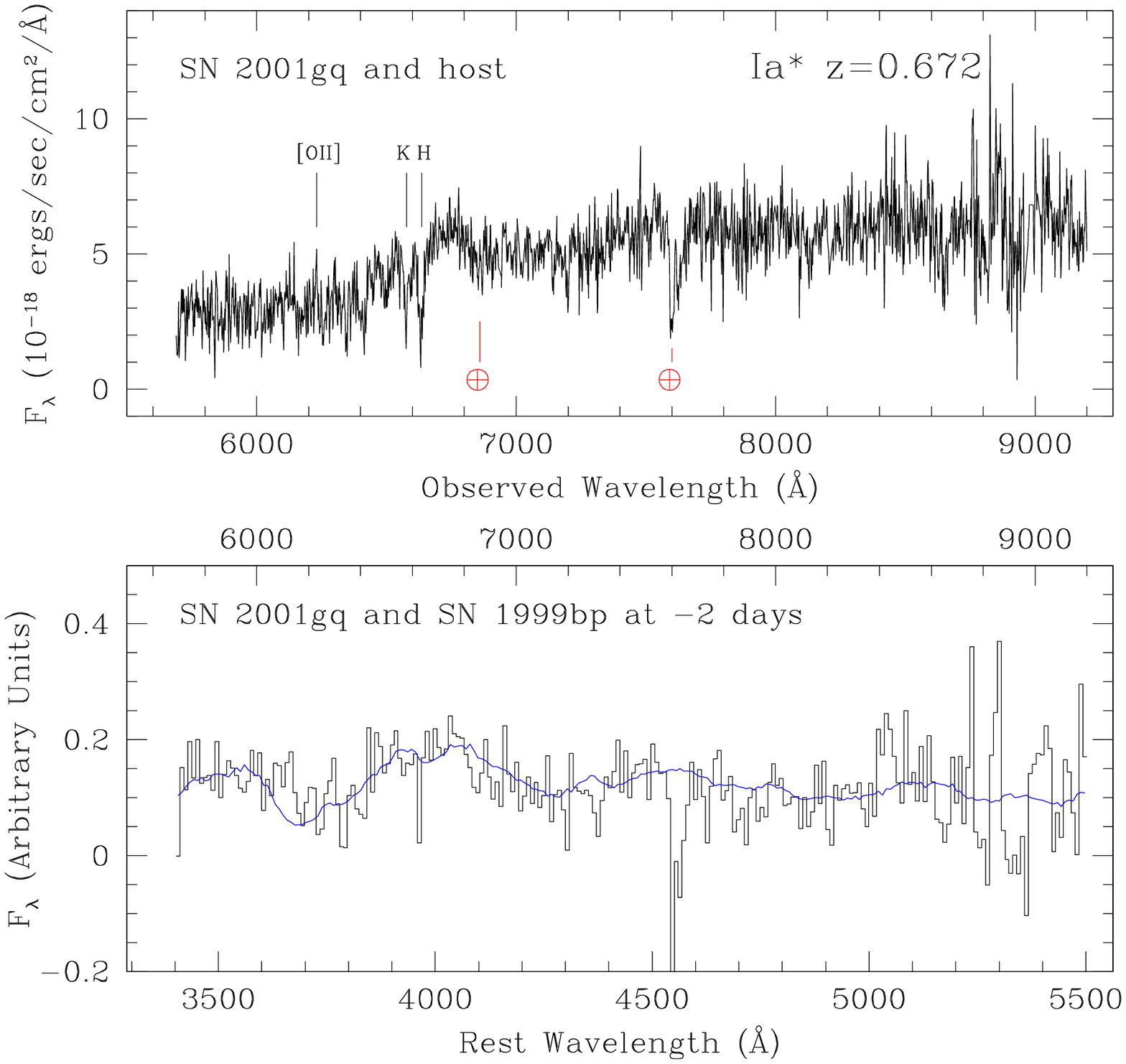}
\caption{{\bf Upper panel:} The unbinned LRIS spectrum of \sngq, a
  probable \snia at $z=0.672$. {\bf Lower Panel:} The rebinned host
  galaxy subtracted spectrum and SN~1999bp at $-2$ days. %
  \label{fig:01gq}}
\end{figure*}

\clearpage

%\clearpage
%
%\begin{figure}
%\centering
%\resizebox{\hsize}{!}{\includegraphics{figures/SN01gy2.eps}}
%\caption{{\bf Upper panel:} The unbinned \fors spectrum of \sngy, a
%  secure \snia at $z=0.511$. The Si~II feature at 4000\ang\ is clearly
%  detected. {\bf Lower Panel:} The rebinned host galaxy subtracted
%  spectrum (plotted in black) and SN~1990N at $-7$ days. %
%  \label{fig:01gy}}
%\end{figure}
%
%\clearpage
%
%\begin{figure}
%\centering
%\resizebox{\hsize}{!}{\includegraphics{figures/SN01go2.eps}}
%\caption{{\bf Upper panel:} The \fors spectrum of \sngo, a secure
%  \snia at $z=0.552$. {\bf Lower Panel:} The rebinned host galaxy
%  subtracted spectrum and SN~1992A at +5 days. %
%  \label{fig:01go}}
%\end{figure}

\section{Light curve data}
Tabulated below are light curve data for the six \sne presented in
this paper. The data are listed in chronological order for one \sn at a
time, and correspond to data points plotted in
Figure~\ref{fig:lightcurves}.

All photometry is given as the measured flux, $f$, in the instrumental
system, \ie no color correction ($S$-correction) to a standard system
has been applied. The instrumental Vega magnitude may be calculated
using the standard formula: $m=-2.5\log f + \mathrm{ZP}$, where
$\mathrm{ZP}$ is the associated zeropoint. For the \hst data these are
the updated values from \citet{2000PASP..112.1397D} given on Andrew
Dolphin's webpage%
\footnote{\url{\texttt{http://purcell.as.arizona.edu/wfpc2\_calib/}}}. %
The zeropoints have then been converted to $e/s$ by using a gain of
$7.12$. The zeropoints for the ground-based data on the other hand are
always given for the photometric reference image, and therefore
include exposure time, gain, and for the \psf fitted data, the
aperture correction. Note that there are correlated errors between the
data points that come from the same light curve build, since the same
host galaxy model and zeropoint has been used. This is also reflected
by small offsets between different data sets in
Figure~\ref{fig:lightcurves}. The covariances are taken into account
for the light curve fitting and the full covariance matrices used will
be available from the \scp Web site.

The instruments used are as given below. \mosaic is the multichip
imager on the \ctio 4~m telescope. \cfhk is the multichip imager on
the 3.6~m \cfht telescope on Mauna Kea in Hawaii. \suprimecam is the
wide-field imager on the 8.2~m \subaru telescope. \susi is the imager on
the NTT 3.6~m telescope at ESO. \fors is the imager on the 8~m UT2
(Kueyen) at \vlt, ESO. \isaac and \niri are the near-infrared imagers
on the 8~m UT1 (Antu) at \vlt, ESO and Gemini North telescopes
respectively. Finally, \wfpc indicates data obtained with the
Planetary Camera CCD and and \acs is the Advanced Camera for Surveys,
both on \hst.

%\begin{longtable}{l r c r c r@{\ }r r c}
%  \hline\hline
%  \multicolumn{1}{c}{\sn} &
%  \multicolumn{1}{c}{MJD} &  
%  \multicolumn{1}{c}{Inst.} &
%  \multicolumn{1}{c}{Exp.} &
%  \multicolumn{1}{c}{Band} &  
%  \multicolumn{2}{c}{Flux} &  
%  \multicolumn{1}{c}{$ZP$} &
%  \multicolumn{1}{c}{I.Q.}\\
%  \hline
%  \endfirsthead
%  \hline
%  \multicolumn{1}{c}{\sn} &
%  \multicolumn{1}{c}{MJD} & 
%  \multicolumn{1}{c}{Inst.} &
%  \multicolumn{1}{c}{Exp.} &
%  \multicolumn{1}{c}{Band} &     
%  \multicolumn{2}{c}{Flux} &
%  \multicolumn{1}{c}{$ZP$} &
%  \multicolumn{1}{c}{I.Q.}\\
%  \hline  
%  \endhead
%  \hline\multicolumn{7}{r}{\it continues on next page}
%  \endfoot
%  \hline\hline
%  \endlastfoot
%  % \\[0.5ex]
\begin{deluxetable}{l r c r c r@{\ }r r c}
  \tabletypesize{\small}
  \tablecaption{%
    Follow-up photometry of the six \sne presented in this paper. The
    columns list the \sn name, the weighted average MJD of the
    observation, exposure time, filter pass band, measured flux,
    instrumental ZP, and image quality in arc seconds. The ZPs are in
    the Landolt systems, and the instrumental magnitude, $m$, is
    obtained as $m = -2.5\log_{10} f + \mathrm{ZP}$, where $f$ can be obtained
    from the Flux column.
    \label{tb:photometry}}
  \tablehead{
    \colhead{\sn} &
    \colhead{MJD} &
    \colhead{Instrument} &
    \colhead{Exp.} &
    \colhead{Band} &  
    \multicolumn{2}{c}{Flux} &  
    \colhead{$\mathrm{ZP}$} &
    \colhead{I.Q.}}
  \startdata
  %% automatically generated by 'lightcurves.pl'
%%
\sncw & $52024.43$ & \suprimecam & 3600 & $i$ & $1.6\textrm{E}+00$ & ($5.0\textrm{E}-01$) & $27.30$ ($0.03$) & $1.05$\\
\sncw & $52049.50$ & \suprimecam & 3240 & $i$ & $1.3\textrm{E}+01$ & ($4.6\textrm{E}-01$) & $27.30$ ($0.03$) & $1.47$\\
\sncw & $52079.88$ & \wfpc & 2300 & F814W & $8.3\textrm{E}-01$ & ($4.3\textrm{E}-02$) & $23.72$ ($0.02$) & $0.10$\\
\sncw & $52079.94$ & \wfpc & 2400 & F850LP & $2.1\textrm{E}-01$ & ($4.5\textrm{E}-02$) & $22.03$ ($0.02$) & $0.10$\\
\sncw & $52088.77$ & \wfpc & 2300 & F814W & $5.6\textrm{E}-01$ & ($4.2\textrm{E}-02$) & $23.72$ ($0.02$) & $0.10$\\
\sncw & $52098.81$ & \wfpc & 2300 & F814W & $3.4\textrm{E}-01$ & ($4.1\textrm{E}-02$) & $23.72$ ($0.02$) & $0.10$\\
\sncw & $52108.91$ & \wfpc & 2300 & F814W & $1.8\textrm{E}-01$ & ($3.7\textrm{E}-02$) & $23.72$ ($0.02$) & $0.10$\\
\sncw & $52118.81$ & \wfpc & 4400 & F814W & $1.4\textrm{E}-01$ & ($2.9\textrm{E}-02$) & $23.72$ ($0.02$) & $0.10$\\
\hline
\sngn & $51998.47$ & \cfhk & 9240 & $I$ & $-2.3\textrm{E}+02$ & ($4.0\textrm{E}+02$) & $33.08$ ($0.03$) & $0.80$\\
\sngn & $52015.50$ & \cfhk & 9900 & $I$ & $2.5\textrm{E}+03$ & ($4.5\textrm{E}+02$) & $33.08$ ($0.03$) & $0.70$\\

  \enddata
\tablecomments{This table is published in its entirety in the
  electronic edition of the Astrophysical Journal. A portion is shown
  here for guidance regarding its form and content.}
\end{deluxetable}

\begin{deluxetable}{l l r r r r c c}
  \tabletypesize{\small}
  \tablecaption{%
    All \sne included in this compilation. The
    columns represent the \sn name, redshift, peak $B$-band magnitude,
    \saltii parameters $x_1$ and $c$, distance modulus, sample and
    failed cuts (if any).
    \label{tb:unionsne}}
  \tablehead{
    \colhead{\sn} &
    \colhead{$z_{\mathrm{CMB}}$} &
    \colhead{$m^{\mathrm{max}}_B$} &
    \colhead{$x_1$} &
    \colhead{$c$} &
    \colhead{$\mu$} &
    \colhead{sample} &
    \colhead{cut}}
  \startdata
  1993ah & 0.0285 & 16.86(0.19) & -2.26(0.93) & 0.23(0.09) & 34.61(0.23) & 1 & \nodata\\ 
1993ag & 0.0500 & 17.79(0.05) & -1.09(0.24) & 0.12(0.02) & 35.95(0.17) & 1 & \nodata\\ 
1993o & 0.0529 & 17.60(0.05) & -1.03(0.14) & -0.01(0.02) & 36.09(0.16) & 1 & \nodata\\ 
1993b & 0.0701 & 18.43(0.04) & -0.53(0.21) & 0.09(0.02) & 36.71(0.16) & 1 & \nodata\\ 
1992bs & 0.0627 & 18.25(0.05) & -0.27(0.23) & 0.02(0.02) & 36.75(0.16) & 1 & \nodata\\ 
1992br & 0.0876 & 19.19(0.11) & -2.97(0.38) & -0.04(0.07) & 37.50(0.19) & 1 & \nodata\\ 
1992bp & 0.0786 & 18.27(0.04) & -1.27(0.20) & -0.02(0.02) & 36.76(0.16) & 1 & \nodata\\ 
1992bo & 0.0172 & 15.75(0.13) & -2.68(0.18) & 0.03(0.02) & 33.93(0.20) & 1 & \nodata\\ 
1992bl & 0.0422 & 17.30(0.08) & -1.95(0.24) & 0.02(0.04) & 35.61(0.17) & 1 & \nodata\\ 
1992bh & 0.0453 & 17.58(0.05) & -0.02(0.25) & 0.10(0.02) & 35.91(0.17) & 1 & \nodata\\ 
1992bg & 0.0365 & 16.71(0.09) & -0.66(0.22) & 0.01(0.03) & 35.18(0.17) & 1 & \nodata\\ 
1992bc & 0.0196 & 15.07(0.11) & 0.51(0.10) & -0.05(0.01) & 33.86(0.18) & 1 & \nodata\\ 
1992aq & 0.1009 & 19.27(0.05) & -1.36(0.43) & -0.02(0.03) & 37.73(0.17) & 1 & \nodata\\ 
1992al & 0.0135 & 14.44(0.16) & -0.41(0.13) & -0.05(0.01) & 33.10(0.22) & 1 & z\\ 
1992ag & 0.0273 & 16.26(0.09) & -0.13(0.21) & 0.19(0.02) & 34.35(0.18) & 1 & \nodata\\ 
1992ae & 0.0746 & 18.39(0.04) & -0.68(0.21) & 0.01(0.03) & 36.86(0.16) & 1 & \nodata\\ 
1992p & 0.0265 & 16.03(0.09) & 0.70(0.58) & -0.02(0.02) & 34.75(0.19) & 1 & \nodata\\ 
1990af & 0.0499 & 17.75(0.05) & -2.65(0.26) & 0.07(0.02) & 35.84(0.16) & 1 & \nodata\\ 
1990o & 0.0306 & 16.19(0.08) & 0.39(0.29) & -0.00(0.03) & 34.82(0.17) & 1 & \nodata\\ 
1992bk & 0.0589 & 18.04(0.13) & -2.11(0.32) & -0.04(0.08) & 36.35(0.17) & 1 & p\\ 
1992au & 0.0603 & 18.12(0.11) & -1.80(0.53) & 0.01(0.09) & 36.33(0.17) & 1 & p\\ 
1992j & 0.0461 & 17.70(0.09) & -1.86(0.27) & 0.15(0.05) & 35.55(0.16) & 1 & p\\ 
1991ag & 0.0139 & 14.37(0.17) & 0.77(0.22) & 0.02(0.03) & 32.88(0.22) & 1 & p,z\\ 
1991u & 0.0324 & 16.40(0.09) & 0.23(0.22) & 0.11(0.03) & 34.62(0.17) & 1 & p\\ 
1991s & 0.0561 & 17.71(0.07) & 0.24(0.23) & 0.02(0.03) & 36.17(0.15) & 1 & p\\ 
1990y & 0.0387 & 17.38(0.08) & -0.33(0.26) & 0.25(0.03) & 35.17(0.17) & 1 & p\\ 
1990t & 0.0397 & 17.18(0.07) & -0.59(0.16) & 0.08(0.03) & 35.38(0.16) & 1 & p\\ 
2001cz & 0.0163 & 15.02(0.14) & 0.04(0.14) & 0.12(0.01) & 33.31(0.14) & 2 & \nodata\\ 
2001cn & 0.0154 & 15.26(0.14) & -0.79(0.12) & 0.22(0.01) & 33.20(0.15) & 2 & \nodata\\ 
2001bt & 0.0144 & 15.26(0.15) & -1.14(0.11) & 0.26(0.01) & 33.06(0.16) & 2 & z\\ 
2001ba & 0.0305 & 16.19(0.07) & 0.00(0.14) & -0.04(0.01) & 34.87(0.09) & 2 & \nodata\\ 
2000ca & 0.0245 & 15.53(0.09) & 0.35(0.20) & -0.07(0.01) & 34.33(0.10) & 2 & \nodata\\ 
2000bh & 0.0240 & 15.92(0.10) & -0.55(0.17) & 0.08(0.03) & 34.24(0.11) & 2 & \nodata\\ 
1999gp & 0.0260 & 16.00(0.09) & 1.64(0.21) & 0.06(0.02) & 34.63(0.11) & 2 & \nodata\\ 
1999dk & 0.0139 & 14.81(0.16) & 0.29(0.27) & 0.11(0.02) & 33.14(0.17) & 2 & z\\ 
1999cp & 0.0104 & 13.93(0.21) & -0.05(0.25) & -0.01(0.02) & 32.54(0.22) & 2 & z\\ 
1999cl & 0.0087 & 14.81(0.25) & -0.36(0.20) & 1.20(0.01) & 30.35(0.26) & 2 & z\\ 
2000ce & 0.0165 & 17.02(0.14) & 0.65(0.24) & 0.61(0.03) & 34.03(0.15) & 2 & p\\ 

  \enddata
\tablecomments{This table is published in its entirety in the
  electronic edition of the Astrophysical Journal. A portion is shown
  here for guidance regarding its form and content.}
\tablenotetext{p}{first phase greater than six days after $B$-band maximum}
\tablenotetext{z}{redshift $< 0.015$}
\tablenotetext{o}{outlier}
\tablenotetext{d}{number of observations $<5$}
\tablenotetext{f}{bad light curve fit}
%\tablenotetext{1}{\citet{1996AJ....112.2398H}} % Hamuy et al. (1996)
%\tablenotetext{2}{\citet{2005AJ....130.2453K}} % Krisciunas et al. (2005)
%\tablenotetext{3}{\citet{1999AJ....117..707R}} % Riess et al. (1999)
%\tablenotetext{4}{\citet{2006AJ....131..527J}} % Jha et al. (2006)
%\tablenotetext{5}{\citet{2008ApJ...686..749K}} % Kowalski et al. (2008)
%\tablenotetext{6}{\citet{2009ApJ...700..331H}} % Hicken et al. (2009)
%\tablenotetext{7}{\citet{2008AJ....136.2306H}} % Holtzman et al. (2009)
%\tablenotetext{8}{\citet[+ HZT]{1998ApJ...507...46S}} % Riess et al. (1998) + HZT
%\tablenotetext{9}{\citet{1999ApJ...517..565P}} % Perlmutter et al. (1999)
%\tablenotetext{10}{\citet{2004ApJ...602..571B}} % Barris et al. (2003)
%\tablenotetext{11}{\citet{2008A&A...486..375A}} % Amanullah et al. (2008)
%\tablenotetext{12}{\citet{2003ApJ...598..102K}} % Knop et al. (2003)
%\tablenotetext{13}{\citet{2006AA...447...31A}} % Astier et al. (2006)
%\tablenotetext{14}{\citet{2007ApJ...666..674M}} % Miknaitis et al. (2007)
%\tablenotetext{15}{\citet{2003ApJ...594....1T}} % Tonry et al. (2003)
%\tablenotetext{16}{\citet{2007ApJ...659...98R}} % Riess et al. (2007)
%\tablenotetext{17}{This Paper}
\tablenotetext{1}{\citet{1996AJ....112.2398H},
$^{2}$~\citet{2005AJ....130.2453K}, % Krisciunas et al. (2005)
$^{3}$~\citet{1999AJ....117..707R}, % Riess et al. (1999)
$^{4}$~\citet{2006AJ....131..527J}, % Jha et al. (2006)
$^{5}$~\citet{2008ApJ...686..749K}, % Kowalski et al. (2008)
$^{6}$~\citet{2009ApJ...700..331H}, % Hicken et al. (2009)
$^{7}$~\citet{2008AJ....136.2306H}, % Holtzman et al. (2009)
$^{8}$~\citet[+ HZT]{1998ApJ...507...46S}, % Riess et al. (1998) + HZT
$^{9}$~\citet{1999ApJ...517..565P}, % Perlmutter et al. (1999)
$^{10}$~\citet{2004ApJ...602..571B}, % Barris et al. (2003)
$^{11}$~\citet{2008A&A...486..375A}, % Amanullah et al. (2008)
$^{12}$~\citet{2003ApJ...598..102K}, % Knop et al. (2003)
$^{13}$~\citet{2006AA...447...31A}, % Astier et al. (2006)
$^{14}$~\citet{2007ApJ...666..674M}, % Miknaitis et al. (2007)
$^{15}$~\citet{2003ApJ...594....1T}, % Tonry et al. (2003)
$^{16}$~\citet{2007ApJ...659...98R}, % Riess et al. (2007)
$^{17}$~This Paper}
\end{deluxetable}

\section{Nuisance Parameters as Covariances}
\label{sec:statistics}
\newcommand{\bF}{\boldsymbol{F}}
\newcommand{\by}{\boldsymbol{y}}
\newcommand{\bth}{\boldsymbol{\theta}}

Suppose that observations $\by$ are modeled by $\bF(\bth)$, where some
of the parameters enter into the model linearly ($\bth^L$) while some
enter non-linearly ($\bth^N$). The $\chi^2$ can be written as
\begin{equation}
\chi^2(\bth) = (\by - (\bF_N(\bth^N) + H\bth^L))^T V^{-1} (\by - (\bF_N(\bth^N) + H\bth^L))
\end{equation}
where $V$ is the covariance matrix of the observations, and $H$ is the
Jacobian matrix of the model with respect to the $\bth^L$. Taking the
derivative of the $\chi^2$ and setting it to zero gives the analytic
formula for the best-fit $\bth^L$
\begin{equation}
\hat{\bth^L} = (H^T V^{-1} H)^{-1} H^T V^{-1} (\by - \bF_N(\bth^N)) \equiv D(\by - \bF_N(\bth^N))\;.
\end{equation}
The likelihoods for the $\bth^N$ can be found from this restricted
parameter space where $\bth^L = \hat{\bth^L}$. Thus, the restricted
$\chi^2$ is given by:
\begin{eqnarray}
  \chi^2  & = & ((I - HD)(\by - \bF_N(\bth^N)))^T V^{-1}(I - HD)(\by - \bF_N(\bth^N)) \\
  & \equiv & \boxed{(\by - \bF_N(\bth^N))^T U^{-1}(\by - \bF_N(\bth^N))} \label{eq:newchi2}
\end{eqnarray}
with $U^{-1}$ given by 
\begin{eqnarray}
U^{-1} & = & (I - HD)^T V^{-1} (I - HD)\\
 & = & \boxed{V^{-1} - V^{-1} H (H^T V^{-1} H)^{-1} H^T V^{-1}}\;.
\end{eqnarray}

In this way, all of the $\bth^L$ do not have to be explicitly included in the $\chi^2$, as long as the weight matrix, $U^{-1}$, is updated appropriately. Note that $U^{-1}$ may depend on $\bth^N$.

\newcommand{\xt}{x_1^{\mathrm{true}}}
\newcommand{\ct}{c^{\mathrm{true}}}
\subsection{Minimization over $\xt$ and $\ct$}
When errors in the independent variable are present, the true values
must be solved for as part of the fit (see discussion in K08). For one
supernova, $\by= (m_B, x_1, c)$, while the model is given by
\begin{equation}
  (M_B + \mu(z, \mathrm{cosmology}) - (\alpha \xt - \beta \ct), \xt, \ct)\;.
\end{equation}
Using $\bth^L = (\xt, \ct)$ ($M_B, \alpha$ and $\beta$ are global
parameters and cannot be handled one supernova at a time), we have
\begin{equation}
  H = \left( \begin{array}{ccc}
      -\alpha & \beta \\
      1 & 0\\
      0 & 1
    \end{array} \right)\;.
\end{equation}
The new $\chi^2$ for this specific example (equation
\eqref{eq:newchi2}) is given by equations \eqref{eq:cosmochi2} and
\eqref{eq:lcchi2}.

\subsection{Minimization over Systematic Errors}
\label{sec:minsyserr}
\newcommand{\Vmu}{V_{\mu}}
\newcommand{\Hmu}{H_{\mu}}
\newcommand{\der}[2]{\frac{\partial #1}{\partial #2}}
\newcommand{\Dzp}{\Delta \mathrm{ZP}}
\newcommand{\Vzp}{V_{\Dzp}}
Suppose that we have two kinds of measurements: supernova distance
measurements ($N$ supernovae), and zeropoint measurements ($M$
zeropoints). The true value of each zeropoint is given by
$\mathrm{ZP_{true}} = \mathrm{ZP_{observed}} + \Dzp$. Because the
uncertainties on the supernova distances are unrelated to the
uncertainties on the zeropoints,
\[
 V = \left( \begin{array}{cc}
     \Vmu &\\
    & \Vzp
  \end{array} \right)\,,
\]
where $\Vmu$ is an $N\times N$ block and $\Vzp$ is an $M\times M$
block. Similarly, $\by$ can be split into distances and zeropoints
$\by = (\by_{\mu}, \by_{\Dzp} = \boldsymbol{0})$. The model for each
supernova distance is given by
\begin{equation}
  M_B + \mu(z, \mathrm{cosmology}) - \sum_{\lambda}
  \der{(m_B + \alpha x_1 - \beta c)}{\mathrm{ZP}_{\lambda}}\Dzp_{\lambda}\;.
\end{equation}
It is derived in the same way as in the previous section, with the
substitutions $x_1 \rightarrow x_1 + \sum_{\lambda}
\der{x_1}{\mathrm{ZP}_{\lambda}} \Dzp_{\lambda}$ and $c \rightarrow c
+ \sum_{\lambda} \der{c}{\mathrm{ZP}_{\lambda}} \Dzp_{\lambda}$. The
model for the true offset of each zeropoint is simply $\Dzp$. Thus,
\[
H = \left( \begin{array}{c}
    \Hmu\\
    I
\end{array} \right)\,.
\]
The upper block of $U^{-1}$ is given by
\begin{equation}
  U^{-1}_{\mathrm{upper\;block}} = \Vmu^{-1} - \Vmu^{-1}\Hmu(\Vzp^{-1}
  + \Hmu^{T}\Vmu^{-1} \Hmu)^{-1} \Hmu^{T}\Vmu^{-1}\; ;
\end{equation}
none of the other blocks enter into the $\chi^2$. Inverting this
block (see \cite{Hager75570}) gives:
\begin{equation}
  U = \Vmu + \Hmu \Vzp \Hmu^{T}
\label{eq:simplecov}
\end{equation}
Equation~\eqref{eq:simplecov} gives us the terms we must add to the
supernova distance modulus errors to correctly take the zeropoints
into account.

\end{document}